\documentclass[aps,prd,twocolumn,superscriptaddress,amsmath,preprintnumbers,amssymb,nofootinbib,longbibliography]{revtex4-2}

\usepackage{graphicx}
\usepackage[dvipsnames]{xcolor}
\usepackage{bbm}
\usepackage[colorlinks]{hyperref}
\usepackage[normalem]{ulem}

\usepackage{physics}
\usepackage{xspace}
\usepackage{tabstackengine}
\usepackage{xparse}

\usepackage{siunitx}
\stackMath

\usepackage{amsmath,amstext}
\usepackage[T1]{fontenc}
\usepackage[figure,figure*]{hypcap}
\usepackage{comment}
\usepackage{graphicx}
\usepackage{natbib}
\usepackage{epsfig}
\usepackage{epstopdf}
\usepackage{enumerate}
\usepackage{tabularx}
\usepackage{multirow}
\usepackage{booktabs}
\usepackage{color}
\usepackage[normalem]{ulem}  
\usepackage{xspace}
\usepackage{physics}
\usepackage{tabstackengine}
\usepackage{xparse}
\usepackage{aas_macros}

\hypersetup{linkcolor=BrickRed,citecolor=Green,
filecolor=Mulberry,
urlcolor=NavyBlue,
menucolor=BrickRed,
runcolor=Mulberry
}

\allowdisplaybreaks

\graphicspath{{./}{Figures_final}}

\newcommand{\pd}[3][]{\frac{\partial^{#1} #2}{\partial #3^{#1}}}
\newcommand{\msun}{\ensuremath{{\rm M}_{\odot}}\xspace}

\newcommand{\rot}{\mathrm{rot}}
\newcommand{\kin}{\mathrm{kin}}
\newcommand{\turb}{\mathrm{TKE}}

\usepackage{tikz}
\usetikzlibrary{tikzmark}
\usetikzlibrary{positioning}
\usepackage{float}

\definecolor{lime}{HTML}{A6CE39}
\DeclareRobustCommand{\orcidicon}{\hspace{-1mm}
	\begin{tikzpicture}
	\draw[lime, fill=lime] (0,0) 
	circle [radius=0.16] 
	node[white] {{\fontfamily{qag}\selectfont \tiny \,ID}};
	\draw[white, fill=white] (-0.0525,0.095) 
	circle [radius=0.007];
	\end{tikzpicture}
	\hspace{-3mm}
}

\foreach \x in {A, ..., Z}{\expandafter\xdef\csname orcid\x\endcsname{\noexpand\href{https://orcid.org/\csname orcidauthor\x\endcsname}
			{\noexpand\orcidicon}}
}

\begin{document}

\title{The Interplay of Magnetic Fields, Turbulence and Vorticity in Core-Collapse Supernovae}

\newcommand{\orcidauthorA}{0009-0000-5846-5984}
\author{David Calvert\orcidA{}} 
\affiliation{Department of Physics, North Carolina State University, Raleigh, NC 27606 USA}

\newcommand{\orcidauthorB}{0000-0002-4099-2978}
\author{Michael Redle\orcidB{}}
\affiliation{Department of Mathematics, North Carolina State University, Raleigh, NC 27606 USA}
\affiliation{Institute of Mathematics, Clausthal University of Technology, 38678 Clausthal-Zellerfeld, Germany}

\newcommand{\orcidauthorC}{0000-0002-3211-3427}
\author{Bibek Gautam\orcidC{}}
\affiliation{Department of Physics, North Carolina State University, Raleigh, NC 27606 USA}

\newcommand{\orcidauthorD}{0000-0002-5667-7577}
\author{Charles J.\ Stapleford\orcidD{}}
\affiliation{Oak Ridge National Laboratory, Oak Ridge, TN 37830 USA}

\newcommand{\orcidauthorE}{0000-0003-0191-2477}
\author{Carla Fr\"{o}hlich\orcidE{}}
\email{cfrohli@ncsu.edu}
\affiliation{Department of Physics, North Carolina State University, Raleigh, NC 27606 USA}

\newcommand{\orcidauthorF}{0000-0002-3502-3830}
\author{James P.\ Kneller\orcidF{}}
\email{jpknelle@ncsu.edu}
\affiliation{Department of Physics, North Carolina State University, Raleigh, NC 27606 USA}

\newcommand{\orcidauthorG}{0000-0001-8523-9528}
\author{Matthias Liebend\"{o}rfer\orcidG{}}
\affiliation{Department of Physics, University of Basel, Basel, Switzerland}

\date{\today}

\begin{abstract}
The convective turbulent motion of the fluid below the shock in a core-collapse supernova stretches and amplifies the magnetic field of the progenitor star. The energy contained in the field is sourced from the work done by the fluid on the field which comes at the expense of the fluid's kinetic and internal energy. In addition to the energy exchange with the fluid, the magnetic field also has a back-reaction upon the fluid via a contribution to the baroclinic vector and thus affects the fluid vorticity. In this paper we explore the interaction of the magnetic field with the fluid in core-collapse supernovae using the \texttt{ELEPHANT} code with solar-metallicity stars of 15 and 20 \msun zero-age main sequence mass and purely toroidal initial magnetic fields of $B_0 = 0, 10^{10}, and 10^{12}$~G. 
We find that the magnetic field in the gain region does not become so large that it alters the global dynamics above the 10\% level in the early post-bounce evolution.
The turbulent kinetic energy of the fluid in the gain region is smaller in simulations with a strong initial magnetic field, but the exact amount of reduction is uncertain due to limitations of the methods for measuring turbulent kinetic energy. 
The structure of the field in the simulations quickly becomes a tangled mass of flux ropes as soon as convection begins, which leads to a large magnetic field contribution to the baroclinic vector that dominates over the hydrodynamic contribution. Although governed by very similar transport equations to the magnetic field, the fluid vorticity and magnetic field are always close to being randomly aligned at every stage of the evolution. The enstrophy, which is seen to be closely associated with the turbulent kinetic energy, is found to be reduced in simulations with a strong initial magnetic field supporting the inference that the magnetic field reduces the amount of turbulence in the fluid. 
\end{abstract}

\maketitle


\section{Introduction}
\label{sec:intro}

At the present time a general consensus has been reached that the means by which most massive stars explode as core-collapse supernovae (CCSNe) is due to the `neutrino heating mechanism' \cite{1985ApJ...295...14B,1995ApJ...450..830B,2025arXiv250214836J}: the stalled shock is revived due to energy deposition by neutrinos into material located in a `gain region' below the shock. While simulations of CCSN in spherical symmetry consistently fail to explode -- except for the lightest progenitors \cite{Fischer_2010,2010PhRvL.104y1101H} -- via the neutrino heating mechanism \cite{2001PhRvL..86.1935M,2002A&A...396..361R,2003ApJ...592..434T,2004ApJS..150..263L,2005ApJ...629..922S}, multi-dimensional simulations have achieved unaided explosions \cite{
2007PhR...442...38J,2017hsn..book.1095J,2017RSPTA.37560271C,
Kuroda_2020,2021ApJ...915...28B,2022MNRAS.514.3941N,2022MNRAS.510.4689V,2025PhRvD.111f3042S,2024MNRAS.531.3732S,2024ApJ...962...71W}. This difference indicates that, in addition to the heating by the neutrinos, multi-dimensional effects must aid the explosions. Among these multi-dimensional effects are turbulence and magnetic fields. 

Turbulence in the fluid is expected due to the low viscosity, large shear stress, and vigorous heating. High spatial resolution 3D simulations of CCSNe are now able to observe some portion of the inertial scales of the turbulence \cite{2011ApJ...742...74M,2012ApJ...749..142F,2013ApJ...765..110D,2013PhST..155a4022E,2014ApJ...783..125H,2015ApJ...808...70A,2015ApJ...799....5C,2016ApJ...820...76R,2018JPhG...45e3003R,2020PhyS...95f4005C,2025ApJ...995..109C} although they are still very far from the Kolmogorov microscale. 
Despite many studies, no definitive consensus of the relative importance of turbulence and large-scale non-radial motion in aiding the explosion has been reached
\cite{2015ApJ...799....5C,2015ApJ...801L..24M,2015MNRAS.448.2141M,2016MNRAS.461.3864A,2016ARNPS..66..341J,2017MNRAS.472..491M,2018ApJ...856...22M,2018JPhG...45e3003R,2021MNRAS.502.4125R,2025ApJ...995..109C}.

Many 3D simulations of CCSNe do not regularly include magnetic fields, even though the formation of pulsars and/or magnetars in CCSNe would require them. Long-duration gamma-ray bursts are associated with CCSNe \cite{1999Natur.401..453B,2006ARA&A..44..507W} and the relativistic jets that pierce the outer layers of the star are expected to possess strong magnetic fields. 
Magnetorotationally-driven supernovae are often proposed as alternatives to the neutrino heating mechanism for fast spinning, highly magnetized stars~\cite{Burrows_2007,2012ApJ...750L..22W,Kuroda_2020}. To date the few simulations that have included magnetic fields have usually found that the field aids (but does not dominate) the explosion - for a review see \cite{2024arXiv240318952M}. Obergaulinger and Aloy \cite{2021MNRAS.503.4942O} showed that explosion energies can vary by over an order of magnitude depending upon the strength and morphology of the initial magnetic field. In some cases, they also found the supernova explosion to be caused primarily by the pressure from the magnetic fields rather than the neutrino heating. The inclusion of magnetic fields and rotation in simulations of a $20\;M_{\odot}$ star was found to be the difference between achieving an explosion or not by both Kuroda \emph{et al.} ~\cite{Kuroda_2020} and M{\"u}ller and Varma \cite{2020MNRAS.498L.109M}. 
More recently Matsumoto, Takiwaki and Kotake \cite{2024MNRAS.528L..96M} found that the shocks in their simulations were revived both sooner and with much more energy when they included magnetic fields compared to the unmagnetized simulations \cite{2025MNRAS.536..280N,Matsumoto-2022}. 
Varma \emph{et al} \cite{2023MNRAS.518.3622V} also see earlier explosions in a magnetized simulation of a 16.9 \msun progenitor compared to the weak-field control. 
However, in a set of simulations where the initial field of the progenitor was not as strong, Sykes and M{\"u}ller \cite{2025PhRvD.111f3042S} and Varma and M{\"u}ller \cite{2026MNRAS.tmp..593V} found that the field had little effect upon the dynamics. 

The magnetic field in a CCSN is the highly twisted and amplified field of the progenitor star. The magnetic fields of progenitors are expected to be large due to the action of a small-scale turbulent dynamo that operates during the late burning stages (e.g. oxygen burning) \cite{2023A&A...679A.132L}. Field strengths of $10^{10}\;{\rm G}$ are readily achieved in models. Once the star begins to collapse, the magnetic field will amplify due to compression and stretching even before the fluid passes through the shock. Once inside the shock, the evolution is more dynamic and complicated. The growth of the field energy inside the shock has been attributed to instabilities such as the magnetorotational instability \cite{2003ApJ...584..954A} which is typically unresolved in ``global'' models, by field stretching due to turbulence \cite{1996JFM...306..325B} and/or a Standing Accretion Shock Instability \cite{2010ApJ...713.1219E,Endeve_2012,2013PhST..155a4022E}, and/or a small-scale turbulent dynamo \cite{2020MNRAS.498L.109M,2024MNRAS.528L..96M,2026MNRAS.tmp..593V}. 
However it grows, if the field becomes large enough, it affects the fluid through well-known effects such as magnetic buoyancy~\cite{1966ApJ...145..811P}, and via the field's contribution to the baroclinic vector which generates vorticity---often taken to be the defining characteristic of turbulence \cite{bradshaw,2004iit..book.....T}.

The purpose of our paper is to investigate the interplay of the magnetic field, turbulence and vorticity in a CCSN and attempt to determine whether magnetic fields change the amount of turbulence in the gain region. Our study is a continuation of \cite{2025ApJ...995..109C}. We explore the interplay of magnetic field, turbulence and vorticity by simulating six core-collapse supernovae using the \texttt{ELEPHANT} code~\cite{Cabezon-3Dcomparison,kappeli2011fish,Liebendoerfer.IDSA:2009}. 

Our paper is organized as follows. In section \S\ref{sec:sims}, we describe the code, the progenitors, the initial magnetic field and the initial rotation profiles, and present the general properties of our simulations. 
We then present (in sections \S\ref{sec:Bfield-results} and \S\ref{sec:fluid_energies}) our results which we break down into looking more closely at the magnetic field, at the energy exchanged between the field and fluid, at the turbulent kinetic energy, and at the vorticity and enstrophy. 
We summarize and present our conclusions in section \S\ref{sec:conclusions}.

\section{The Simulations}
\label{sec:sims}

\subsection{The \texttt{ELEPHANT} code} 
\label{sec:code}

The \texttt{ELEPHANT} code, described in detail in ~\cite{Liebendoerfer.IDSA:2009,kappeli2011fish,Cabezon-3Dcomparison}, solves the equations of ideal magnetohydrodynamics (MHD) coupled to neutrino transport as a function of position $\vec{x}$ and time $t$ in a modified Newtonian gravitational potential. 
The ideal MHD equations with gravitational source terms and neutrino source/sink terms are:
\begin{eqnarray}
    \pd{\rho}{t}+ \nabla \cdot (\rho\, \vec{v} ) &=& 0, \label{eq:mass}\\
     \pd{\rho \vec{v}}{t} + \nabla \cdot \left[ \vec{v}\, \rho \vec{v} - \vec{b}\, \vec{b} \right] + \nabla P &=& - \rho\, \nabla \phi 
     + \vec{\mathcal{S}}_{M} ,
     \label{eq:momentum}\\
    \pd{E}{t} + \nabla \cdot \left[(E+P)\,\vec{v} - (\vec{v}\cdot\vec{b}) \,\vec{b}\right] &=& - \rho \,\vec{v} \cdot \nabla \phi
    + \vec{\mathcal{S}}_{E} 
    + \vec{v} \cdot \vec{\mathcal{S}}_{M} 
    , \nonumber \\  & &   \label{eq:energy}\\
    \pd{\vec{b}}{t} - \nabla\times\left(\vec{v}\times \vec{b}\right) &=& \vec{0}. \label{eq:magnetic}
\end{eqnarray}
Here, $\rho$ is the baryonic mass density and $\vec{v}$ is the fluid velocity. 
The quantity $P$ represents the sum of material pressure and `magnetic pressure' (i.e.\ the isotropic component of the Lorentz force, $P_{\mathrm{mag}} =  ( \vec{b} \cdot \vec{b} ) / 2$). 
Similarly, $E$ represent the sum of internal energy density, kinetic energy density, and magnetic field energy density: 
$E = \rho\, e + \rho\, v^2/2 +  ( \vec{b} \cdot \vec{b} ) / 2$. 
The momentum exchange between the fluid and neutrinos is denoted by $\mathcal{S}_{M}$, and $\mathcal{S}_{E}$ denotes the energy exchange via neutrino heating.
They are combined with the advection of the electron fraction $Y_e$, the trapped neutrino fractions $Y_{\nu}^{\mathrm{t}}$, and a multiple of the neutrino entropies $(\rho Z_{\nu}^{\mathrm{t}})^{3/4}$:
\begin{eqnarray}
    \pd{}{t}(\rho\, Y_e)+ \nabla \cdot (\rho \,Y_e\, \vec{v} ) &=& \mathcal{S}_{e^-} ,
    \label{eq:electron_fraction}\\
    \pd{}{t}(\rho\, Y_{\nu}^{\mathrm{t}})+ \nabla \cdot (\rho \,Y_{\nu}^{\mathrm{t}}\, \vec{v} ) &=& \mathcal{S}_{\nu} ,\label{eq:neutrino_fraction}\\
    \pd{}{t}((\rho Z_{\nu}^{\mathrm{t}})^{3/4})+ \nabla \cdot ((\rho Z_{\nu}^{\mathrm{t}})^{3/4}\, \vec{v} ) &=& \mathcal{S}_{s} .\label{eq:neutrino_entropy}
\end{eqnarray}
where $Z_{\nu}^{\mathrm{t}}$ is the neutrino energy per baryon and $\nu= \{\nu_e$, $\bar{\nu}_e\}$. 
The $\mathcal{S}_{s}$ is the neutrino entropy source term, which is a function of $\mathcal{S}_{E}$,
$\mathcal{S}_{M}$, and 
$\mathcal{S}_{\nu}$. 
The quantity $\mathcal{S}_{e^-}$ represents the net change in electron number density due to neutrino emission / absorption. 
The mu/tau neutrinos/antineutrinos are treated by a simple leakage scheme \cite{2003MNRAS.342..673R}.
The vector $\vec{b}$ is the reduced magnetic field, which obeys
\begin{equation}
    \nabla\cdot \vec{b} = 0,\label{eq:divfree}
\end{equation}
and is related to the true magnetic field $\vec{B}$ in Gaussian units by $\vec{B}=\sqrt{4\pi}\, \vec{b}$. 
We close the system of equations with the Lattimer-Swesty \cite{lattimer1991generalized} equation of state (EOS) with an incompressibility parameter $K=220$~MeV. 
The Poisson equation governing the gravitational potential $\phi$ is:
\begin{equation}
    \nabla^2 \phi = 4\,\pi\, G\, \rho, \label{eq:gravity} 
\end{equation}
where  $G$ is the gravitational constant as usual. 
The gravitational potential is modified from its purely Newtonian formulation to include general relativistic corrections following \cite{marek2006exploring}.

The equations are solved using operator splitting into the MHD equations and the neutrino transport equations. 
In the MHD step, the neutrino source terms ($\mathcal{S}_{M}$, $\mathcal{S}_{E}$, $\mathcal{S}_{s}$, and $\mathcal{S}_{e^-}$) are set to zero. 
These terms are computed within the neutrino transport scheme, for which we use the Isotropic Diffusion Source Approximation (IDSA) from \cite{Liebendoerfer.IDSA:2009}, and used to update the hydrodynamic quantities before the next MHD step is taken.
In addition, the staggered constrained transport (CT) method proposed by \cite{pen2003free} is applied to ensure that the constraint $\nabla \cdot \vec{b} = 0$ in \eqref{eq:divfree} is fulfilled (within numerical error) for all time (see \cite{kappeli2011fish}). 
The MHD equations \eqref{eq:mass}-\eqref{eq:magnetic} and \eqref{eq:electron_fraction} are solved using the second-order relaxation scheme of \cite{jin1995relaxation}. 
To enforce that the scheme is total variation diminishing (a nonlinear constraint to ensure stability), \texttt{ELEPHANT} implements the minmod limiter in supersonic flow regimes and the van Leer limiter in subsonic regimes \cite{kappeli2011fish}. 
For the time integration, a second-order predictor-corrector method is used.

The computational domain of \texttt{ELEPHANT} consists of a 3D Cartesian cuboid for the innermost region of the supernova surrounded by a 1D spherically-symmetric domain for the outer layers of the progenitor star. 
More details on the implementation of the hydro equations in \texttt{ELEPHANT} on a 3D uniform Cartesian mesh, together with test cases, can be found in \citet{kappeli2011fish}. 
The larger, spherically-symmetric domain surrounding the 3D cuboid is evolved on an adaptive grid using the \texttt{AGILE}-IDSA solver \cite{Liebendoerfer.IDSA:2009}.
The spherically-symmetric solution is used primarily for the properties of the material falling through the boundary of the 3D mesh, but since it extends into the inner part of the Cartesian grid, it also provides a second solution of the equations that is particularly useful within the proto-neutron star (PNS) where it serves as a check of the entropy evolution.
The 1D solution also gives us the ability to enlarge the 3D Cartesian cuboid when the shock expands and reaches the boundary of the 3D cuboid. 
With this setup of the computational domain, the \texttt{ELEPHANT} code has high 3D spatial fidelity in the region below the shock without the burden of a large number of grid zones where the solution is spherically symmetric. Hence, it is ideally suited for this study of the shocked material in the post-bounce phase.


\subsection{The Initial Models} \label{sec:progenitor}

For our study we adopt two solar-metallicity progenitor stars of 15 and 20 \msun zero-age main sequence mass from \cite{woosley2007nucleosynthesis}, also used in our previous study \cite{2025ApJ...995..109C}. 
We performed three simulations per progenitor model (two with a magnetic field and rotation plus a control simulation without magnetic field and without rotation) for a total of six simulations. 
We will refer to the simulations by a codename composed of the progenitor mass and magnetic field, sXX-TYY, where `XX' is the mass of the progenitor model and `YY' is the log base ten of the field strength parameter $B_0$ in Gauss, i.e. $B_0 = 10^{\mathrm{YY}}$ Gauss.
For the magnetic field-free cases, the name of the run will instead be ``sXX-Control''. 
The nomenclature is outlined in Table \ref{tab:runs}, which also includes the final simulation time of each run.
The progenitors do not have any rotation nor magnetic field included, and thus we must impose both of these properties onto the progenitor.

\begin{table}
\centering
\begin{tabular}{c c c c c c} 
 \hline \hline
 Run Name & Progenitor & $B_0$ & $\Omega_{\rot}$ & $r_{\rot}$ & End time\\
   & & [G] & [rad/s] & [km] & [ms] \\
 \hline
  s15-control & s15 & 0 & 0 & 250 & 192.76 \\ 
 s15-T10 & s15 & $10^{10}$ & 0.3 & 250 & 195.44 \\
 s15-T12 & s15 & $10^{12}$ & 0.3 & 250 & 183.43 \\
 \hline
 s20-control & s20 & 0 & 0 & 250 & 216.10 \\ 
 s20-T10 & s20 & $10^{10}$ & 0.3 & 250 & 276.99 \\
 s20-T12 & s20 & $10^{12}$ & 0.3 & 250 & 225.86 \\
 \hline
\end{tabular}
\caption{ Initial parameters and final simulation times
\label{tab:runs}
}
\end{table}

The magnetic field we insert is a very simple toroidal field as expected from stellar evolution due to differential winding \cite{Heger_2005}.
We start from the vector potential $\vec{A}_T = B_0\,r\,\sin\theta\,{\bf \hat{k}}$, 
where $B_0$ is a parameter that sets the magnetic field strength. 
Similar to \cite{2011arXiv1101.1198O}, this vector potential is then scaled by the square root of the initial mass density profile so as to avoid the magnetic pressure overwhelming the thermal pressure. 
Thus, our initial magnetic field has the form
\begin{equation}
\vec{
B} = \nabla \times \left(\sqrt{\frac{\rho}{\rho_0}}\vec{A}_T\right)
= \sqrt{\frac{\rho}{\rho_0}} \,\left(\nabla \times \vec{A}_T \right) + \frac{\left( \nabla \sqrt{\rho} \right) \times \vec{A}_T}{\sqrt{\rho_0}},
\label{eq:initialB}
\end{equation}
where the reference density $\rho_0$ is set to $\rho_0 = 10^{10}\;{\rm g/cm^3}$. 
Both terms of this initial magnetic field are toroidal in the sense that the field lines are circles in the $xy$-plane. 
The first term is the density-scaled toroidal field as desired; the second is also a toroidal field, but (i) it winds in the opposite direction because the density decreases with $r$; and (ii) it has a strength that is proportional to $\sin\theta$, i.e., it is strongest in the $xy$-plane. 
On a discrete level, the initial magnetic field is implemented as follows. 
For the magnetic field values inside the cuboid, the curl in~\eqref{eq:initialB} is computed on the cell interfaces using the central differences for each directional derivative. 
These derivatives are calculated using the cell center values of $\sqrt{\rho}\,\vec{A}_T$, resulting in an initial magnetic field on a staggered mesh which is guaranteed to be divergence-free. 
For the advected magnetic field at the boundary of the inner cuboid, we construct a field in the boundary cells using the same algorithm used to build the initial field within the Cartesian grid, and then correct the field so that the divergence of the accreting field vanishes.

Our initial toroidal field configuration minimizes the amount of field growth prior to the passage of the fluid through the shock. Thus the growth of the field energy inside the gain region is almost entirely due to the processes interior to the shock as opposed to accretion of field energy through the shock. In other studies of magnetic fields in supernovae, authors have adopted split-monopole fields \cite{2010ApJ...713.1219E,Endeve_2012,2013PhST..155a4022E}, or pure poloidal fields \cite{10.1093/mnras/stu1969,10.1093/mnras/staa3095,Matsumoto-2022,2024MNRAS.528L..96M}. Mixed toroidal/poloidal field configurations were used in Sykes \& M\"uller \cite{2025PhRvD.111f3042S}. 

In addition to a magnetic field, we expect the supernova progenitor to be rotating \cite{2005ApJ...626..350H,deMink_2013,2026A&A...710L...7O}.
For the initial rotation profile of our simulations, we adopt the prescription found in \cite{Cabezon-3Dcomparison} which is described by two parameters $\Omega_{\rot}$ and $r_{\rot}$ to set the initial tangential velocity of each grid zone in the 3D domain.
On the 1D grid, the specific angular momentum of zone $i$ (at radius $r_i$ and mass coordinate $m$) is computed as
\begin{equation}
\label{eq:rot}
l(m) = \frac{2}{3}\,\Omega_{\rot}\,\frac{r_{i}^2 ~r_{\rot}^2}{(r_{i}^2+r_{\rot}^2)}, 
\end{equation}
where we use $r_{\rot} = 250$ km and $\Omega_{\rot} = 0.3\;{\rm rad/s}$. 
The parameter $r_{\rot}$ sets the radius within which the star is rotating like a solid body while this value of $\Omega_{\rot}$ means the spin period of the core is $21\;{\rm s}$. These values for $r_{\rot}$ and $\Omega_{\rot}$ are to be compared with those in \cite{Cabezon-3Dcomparison} showing that the amount of rotation we add is rather small.
In the 3D domain, the $z$-component of the angular velocity, $\Omega_z$, for each zone at radius $r$ and mass coordinate $m$ is determined according to 
\begin{equation}
\label{eq:rot2}
\Omega_z = \frac{3}{2} \, \frac{1}{r^2}\, l(m). 
\end{equation}
From this, the initial tangential velocity of a grid zone at radius $r$ in the 3D domain is set as 
$v_x = - r_y\,\Omega_z$ 
and 
$v_y =  r_x\,\Omega_z$. 

As the simulation evolves, material that is newly accreted into the 3D domain is assigned velocity components that follow from the conservation of angular momentum.

\begin{figure}
\begin{center}
    \includegraphics[trim=0.4cm 0.0cm 1.5cm 0.8cm, clip, width=\linewidth]{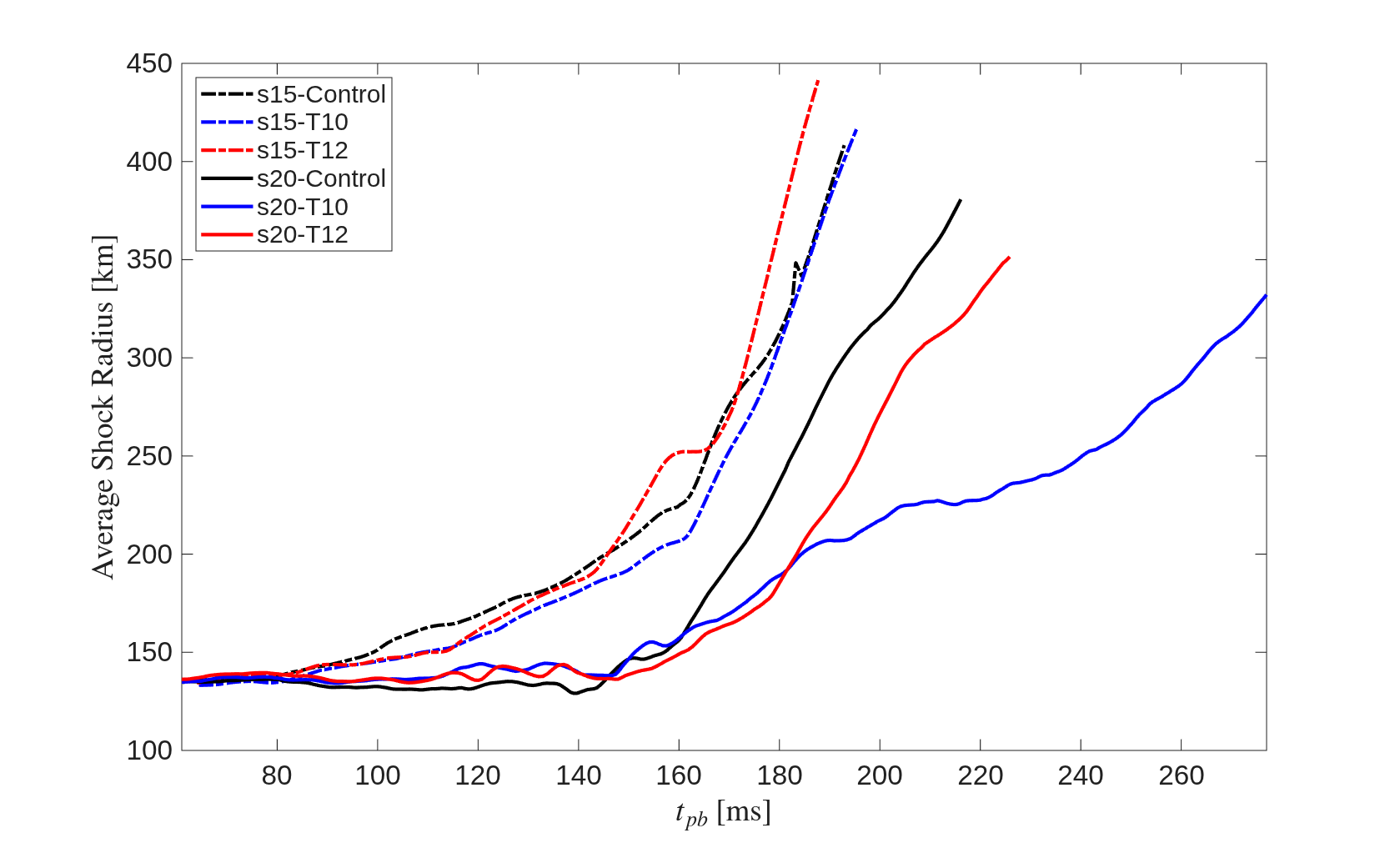}
\end{center}
\caption{Average shock radii as a function of the post-bounce time for all six simulations. 
Dashed lines are for the s15 progenitor, solid lines for the s20 progenitor.
\label{fig:shockradii} 
}
\end{figure}

\subsection{General Properties of the Simulations \label{sec:general_results}}

For each simulation we adopt a uniform fidelity of 1~km throughout the 3D volume; initially, the 3D volume encompasses the innermost 600~km of the star (i.e., along each axis the 3D volume ranges from -300~km to +300~km).
When the shock reaches the edge of the 3D computational domain, we expand the 3D domain by adding 100~km in both the negative and positive directions along each coordinate axis.
We expand the 3D volume twice and stop the simulation when the shock reaches 500 km in any direction. 
A direct comparison of the resolution of the 3D volume with the resolution of the grids adopted elsewhere in the literature for studies of magnetic fields in supernovae, is challenging
because of the wide variety of methods used: 
for example, the grids used in the studies by \cite{2020MNRAS.498L.109M,2023MNRAS.518.3622V,2024MNRAS.528L..96M,2025PhRvD.111f3042S,2026MNRAS.tmp..593V} are spherical while those in \cite{2025arXiv250411537S,2026MNRAS.546ag056S,2026arXiv260518347K} are nested Cartesian or use adaptive mesh refinement (AMR) with the finest resolution being of order $\sim 1\;{\rm km}$ in the innermost tens of kilometers (approximately the proto-neutron star) and of order of a few kilometers in the gain layer. 
The most similar studies to this works are by E. Endeve et al. \cite{2010ApJ...713.1219E,Endeve_2012,2013PhST..155a4022E} which are uniformly Cartesian and where the highest resolution simulation used a cell grid scale of 1.17 km.

We show in Figure \ref{fig:shockradii} the average shock radius (various line styles) for each simulation. The evolution of the average shock radius is the same for all simulations until $t_{\rm pb} \sim 80 \;{\rm ms}$ when the three s15 simulations separate from the s20 simulations. 
Thereafter the average shock radius in the s15 simulations grows approximately linearly until $t_{\rm pb} \sim 150 \;{\rm ms}$ whereupon the average shock radius velocity accelerates noticeably with the s15-T10 and s15-control lagging behind the s15-T12 by a few tens of kilometers after $t_{\rm pb} \sim 150 \;{\rm ms}$.
For the simulations using the s20 progenitor, the shocks stall until $t_{\rm pb} \sim 150 \;{\rm ms}$ whereupon they are revived. 
The acceleration of the shock radius velocity in the s20-control occurs shortly thereafter at $t_{\rm pb} \sim 160 \;{\rm ms}$, the acceleration in the s20-T12 is delayed until $t_{\rm pb} \sim 175 \;{\rm ms}$, and the acceleration in the s20-T10 is delayed until $t_{\rm pb} \sim 250 \;{\rm ms}$. 
We find no evidence for a standing accretion shock instability (SASI) \cite{2003ApJ...584..971B,2007Natur.445...58B} in any of our simulations---we refer the reader to \cite{2010ApJ...713.1219E,Endeve_2012,2013PhST..155a4022E} for studies of the growth of the magnetic field in simulations where the SASI does appear.

\begin{figure*}
\begin{center}
\begin{tabular}{ccc}
\textbf{s15-Control} & \textbf{s15-T10} & \textbf{s15-T12} \\
\includegraphics[trim=0.0cm 1.0cm 2.8cm 0.2cm, clip, height=5cm]{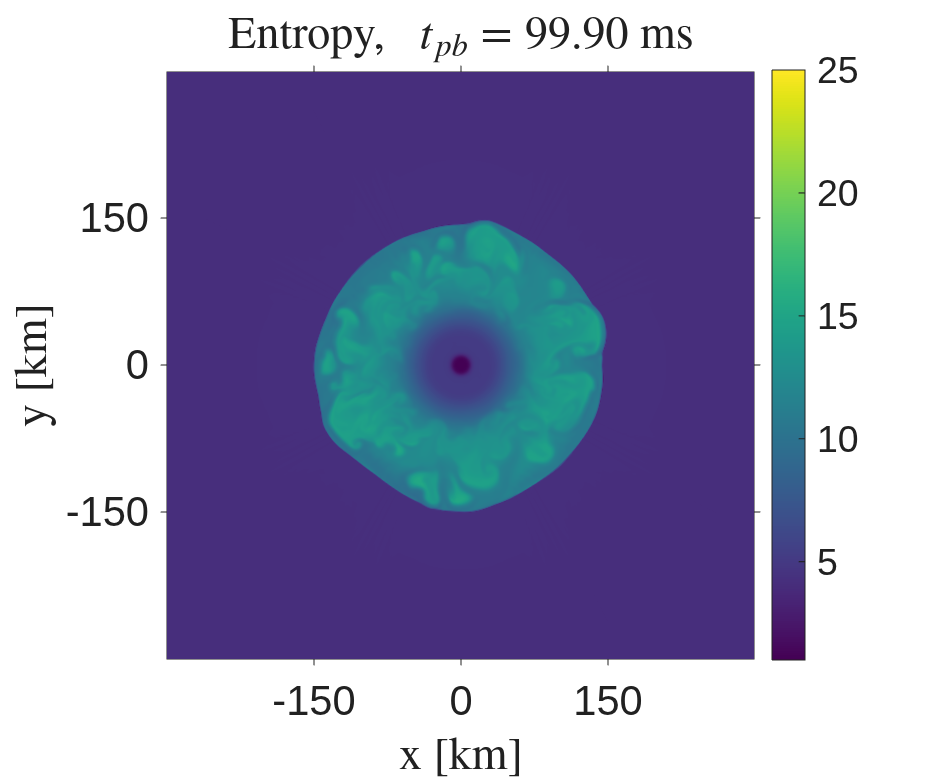}\hspace*{0.0cm}&
\includegraphics[trim=1.0cm 1.0cm 2.8cm 0.2cm, clip, height=5cm]{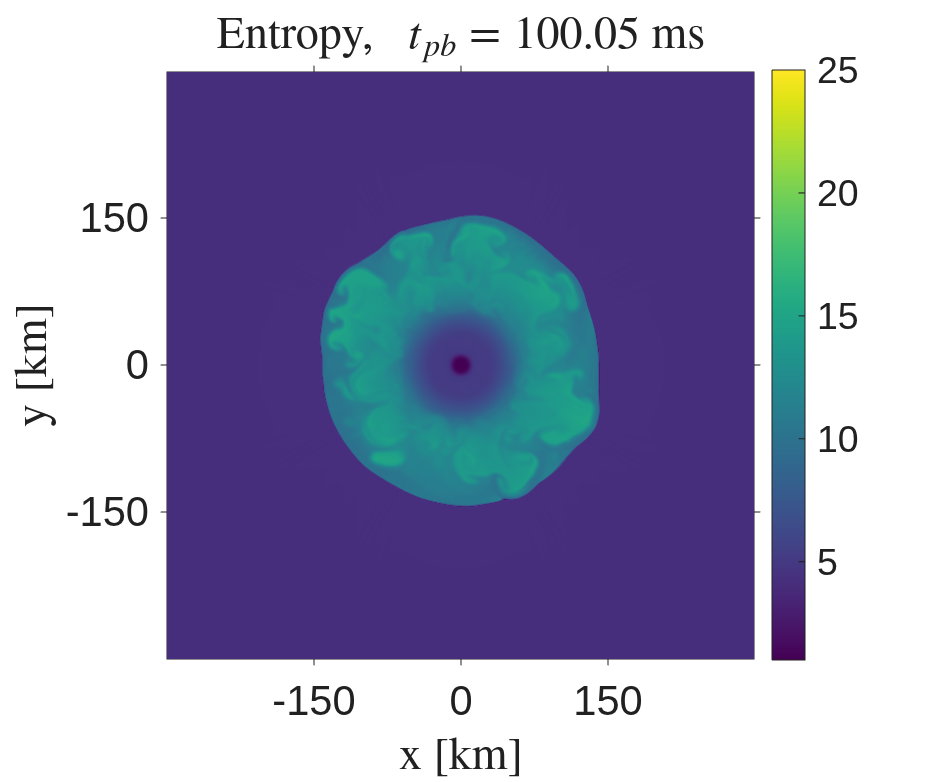}\hspace*{0.0cm}&
\includegraphics[trim=1.0cm 1.0cm 1.1cm 0.2cm, clip, height=5cm]{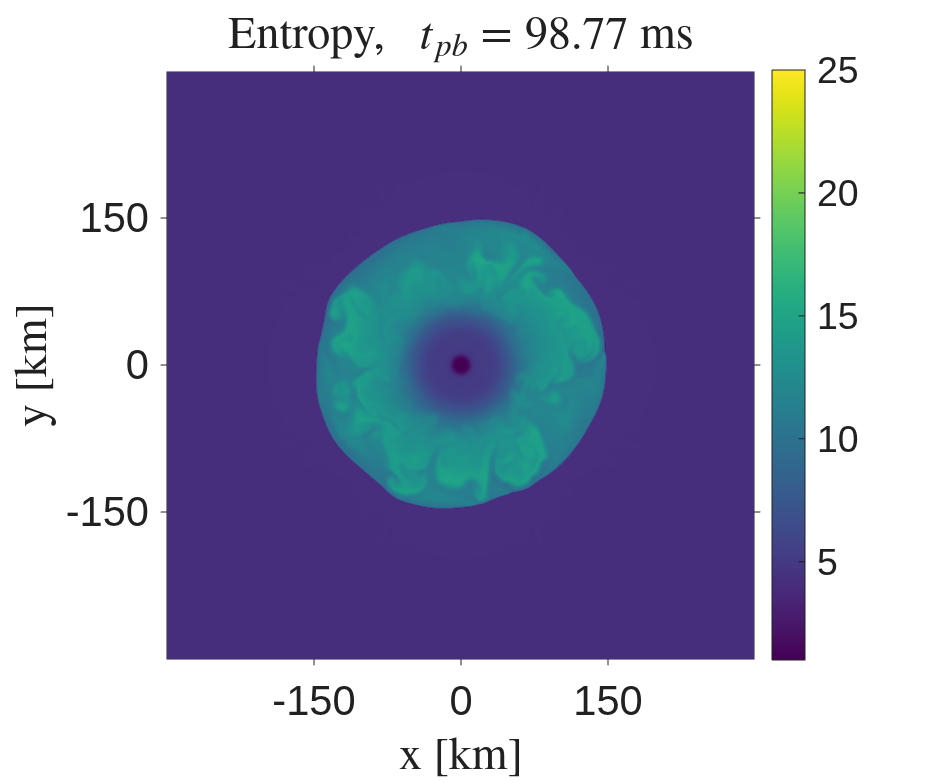}      \\ 
\includegraphics[trim=0.0cm 1.0cm 2.8cm 0.2cm, clip, height=5cm]{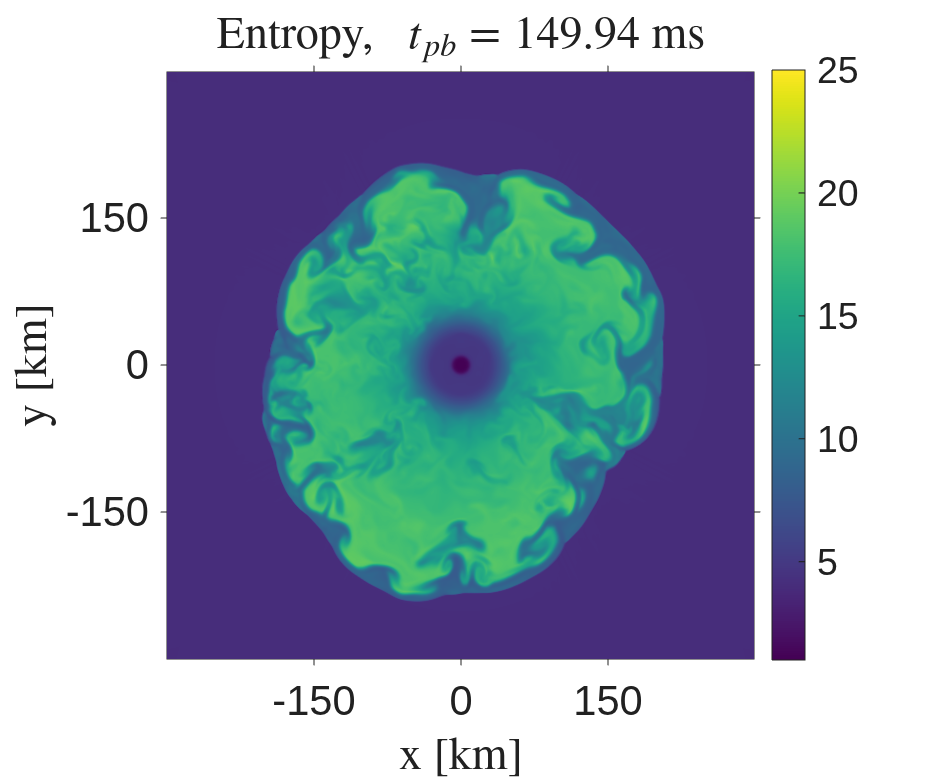}\hspace*{0.0cm}&
\includegraphics[trim=1.0cm 1.0cm 2.8cm 0.2cm, clip, height=5cm]{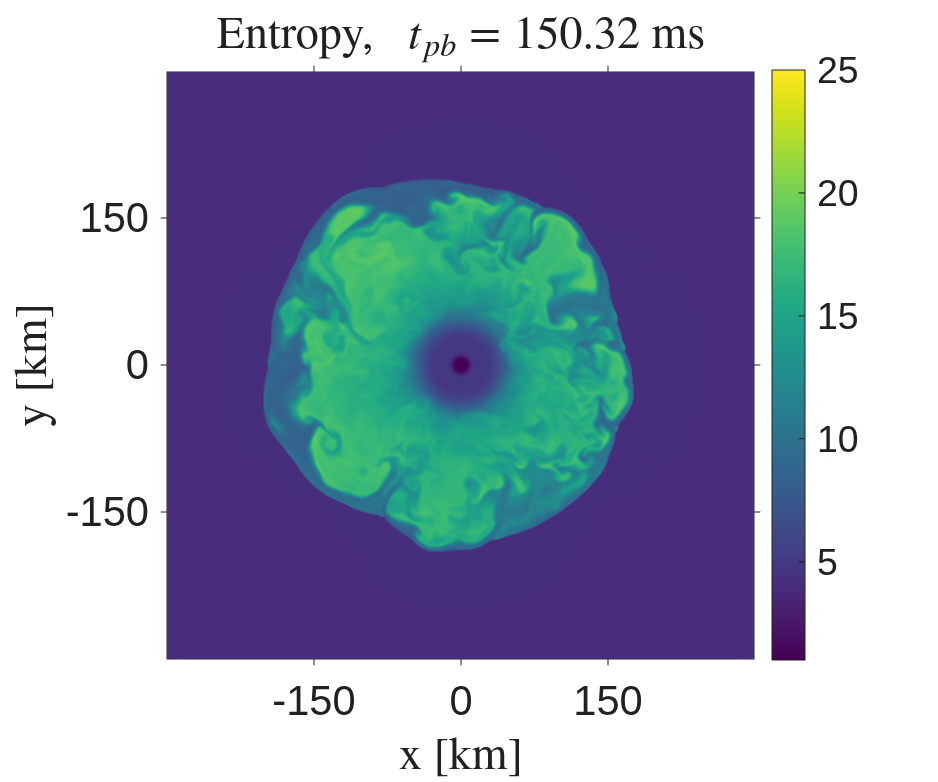}\hspace*{0.0cm}&
\includegraphics[trim=1.0cm 1.0cm 1.1cm 0.2cm, clip, height=5cm]{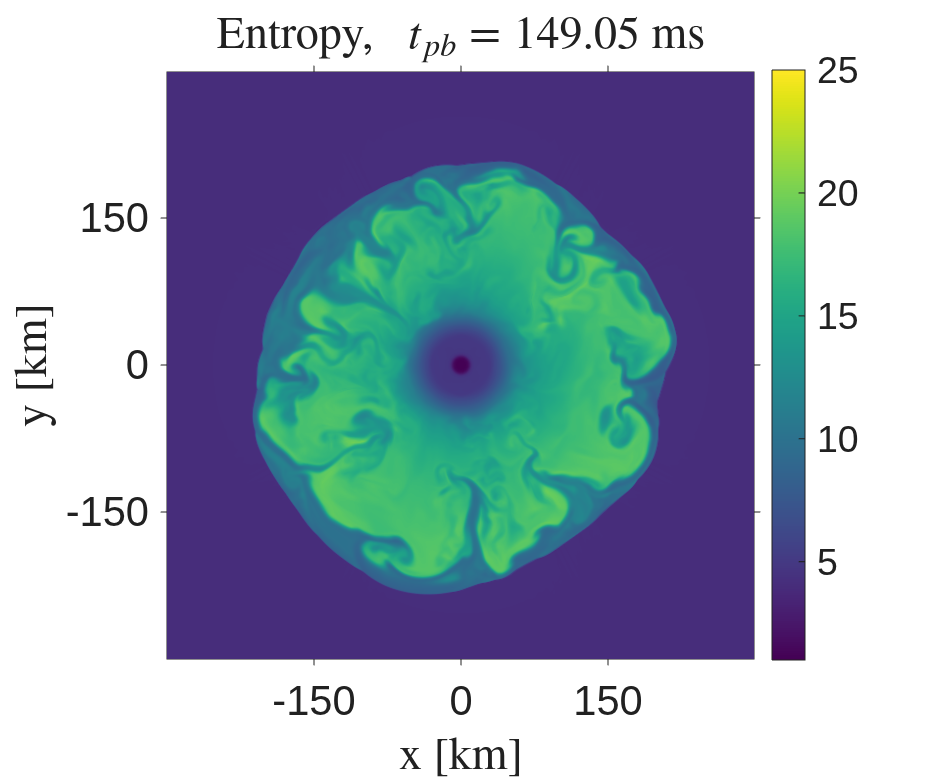} \\
\includegraphics[trim=0.0cm 0.1cm 2.8cm 0.2cm, clip, height=5.4cm]{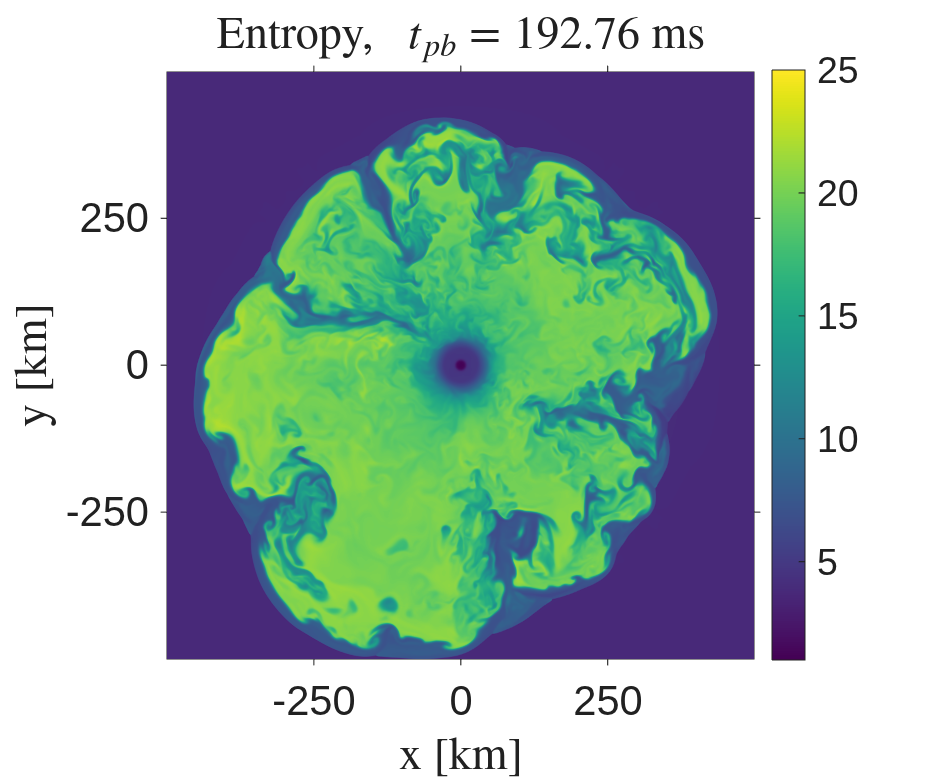}\hspace*{0.0cm} &
\includegraphics[trim=1.0cm 0.1cm 2.8cm 0.2cm, clip, height=5.4cm]{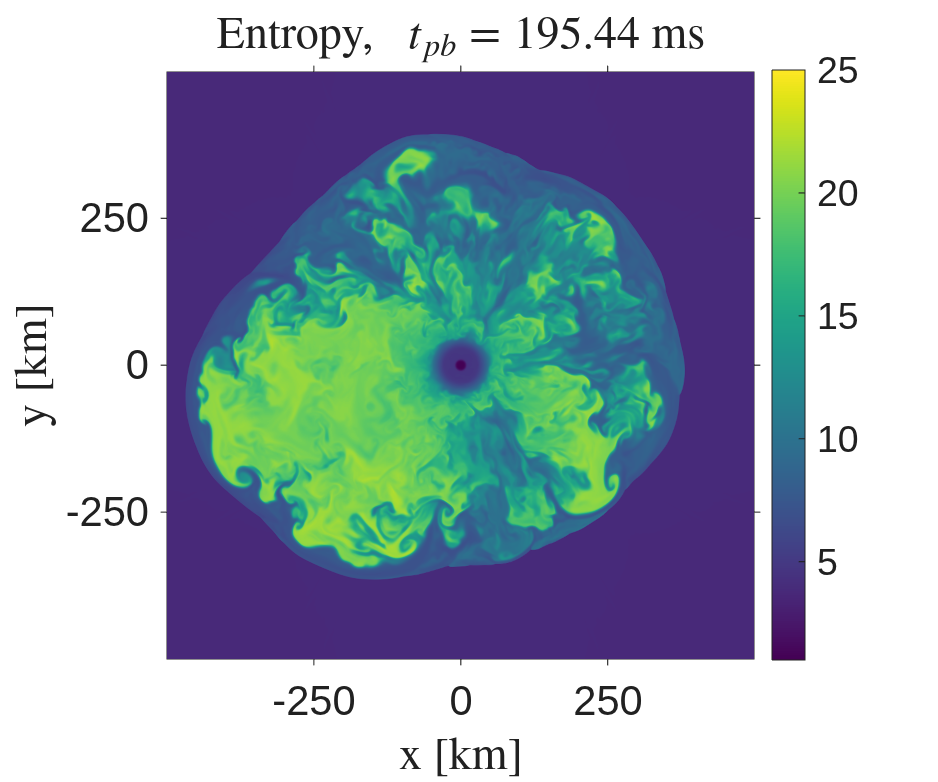}\hspace*{0.0cm}& 
\includegraphics[trim=1.0cm 0.1cm 1.1cm 0.2cm, clip, height=5.4cm]{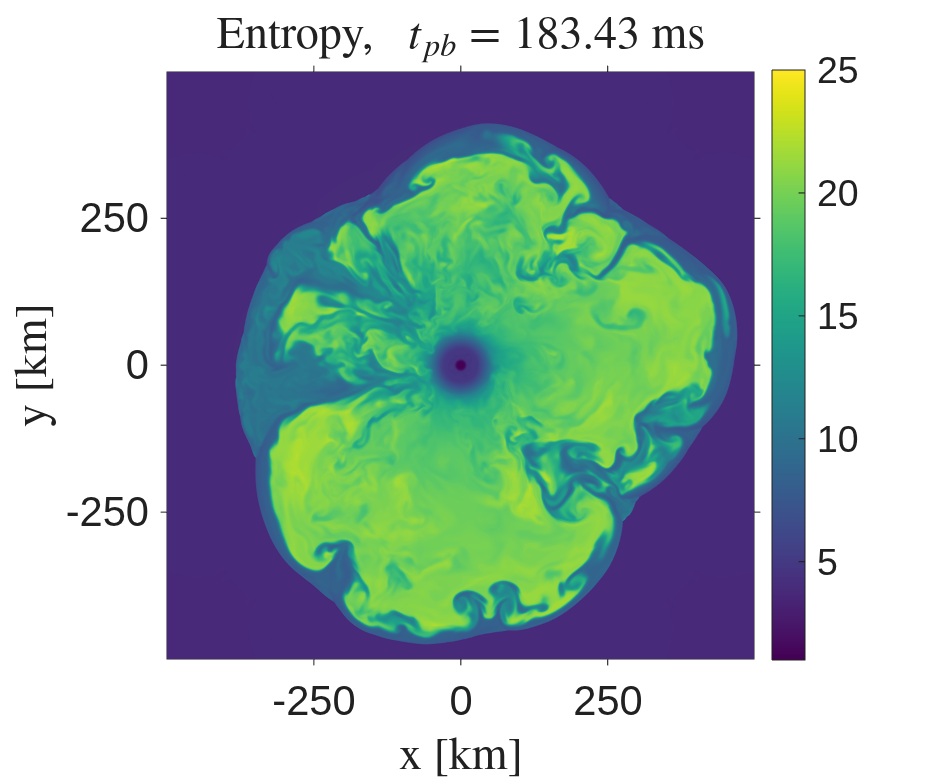}
\end{tabular}
\end{center}
\caption{Maps of the entropy for the s15-control (left), s15-T10 (middle), and s15-T12 (right) simulations at the post-bounce times of $\sim100$~ms (first row), $\sim 150$~ms (second row), and at final time for each simulation (third row). Note that the scales are not the same on each panel due to expansion of the simulation volume.
\label{fig:entropy_s15}
}
\end{figure*}

\begin{figure*}
\begin{tabular}{ccc}
\textbf{s20-Control} & \textbf{s20-T10} & \textbf{s20-T12} \\
\includegraphics[trim=0.0cm 1.0cm 3cm 0.2cm, clip, height=5cm]{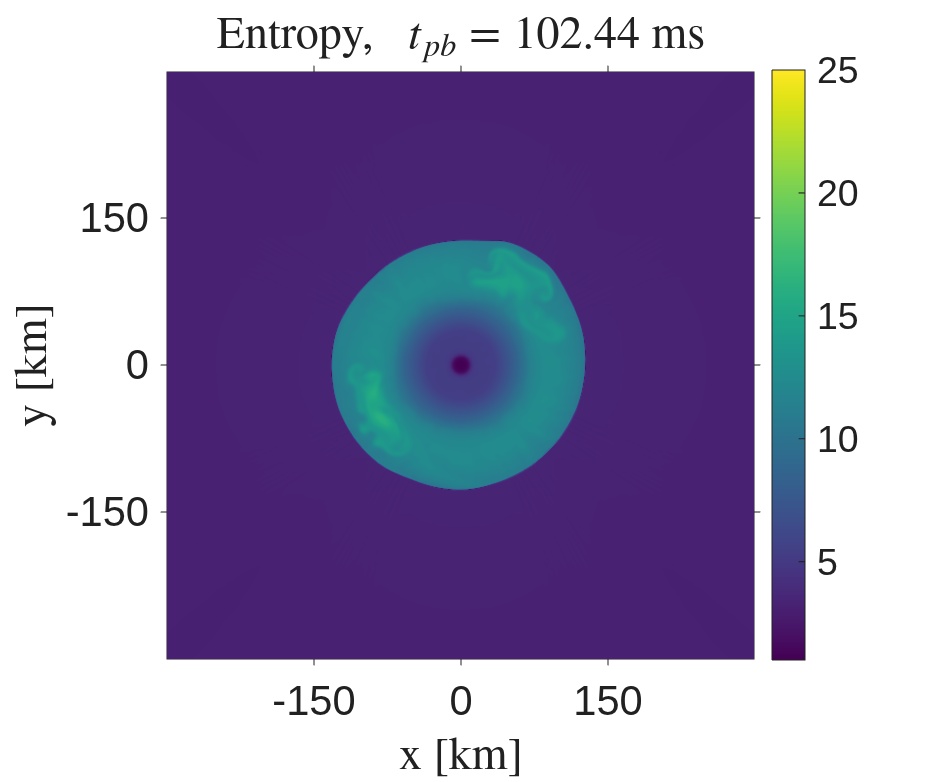}\hspace*{0.0cm}&
\includegraphics[trim=1.0cm 1.0cm 3cm 0.2cm, clip, height=5cm]{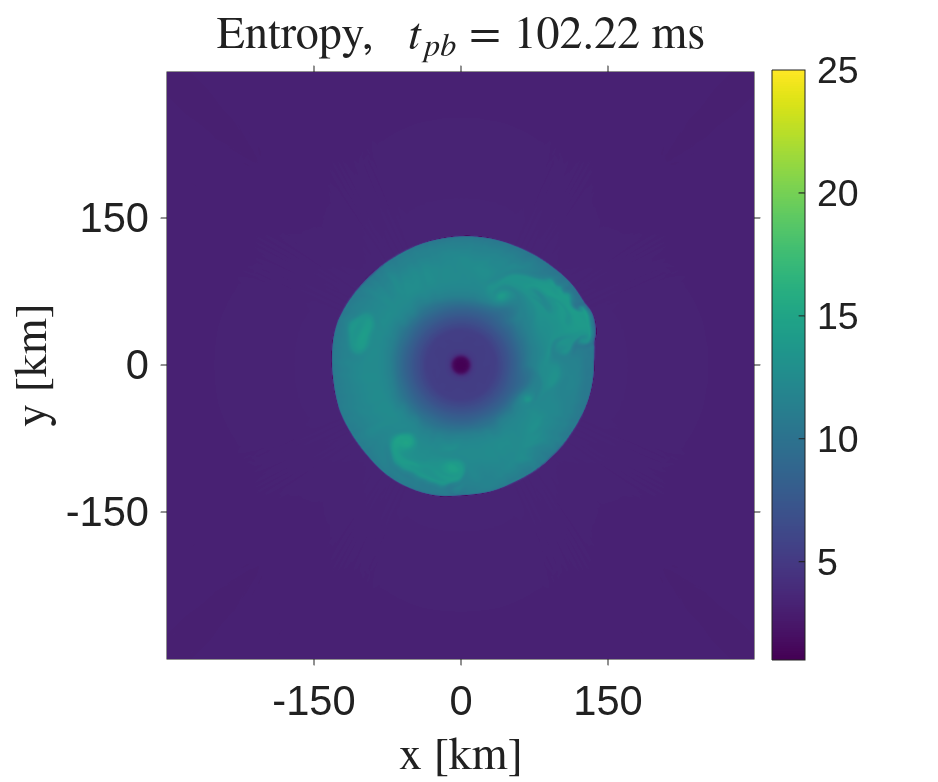}\hspace*{0.0cm}&
\includegraphics[trim=1.0cm 1.0cm 1.3cm 0.2cm, clip, height=5cm]{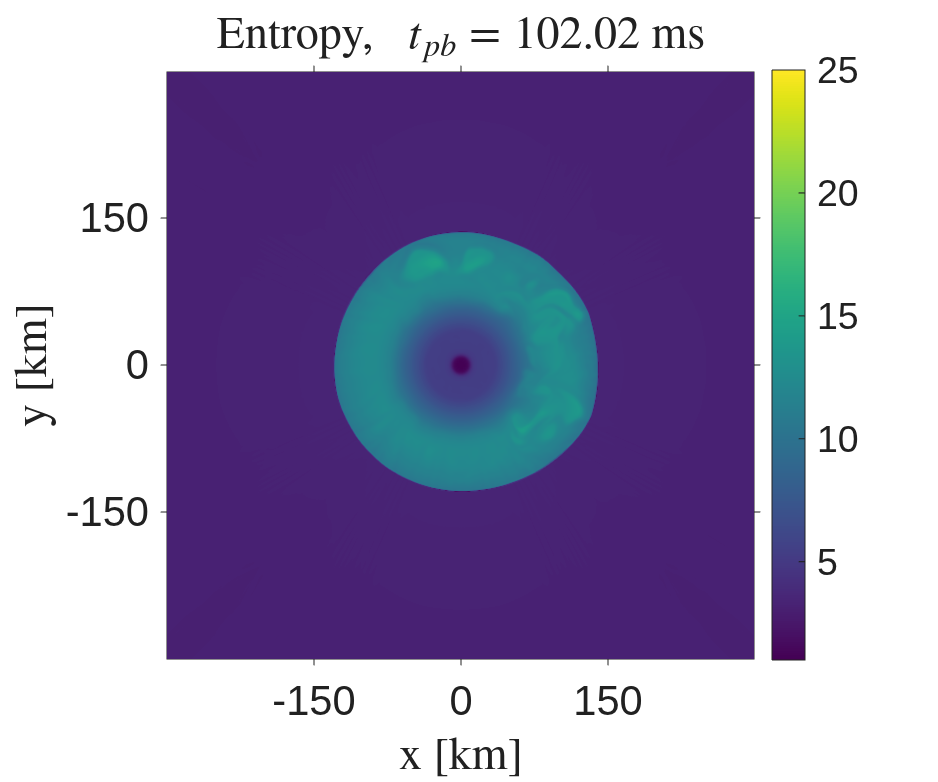}\\
\includegraphics[trim=0.0cm 1.0cm 3cm 0.2cm, clip, height=5cm]{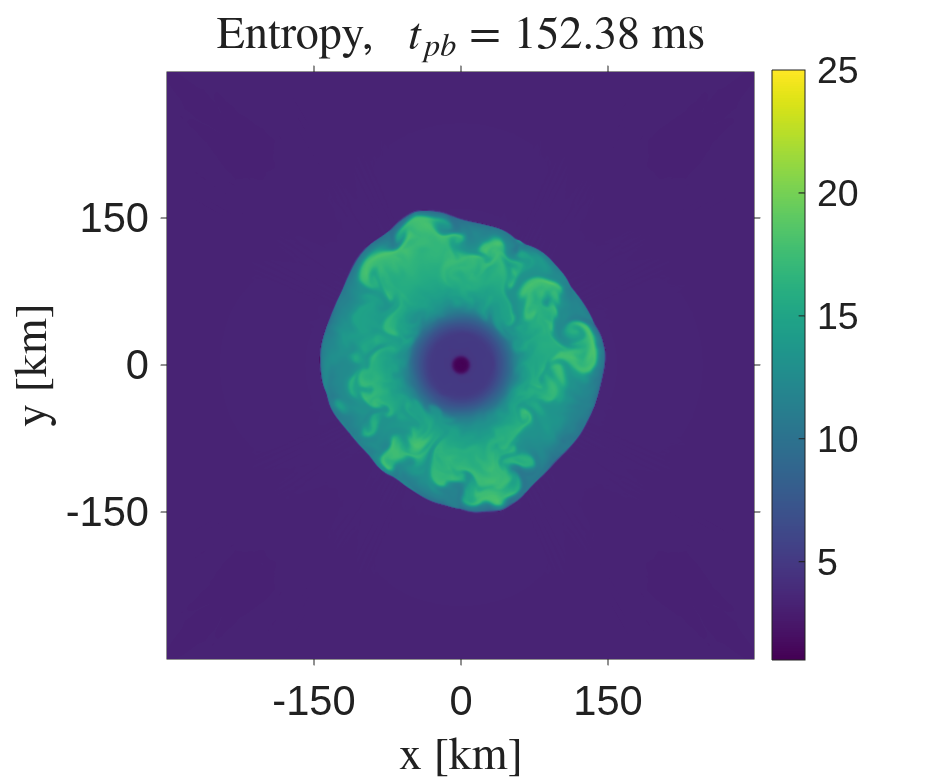}\hspace*{0.0cm}&
\includegraphics[trim=1.0cm 1.0cm 3cm 0.2cm, clip, height=5cm]{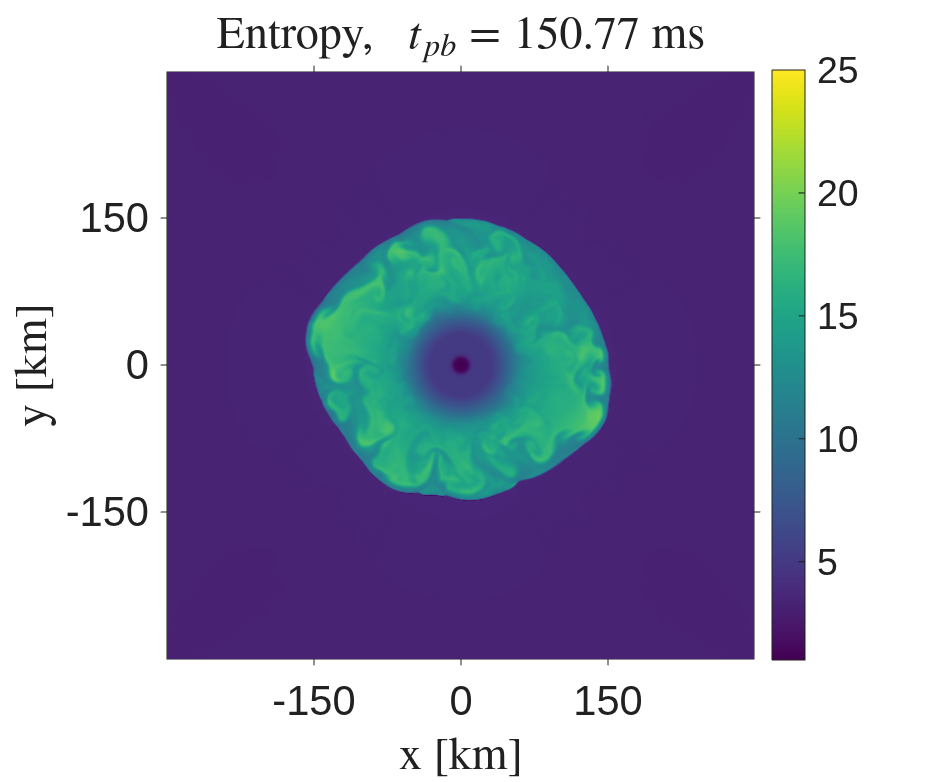}\hspace*{0.0cm}&
\includegraphics[trim=1.0cm 1.0cm 1.3cm 0.2cm, clip, height=5cm]{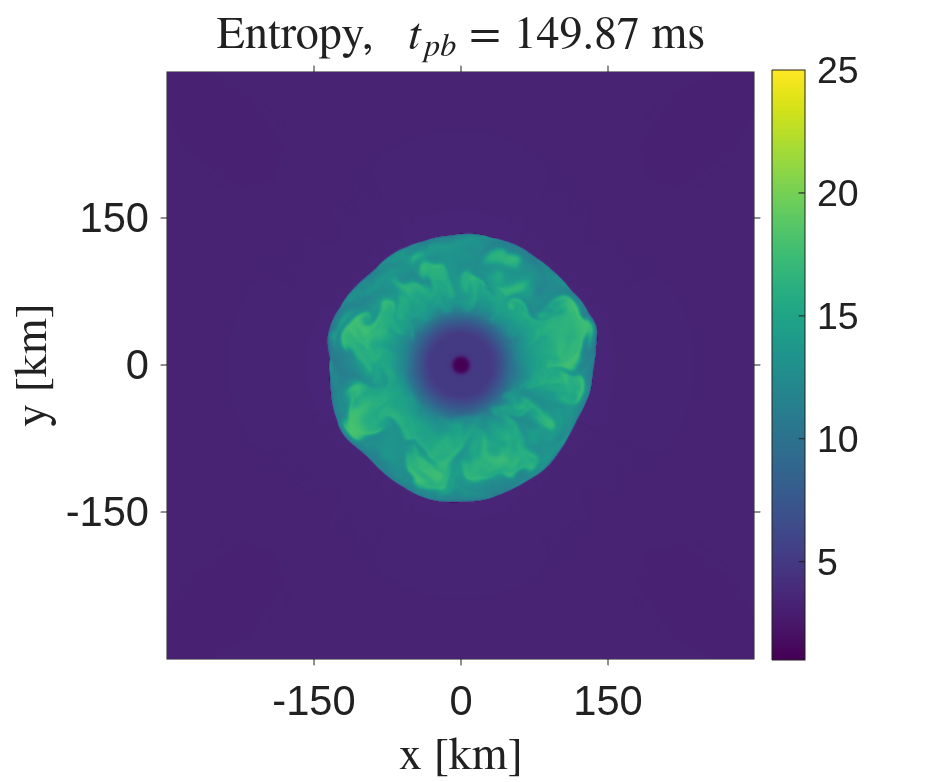} \\          
\includegraphics[trim=0.0cm 1.0cm 3cm 0.2cm, clip, height=5cm]{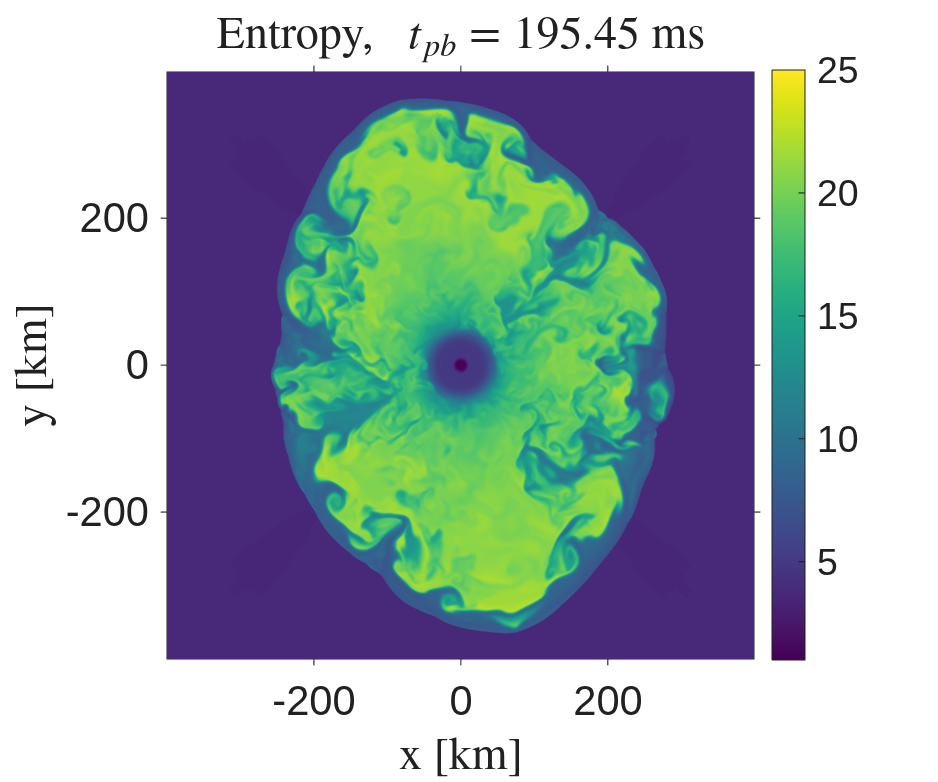}\hspace*{0.0cm}&
\includegraphics[trim=1.0cm 1.0cm 3cm 0.2cm, clip, height=5cm]{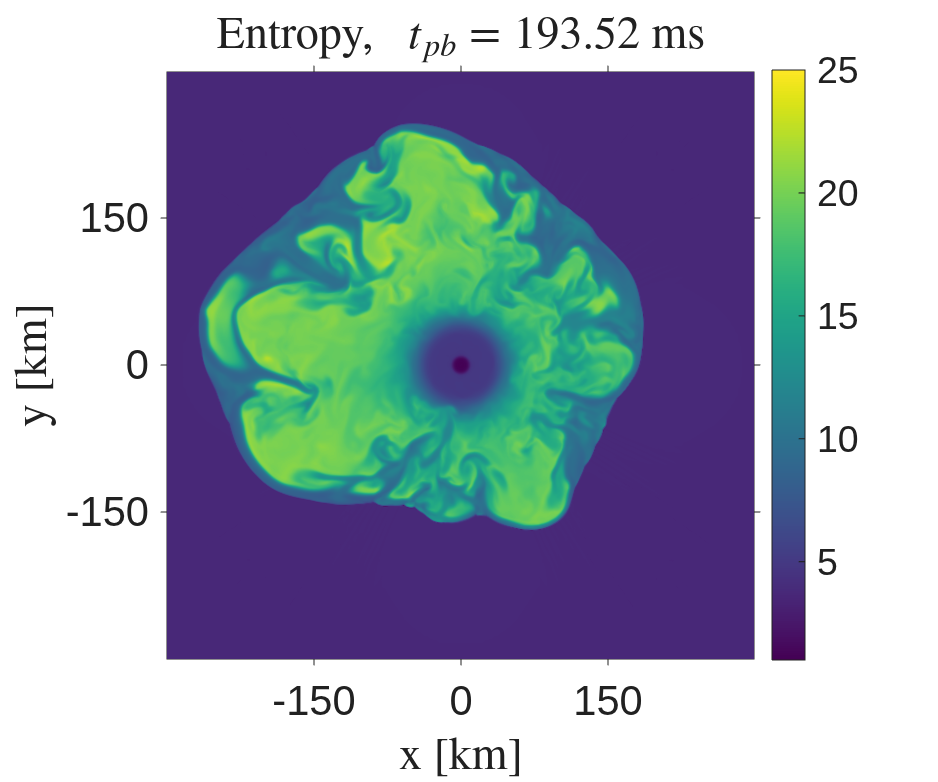}\hspace*{0.0cm}&
\includegraphics[trim=1.0cm 1.0cm 1.3cm 0.2cm, clip, height=5cm]{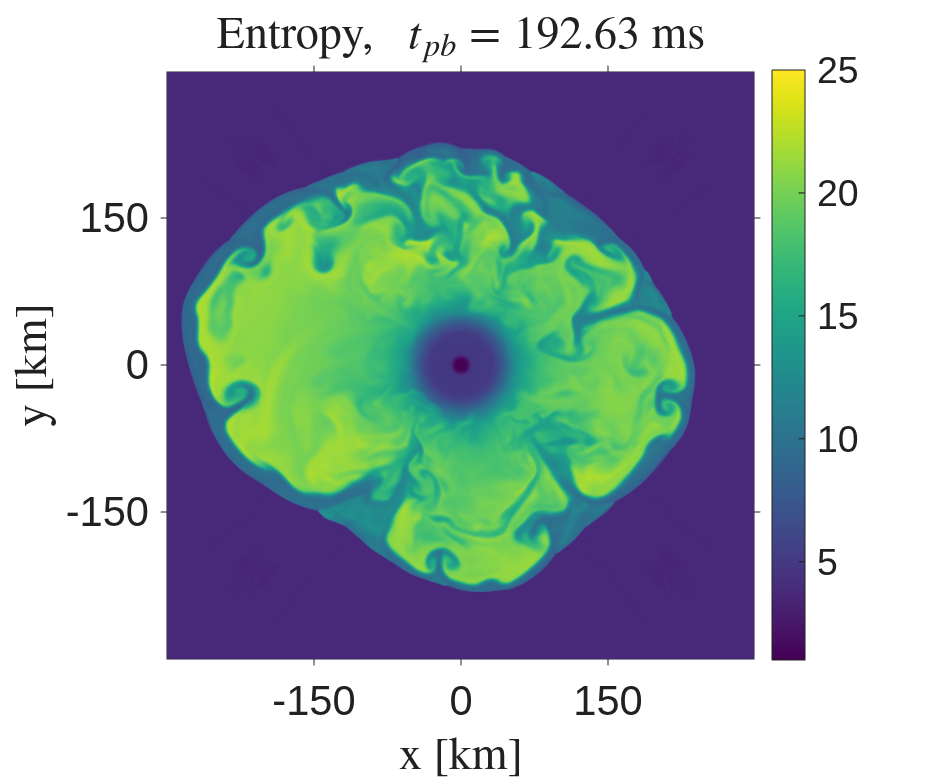}\\
\includegraphics[trim=0.0cm 0.1cm 3.0cm 0.2cm, clip, height=5.4cm]{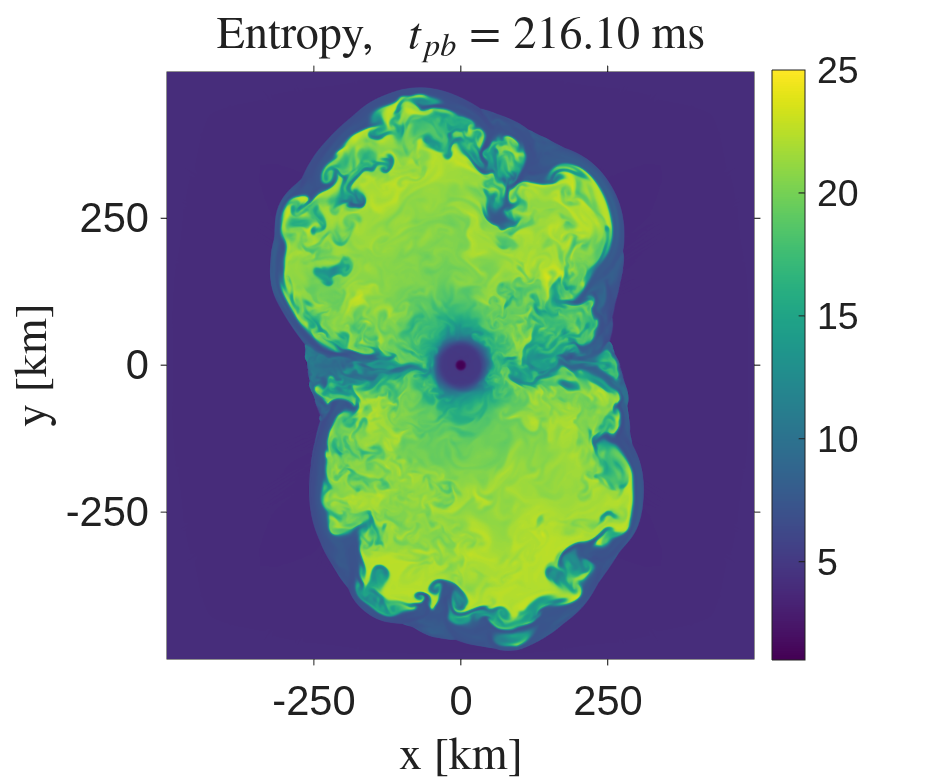}\hspace*{0.0cm}&
\includegraphics[trim=1.0cm 0.1cm 3.0cm 0.2cm, clip, height=5.4cm]{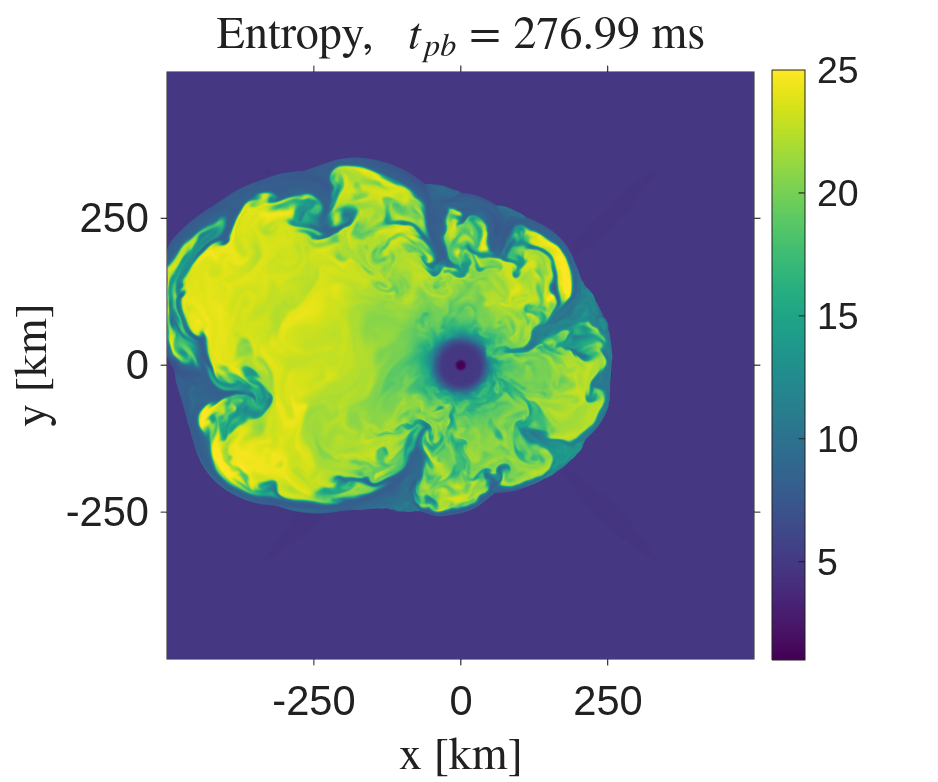}\hspace*{0.0cm}&
\includegraphics[trim=1.0cm 0.1cm 1.3cm 0.2cm, clip, height=5.4cm]{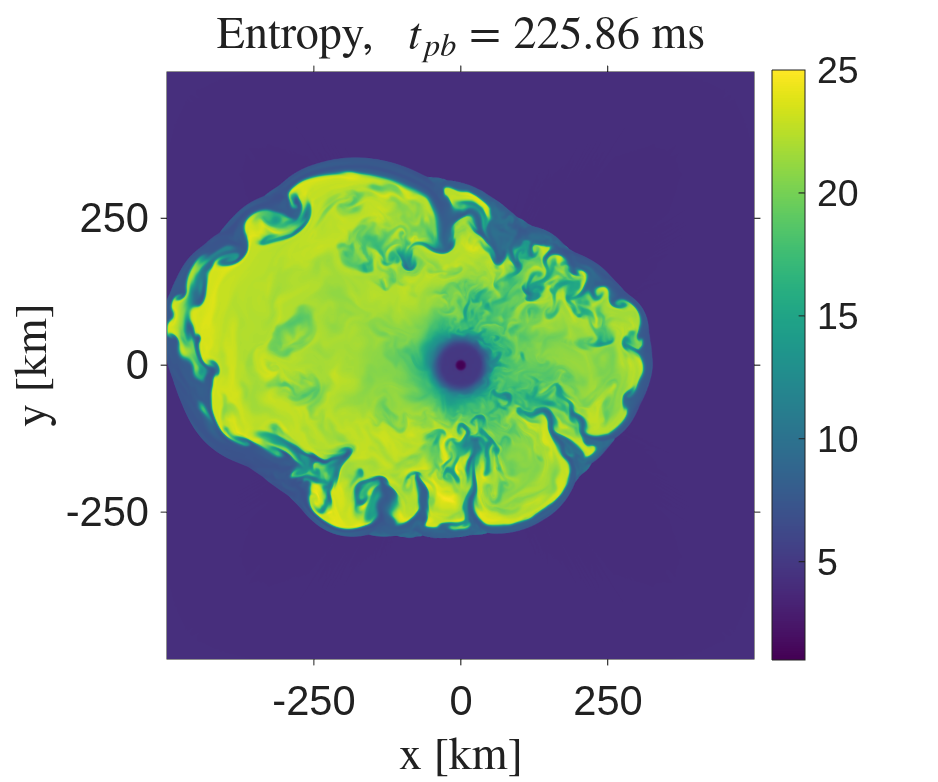}
\end{tabular}
\caption{Maps of the entropy for the s20-control (left), s20-T10 (middle), and s20-T12 (right) simulations at the post-bounce times of $\sim102$~ms (first row), $\sim 150$~ms (second row), $\sim 193$~ms (third row), and at final time (last row). Note that the scales are not the same on each panel due to expansion of the simulation volume.
\label{fig:entropy_s20}
}
\end{figure*}

In Figures \ref{fig:entropy_s15} and \ref{fig:entropy_s20} we plot 2D slices of the entropy for the three s15 and three s20 simulations respectively. The low-entropy, pre-shocked material and the low-entropy PNS form the innermost and outermost layers.
Between the low-entropy, pre-shocked material and the low-entropy PNS is the gain region, which we define as the interior region of the shock where the entropy exceeds $s > 5\,k_{\rm B}/\text{baryon}$, the density is less than $\rho < 10^{10}\;{\rm g/cm^3}$, and the neutrino heating is positive. 
The spherical symmetry of the initial collapse, PNS formation, initial shock formation and expansion is broken by neutrino heating at approximately 80 - 100 ms postbounce in all simulations.
By 150 ms postbounce the shocks are noticeably aspherical, and towards the end of the simulations  the presence of low entropy downflows piercing through large convective plumes are seen. 
By the end of the simulations the shock front has expanded preferentially into one hemisphere for the s20-T10, s20-T12, s15-control, and s15-T10 simulations, while we observe a more symmetric (but not spherical) explosion in the s15-T12 and s20-control simulations.
The evolution of the entropy is similar to that seen previously in multi-D simulations, such as \cite{1995ApJ...450..830B,1998ApJ...495..911M,1995ApJ...448L.109J} to name a few.

\section{The magnetic field}
\label{sec:Bfield-results}

\subsection{Theoretical considerations}
\label{sec:MagneticField}

The magnetic field at a given point and time within the supernova will evolve due to processes which are typically divided into the three categories of `advection', `stretching', and `compression' \cite{landau2013electrodynamics,BRANDENBURG20051}. 
As detailed below, in order to better observe the processes which generate the magnetic field \emph{within} the shock rather than advecting a lot of field energy \emph{through} the shock, we focus on initially pure toroidal magnetic fields in our simulations. 
The field passing through a fluid element begins its evolution due to compression and stretching even before the element reaches the shock.
For radial infall, the solid angle subtended by a spherical wedge-shaped mass element does not change as it falls inwards.
Since the fluid is taken to be a perfect conductor, the magnetic field is entrained by the fluid and the magnetic flux through the faces of the mass element are conserved. This means the components of the field grow as 
\begin{eqnarray}
b_r(r) & = & b_r(r_0) \left( \frac{r_0}{r} \right)^2, \label{eq:Br}\\
b_{\theta,\phi}(r) & = & b_{\theta,\phi}(r_0) \left( \frac{r_0}{r}\,\frac{dr_0}{dr} \right),\label{eq:Bth}
\end{eqnarray}
where $r_0$ is the initial location of the mass element. 
The ratio $dr/dr_0$ is the radial stretch factor. 
For the case of pure freefall of a pressure-less fluid, the stretch factor is given by an analytic, albeit complicated, expression \cite{1969MNRAS.145..457P,1969MNRAS.144..425P}.
As the mass element falls towards the core and $r_0 /r$ increases, both the radial and non-radial (transverse) components of the field grow due to compression. For the radial component, this is the only factor, but for the non-radial components there is the dependence upon the stretch factor $dr / dr_0$. For material that is initially well outside the core and at rest, the radial stretching of a mass element is initially small, resulting in  $dr / dr_0 \approx 1$. 
Thus, if there is a nonzero transverse component of the field at the initial location of the fluid element, this component grows as $1/r$ during the initial phase of collapse compared to the $1/r^2$ growth of the radial component. 
As the mass element falls further, the stretch factor $dr / dr_0$ begins to increase from unity. For the case of pure freefall of a pressure-less gas, eventually $dr / dr_0$ will scale as $1/\sqrt{r}$ indicating that the growth of the transverse components of the magnetic field slows to $1/\sqrt{r}$. If the gas pressure is included, the dependence of the stretch factor upon the radius will change but it is still the case that $dr / dr_0 > 1$.
Therefore, if the initial magnetic field passing through a mass element has both transverse and radial components, the field will become increasingly dominated by the radial field as the fluid collapses. This leads to field configurations which resemble the split monopole as seen in \cite{10.1093/mnras/staa3095,2024MNRAS.528L..96M} and used as the initial magnetic field in the simulations by \cite{2010ApJ...713.1219E,Endeve_2012,2013PhST..155a4022E}. 
Rotation of the star will modify these expectations (a mass element gets squeezed in the transverse direction because the orbital planes defining the faces of the mass element must intersect at some point) but the squeezing is only important if the mass element completes a substantial fraction of an orbit.

As the fluid element falls, the magnetic field energy of the element grows due to the compression regardless of its orientation. However the rate of growth of the field energy density \emph{relative} to the kinetic energy density of the element does depend upon the orientation. The density of the fluid element increases as $\rho(r) \propto (1/r^2) (dr_0/dr)$ and the fluid velocity as $v(r) \propto 1/\sqrt{r}$ for pure freefall in the weak field limit, thus the fluid kinetic energy density scales as $E_{\kin} \propto (1/r^3)\,(dr_0/dr)$. For an initial mass density profile that falls off rapidly with $r_0$ and for $r \ll r_0$, $dr_0/dr \propto \sqrt{r}$ and so $E_{\kin} \propto 1/r^{5/2}$. 

If the field is purely transverse, the magnetic field energy density, $E_b = |\vec{b}|^2/2$, scales as $E_b \propto (1/r^2) (dr_0/dr)^2$ so the ratio of magnetic to kinetic energy density, $E_b / E_{\kin}$ is seen to scale as $E_b / E_{\kin} \propto r (dr_0/dr)$ indicating the ratio decreases as the element falls. For the case of a pure radial field, $E_b \propto (1/r^4)$ and the ratio $E_b / E_{\kin}$ is seen to scale as $E_b / E_{\kin} \propto 1/ ( r\, dr_0/dr)$ meaning it increases as the element falls. It is this scaling during the collapse that motivates our decision to focus upon initially pure toroidal magnetic fields. 

The evolution of the magnetic field becomes much more complicated once the fluid passes through the shock, as seen in \cite{2024MNRAS.528L..96M}.
The standard paradigm for the growth of the magnetic field is the amplification of the field lines due to streamline stretching \cite{landau2013electrodynamics}. This picture emerges by combining the evolution of the magnetic field, given by equation \eqref{eq:magnetic}, with the mass continuity equation (\ref{eq:mass}), to produce 
\begin{equation}
\frac{\partial}{\partial t} \left(\frac{\vec{b}}{\rho} \right) + \left( \vec{v} \cdot \nabla \right) \left(\frac{\vec{b}}{\rho} \right) - \left(\frac{\vec{b}}{\rho} \cdot \nabla \right)\;\vec{v} = 0.
\label{eq:boverrho}
\end{equation}
This combination of time and spatial derivatives is often known as the fluid derivative. 
The concept of streamline stretching emerges from this equation by comparing it with the evolution equation for the vector $\Delta\vec{x}$ which connects two fluid elements. This vector evolves (to linear order in $\Delta\vec{x}$) according to the equation 
\begin{equation}
\frac{\partial}{\partial t} \left(\Delta\vec{x}\right) + \left( \vec{v} \cdot \nabla \right) \left(\Delta\vec{x} \right) - \left(\Delta\vec{x}\cdot \nabla \right)\;\vec{v} = 0.
\end{equation}
Thus if we consider two fluid elements along a field line i.e. they satisfy $\vec{b}/\rho \propto \Delta\vec{x}$, then as the distance between the fluid elements grows, the magnetic field must also grow assuming the fluid is incompressible. However, when we allow for changes in the density, the field can grow even for reductions in the distance between fluid elements (which occurs during the collapse outside the shock as we reasoned previously), thus simultaneous changes in the magnetic field and density complicate the understanding of the growth of the field. 


\begin{figure}
\includegraphics[width=0.45\textwidth]{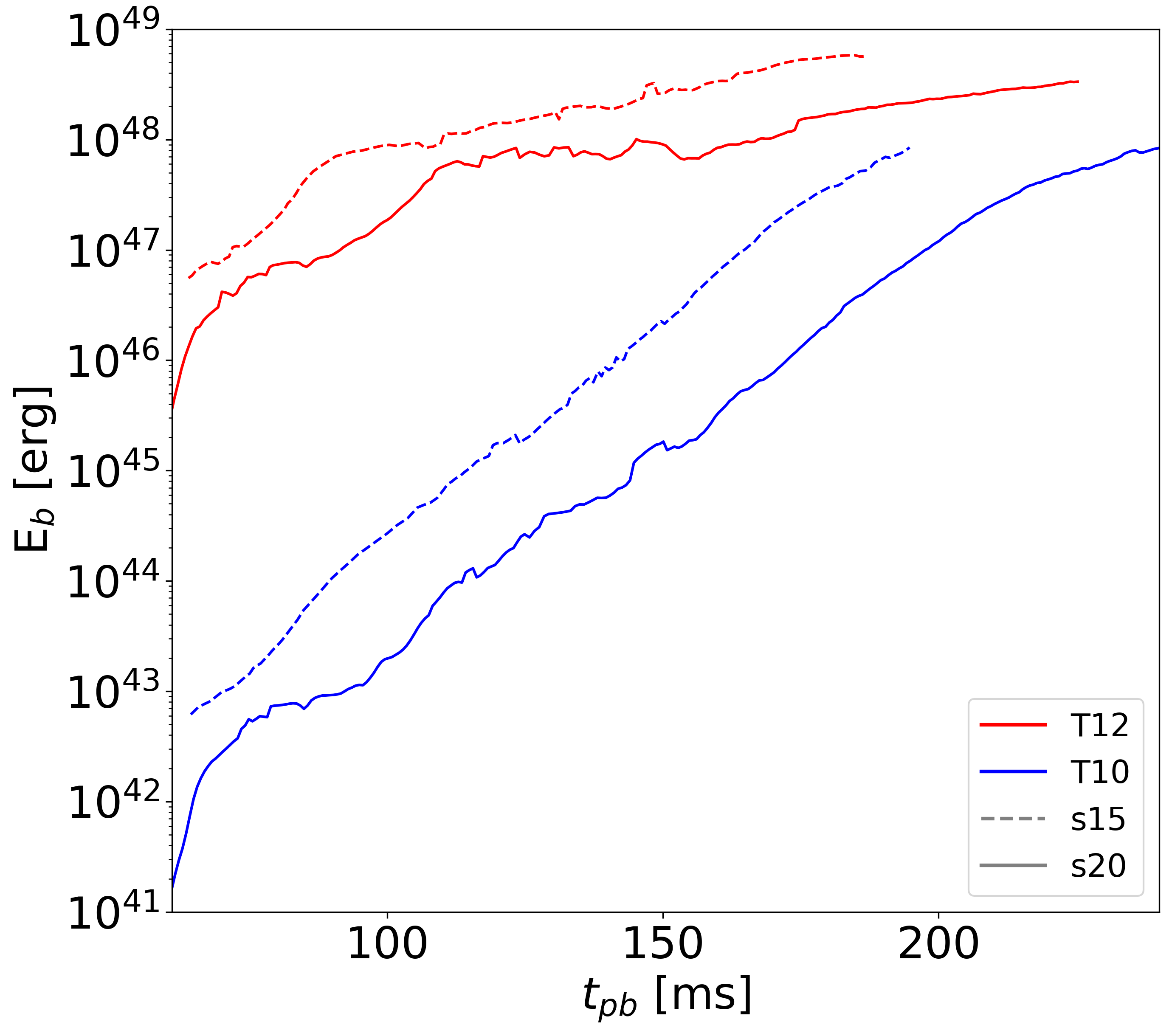}
\includegraphics[width=0.45\textwidth]{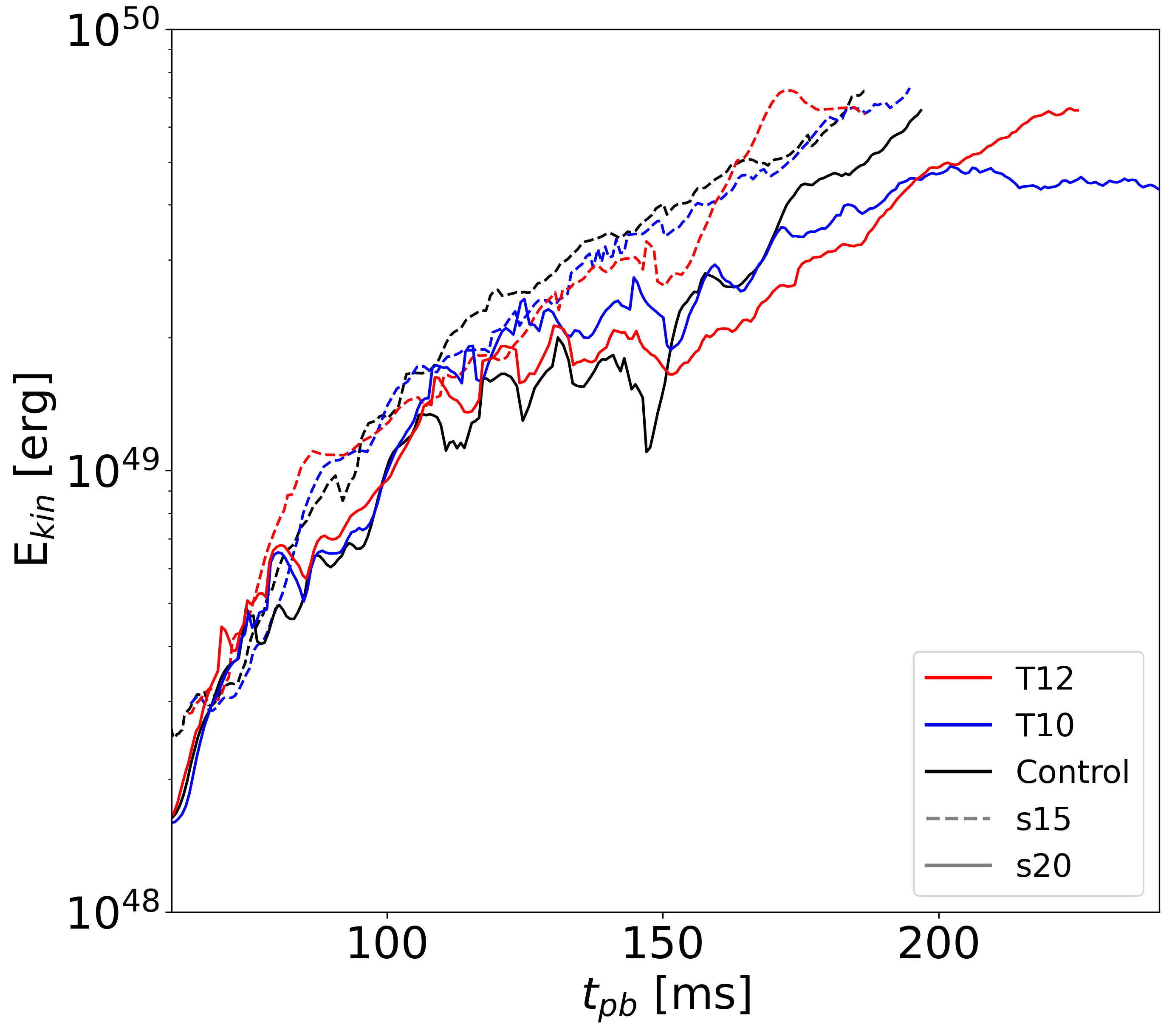}
\includegraphics[width=0.45\textwidth]{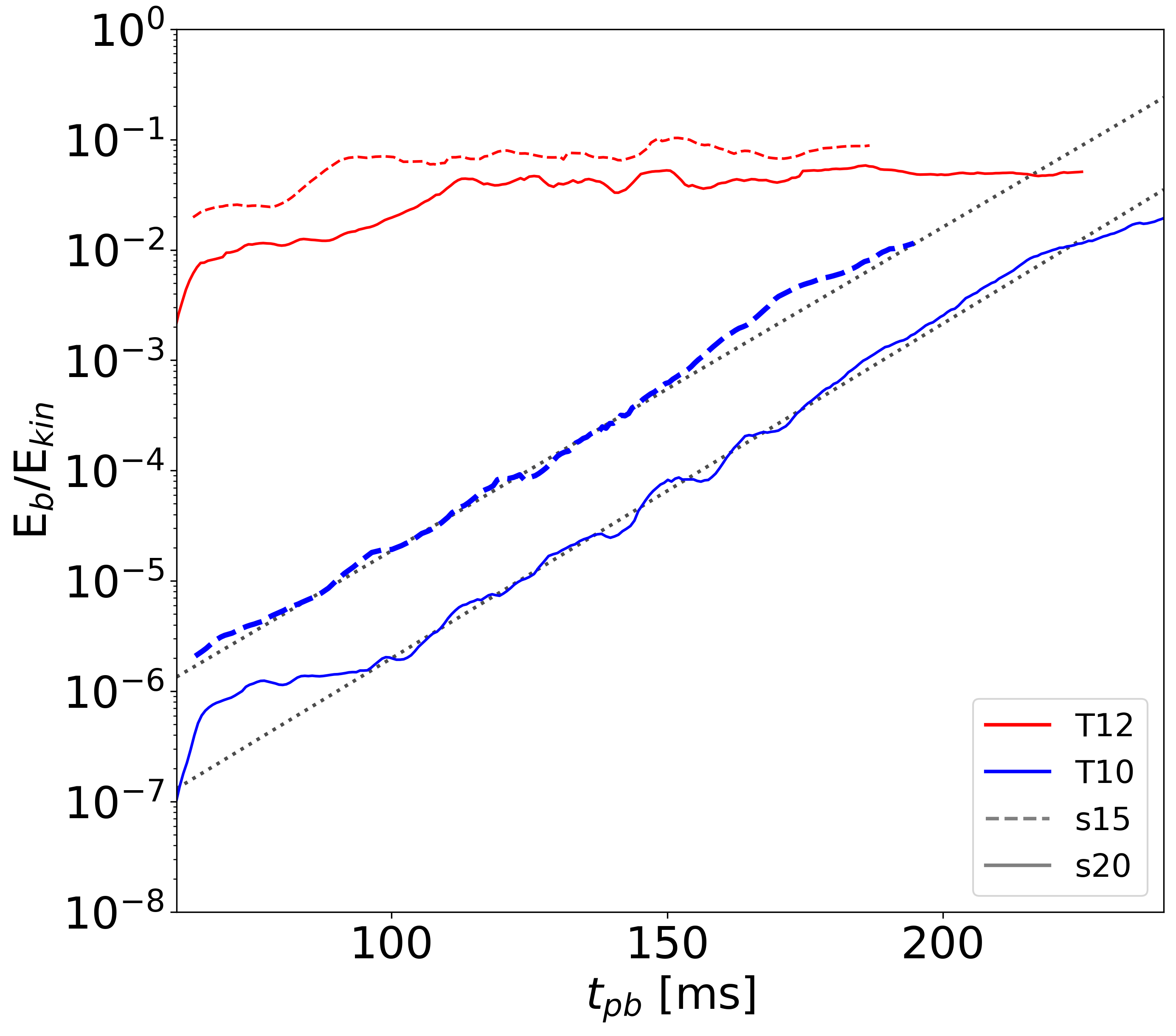}
\caption{
Magnetic field energy (top) and kinetic energy (middle) in the gain region as a function of the post-bounce time for the s20 (dashed) and s15 (solid) simulations. 
Bottom: Ratio of magnetic to kinetic energy with two exponential fits shown as dotted lines. 
\label{fig:Eb}
}
\end{figure}

\subsection{The Total Magnetic Field Energy in the Gain Region}
\label{sec:Bfield-gainlayer}

Figure \ref{fig:Eb} shows the integrated magnetic energy in the gain layer for all four simulations with a magnetic field (top panel), the total kinetic energy in the gain layer for all six simulations (middle panel), and the ratio of magnetic to kinetic energy (bottom panel). 
We observe that for all the simulations with magnetic fields, the total magnetic field energy in the gain region is smaller than the kinetic energy in the gain region by, at least, an order of magnitude, and the s15 models have a larger amount of field energy --- by a factor in the range of 3--10 --- than the s20 simulations. The growth of the field energy in the gain region is due to both an increasing volume of the gain region, the net accumulation of field energy through the shock minus that lost through the inner boundary, and the conversion of fluid energy into field energy inside the gain region 

Before convection begins, the magnetic field energy in the gain region grows only due to net accretion of the field through the shock and due to compression. After convection is initiated at $t_{\rm pb} \approx 80- 100\;{\rm ms}$, there is a temporary  and rapid increase in the growth rate of the magnetic field energy which lasts for $\sim 20\;{\rm ms}$ in the two T12 simulations. This feature is similar to the one seen in Matsumoto, Takiwaki and Kotake \cite{2024MNRAS.528L..96M} which they attribute to the operation of a small-scale (or fluctuation) dynamo formed once the neutrino-driven convection is initiated. 
After $t_{\rm pb} \approx 120\;{\rm ms}$ the field energy in the two T12 simulations grows at a rate similar to the kinetic energy. The field energy in the two T10 simulations also grows exponentially but for a much longer period of time. Unlike \cite{2025PhRvD.111f3042S}, we do not see any change in the exponential growth of the field energy in the in the T10 simulation when the outward motion of the shock is revived. 
In the case of the s15-T10 simulation, the shock reached the boundary of the computational domain before the growth of the field energy shows signs of slowing down. 
For the s20-T10 the figure indicates the growth of the field energy becomes similar to the growth of the kinetic energy at $t_{\rm pb} \approx 250\;{\rm ms}$. Note that the amount of magnetic field energy in the gain region for all our simulations shown in Fig. \ref{fig:Eb} is larger than the energies seen in \cite{2025PhRvD.111f3042S} which barely surpassed $10^{48} \;{\rm erg}$ in any of their models, but are smaller than the field energy in the gain region seen in the simulation by \cite{2024MNRAS.528L..96M} who found a total field energy greater than $10^{49} \;{\rm erg}$ for $t_{pb} \gtrsim 125\;{\rm ms}$. 

The similarity of the growth rates of the field energy and kinetic energy in the T12 simulations after $t_{\rm pb} \sim 120 \;{\rm ms}$ suggests that the ratio of the two may be almost constant. This is verified in the bottom panel of Figure \ref{fig:Eb} where we plot the ratio of magnetic to kinetic energy for the four simulations with magnetic field. 
In the two T12 simulations, we still see the rapid rise in the magnetic field energy around $t_{\rm pb} \approx 80- 100\;{\rm ms}$ and thereafter the ratio of the two energies is almost constant at $E_b / E_{\mathrm{kin}} \approx 5\%$ for the simulation using the s20 progenitor, and 8\% for the simulation using the s15 progenitor. We note that this plateau in the ratio is reached before the shock is revived at $t_{\rm pb}\sim 150\;{\rm ms}$. 
In contrast, the ratio of the field energy to kinetic energy in the gain region in the two T10 simulations grows approximately exponentially until the magnetic field energy is a few percent of the kinetic energy. Thereafter the ratio appears to approach the same, approximately constant, ratio of $E_b / E_{\mathrm{kin}} \approx 5\% - 8\%$. We caution that this value for the ratio is likely to be influenced by the spatial resolution of the simulations --- see, for example, Endeve \emph{et al.} \cite{2010ApJ...713.1219E} and Varma and M{\"u}ller \cite{2026MNRAS.tmp..593V} who both showed how the spatial resolution affected the saturation level of the field in their simulations.
For the T10 simulations this plateau stage does not occur until well after the shock has been revived. During the exponential growth phase between $\sim 120\;{\rm ms}$ and $\sim 200\;{\rm ms}$, the growth rates in the two T10 simulations are $71.5\;{\rm s}^{-1}$, equivalent to a growth timescale of 14 ms. This is longer than the 3 ms timescale found in Matsumoto, Takiwaki and Kotake \cite{2024MNRAS.528L..96M} but smaller than the timescale of 60 ms found by Endeve \emph{et al.} \cite{Endeve_2012} for the growth of the magnetic field energy in their simulations (although the reader should be aware that the two setups are different and may not be directly comparable), and the 43 ms timescale for the growth of the field found by Sykes and M{\"u}ller \cite{2025PhRvD.111f3042S} in their simulations using a different set of progenitors. We have not attempted to determine the timescale for the field growth during the short, $\approx 20$ ms, period around $t_{\rm pb} \approx 80- 100\;{\rm ms}$ in the two T12 simulations.  

The total kinetic energy (see Figure \ref{fig:Eb} middle panel) grows over time from an initial value of order $10^{48}\;{\rm erg}$ at $t_{\rm pb} = 60\;{\rm ms}$ up to $10^{49}-10^{50} \;{\rm erg}$ by $t_{\rm pb} = 200\;{\rm ms}$. This growth is due to net accretion of kinetic energy through the shock minus that lost through the inner boundary of the gain region, as well as work done on the fluid inside the gain region. There is not a distinctive change in the growth of the total kinetic energy at $\approx 80-  100$ ms postbounce when convection begins. We note that the s15 models have a larger amount of kinetic energy in the gain region than the s20 models after $t_{\rm pb} \approx 120\;{\rm ms}$ because the gain region is larger in the s15 simulations.


\subsection{Energy exchange between the fluid and the field}
\label{sec:energy-exchange}

The clear result from Figure \ref{fig:Eb} is that the energy in the magnetic field in the gain region relative to the kinetic energy grows over time until it reaches an equilibrium, but not equipartition. The energy in the field has to come from somewhere. 
From equation \eqref{eq:magnetic} one can derive that the magnetic field energy density $E_{\mathrm{mag}} = ( \vec{b} \cdot \vec{b} )/2$ grows as
\begin{eqnarray}
\frac{\partial E_{\mathrm{mag}}}{\partial t} + \nabla \cdot \vec{S}_{\mathrm{mag}} & = & - \vec{v} \cdot \left[ (\nabla \times \vec{b}) \times \vec{b} \right], \label{eq:dEmagdt} \\
 & = & - (\nabla \times \vec{b}) \cdot \left[  \vec{b} \times \vec{v}\right] \label{eq:dEmagdt2} \\
 & = & \vec{v} \cdot \nabla P_{\mathrm{mag}}  - \vec{v} \cdot (\vec{b} \cdot \nabla)\,\vec{b}
\label{eq:dEmagdt3}
\end{eqnarray}
where $\vec{S}_{\mathrm{mag}} = \vec{b} \times ( \vec{v} \times \vec{b})  = \vec{v}\,( \vec{b} \cdot \vec{b}) - \vec{b}\,( \vec{v} \cdot \vec{b})$ is the Poynting vector. Note that if the velocity and magnetic field are aligned/anti-aligned, the Poynting vector is zero. The term on the right hand side of equation (\ref{eq:dEmagdt}) is the rate at which magnetic field energy is exchanged with the fluid. As equation (\ref{eq:dEmagdt2}) indicates, if the velocity and magnetic field are aligned/anti-aligned then this term is also zero. Thus the fluid can only do work on the magnetic field if the fluid velocity is not parallel/anti-parallel with the magnetic field. The right hand side of equation (\ref{eq:dEmagdt3}) shows that the rate of energy exchanged between the field and the fluid can be split into the power exchanged due to the magnetic pressure gradient, and the power due to the magnetic tension. 
Even though the magnetic field energy grows at the expense of the fluid, simultaneously work is being done on the fluid by gravity and pressure forces plus the fluid in the gain region is being heated by the neutrinos. These additional sources of energy for the fluid can more than compensate for the energy lost to the field.

\begin{figure}
    \includegraphics[width=0.48\textwidth]{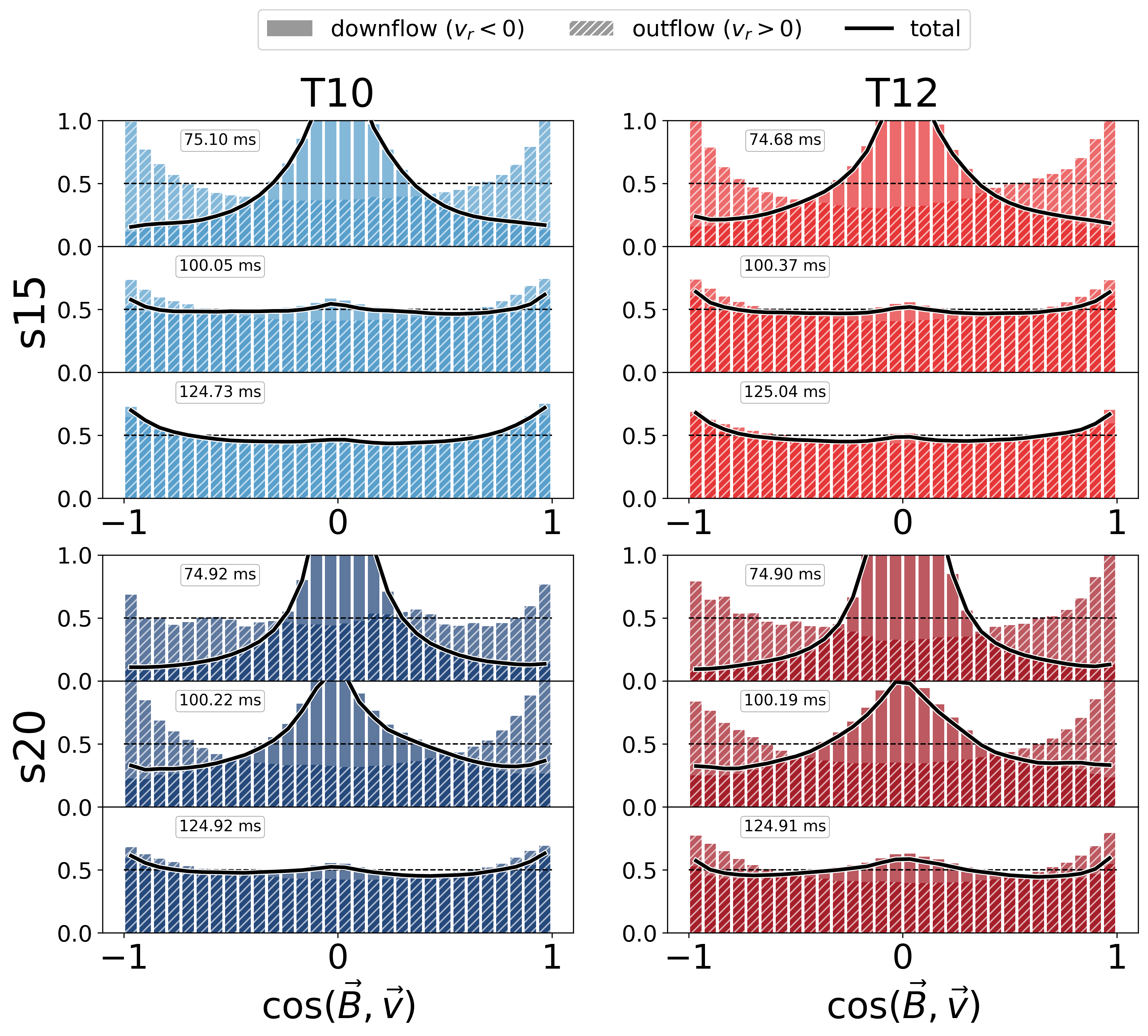}     
    \caption{The angle between the fluid velocity and magnetic field within the gain region at different snapshot times for all four simulations with magnetic field. Top: s15 simulations. Bottom: s20 simulations. Left: T10 simulations. Right: T12 simulations. The two styles of hatching indicate downflows (negative radial velocity; solid) and outflow (positive radial velocity; hatched). The solid black line represent the entire gain region (outflows and downflows). Note that each histogram is normalized separately. 
    The thin black line at 0.5 represents the expected distribution if two vectors are randomly oriented. 
    \label{fig:Bvalignment}
    }
\end{figure}

\begin{figure}
\textbf{$\vec{v}\cdot\left((\nabla\times\vec{b})\times\vec{b}\right)$ \\}
\vspace{0.5em}
    \textbf{\hspace{-1.5em}S20-T10}
    \includegraphics[
        width=\linewidth,
        trim={0cm 2.45cm 0cm 1.7cm},
        clip
    ]{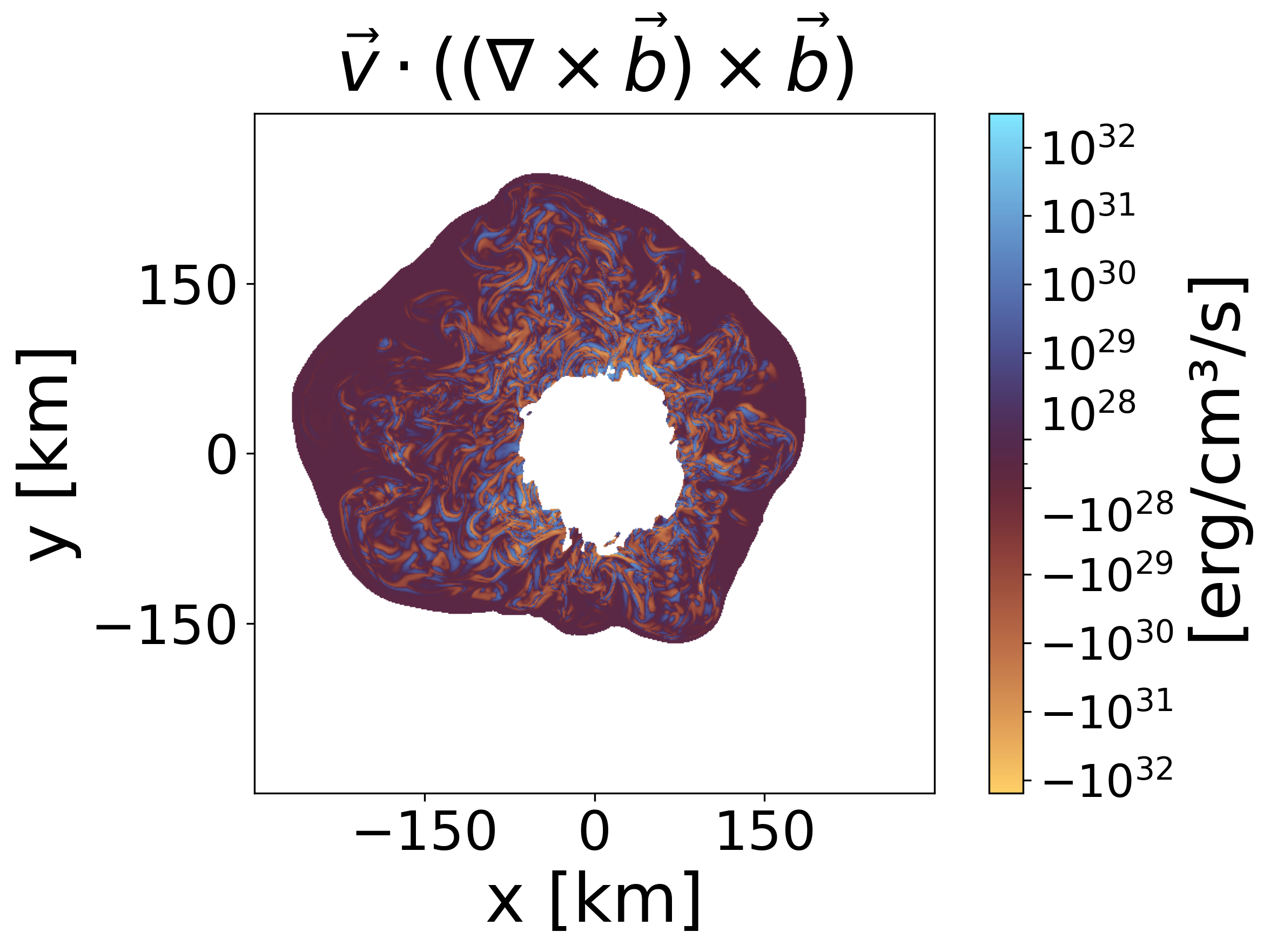}
\vspace{0.3em}
    \textbf{\hspace{-1.5em}S20-T12}
    \includegraphics[
        width=\linewidth,
        trim={0cm 0cm 0cm 1.7cm},
        clip
    ]{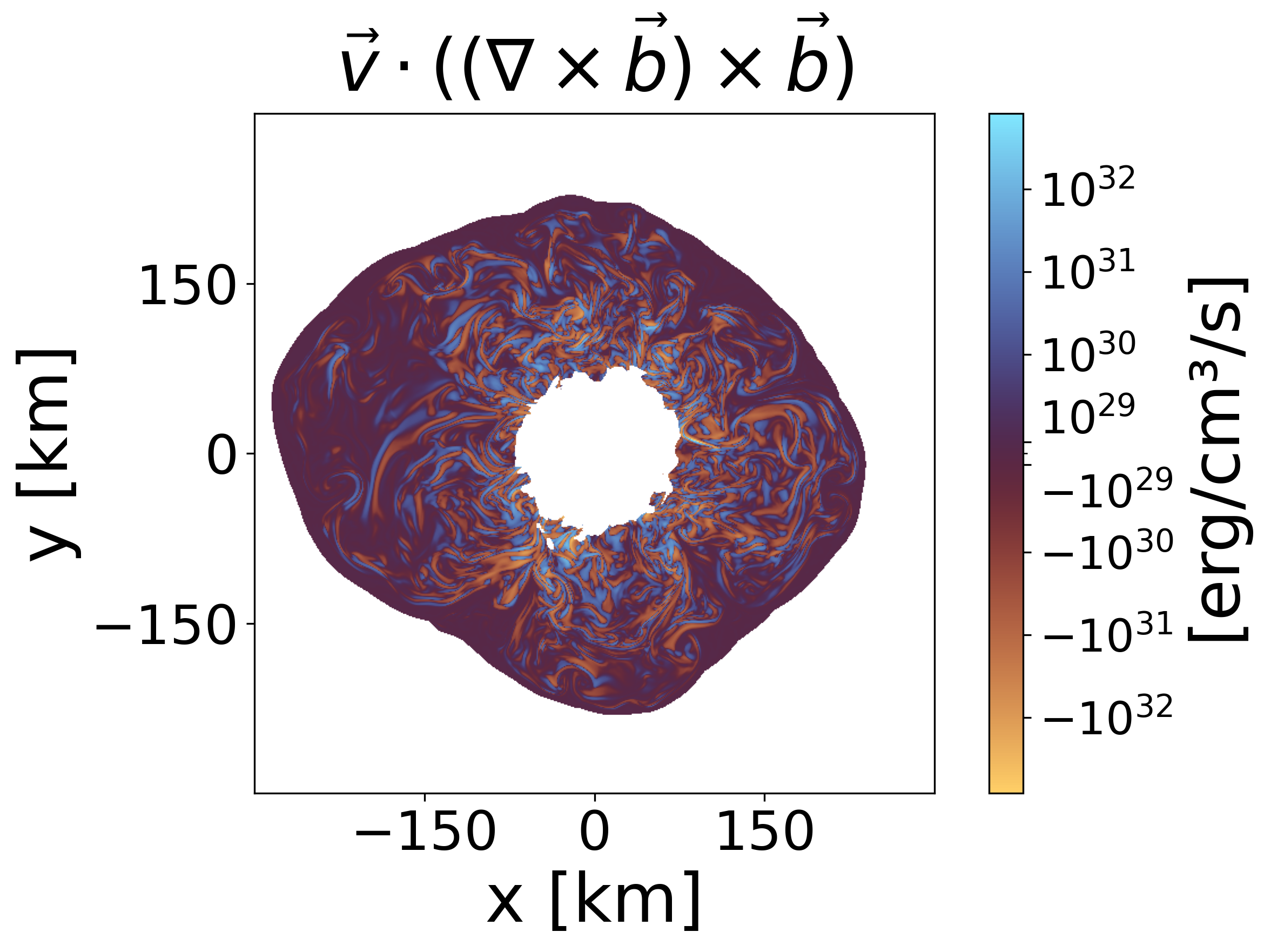}
\caption{
Maps of the rate of energy exchange between the magnetic field and fluid at
$t_{\rm pb} \approx 193\,\mathrm{ms}$ for s20-T10 (top) and s20-T12
(bottom).
\label{fig:energyrates}
}
\end{figure}

Given the rate of energy exchange between fluid and field depends upon the relative orientation of the two vectors, we show in Figure \ref{fig:Bvalignment}  histograms of the alignment of the fluid velocity and magnetic field within the gain region for the four simulations with a field at three snapshots in time. 
We further divide the gain region into downflows (negative radial velocity) and outflows (positive radial velocity), indicated by different hatching patterns. %
Also shown is the expectation if the two vectors are random and have no correlation. At $t_{\rm pb} \approx 75\;{\rm ms}$ all simulations show the two vectors are preferentially orthogonal in the downflows --- recall that the initial field we adopt is toroidal while the velocity will be almost radially inward in the downflows --- while in the outflows the velocity and magnetic field are preferentially parallel / antiparallel. 
At $t_{\rm pb} \approx 100\;{\rm ms}$ we still observe this strong distinction of the alignment in the downflows and outflows of the two s20 simulations but it has almost disappeared in the two s15 simulations. By $t_{\rm pb} \approx 125\;{\rm ms}$ the alignment of the two vectors in the two s20 simulations is now very similar to the two s15 simulations. In all simulations shown the alignment of the velocity and magnetic field at $t_{\rm pb} \approx 125\;{\rm ms}$ exhibits little difference between the downflows and outflows, and in both flows the alignment is almost consistent with a uniform distribution. We also find little distinction between T10 and T12 for a given progenitor. The alignment of the velocity and magnetic field becomes almost uniform in the two T12 simulations by $t_{\rm pb} \approx 125\;{\rm ms}$ is perhaps not surprising because, as seen in the bottom panel of Figure \ref{fig:Eb}, the field reaches an equilibrium with the kinetic energy. 
What \emph{is} surprising is that the almost-uniform alignment of the velocity and magnetic field is also seen in the two T10 simulations at $t_{\rm pb} \approx 125\;{\rm ms}$---even though the magnetic field energy relative to the total kinetic energy is growing exponentially at this time; i.e.\ before any equilibrium is reached.  
An explanation for this observation is if the field and fluid are exchanging energy back and forth in close proximity, and the field energy growth rate is a small fraction of the energy turnover rate between the field and fluid.

To see if this is the case, we show in Figure \ref{fig:energyrates} the field-fluid energy exchange rate $\vec{v} \cdot \left[(\nabla \times \vec{b} ) \times \vec{b} \right]$ at the snapshot time of $t_{\rm pb} \approx 193\;{\rm ms}$ for the two s20 simulations. This time was chosen because it is before the equilibrium of the field energy and fluid kinetic energy in the s20-T10 simulation i.e. the field energy is still growing; see Figure \ref{fig:Eb}. The results shown in  Figure \ref{fig:energyrates} are very similar to those found in \cite{2010ApJ...713.1219E}, and should also be compared with the plots of entropy from the third row of Figure \ref{fig:entropy_s20}. Figure \ref{fig:energyrates} shows the energy exchange between the fluid and field occurs predominantly around the inner boundary of the gain region,
and we do not observe any features coincident with the downflows seen in Figure \ref{fig:entropy_s20}. Figure \ref{fig:energyrates} confirms that the energy exchange occurs in \emph{both} directions around the inner boundary of the gain region and within a random small subvolume, is not preferentially from fluid to field. Instead we observe long ribbons where the fluid is losing energy to the field adjacent to ribbons where the fluid is gaining energy. 
These features are suggestive of magnetoacoustic waves \cite{10.3389/fspas.2019.00020} but we do not attempt to prove this in this paper.


\begin{figure}[htb]
\begin{center}
\begin{tabular}{cc}
\textbf{s20-T10} & \textbf{s20-T12} \\
\includegraphics[trim=0.0cm 0.1cm 1.4cm 0.1cm, clip,width=4.3cm]{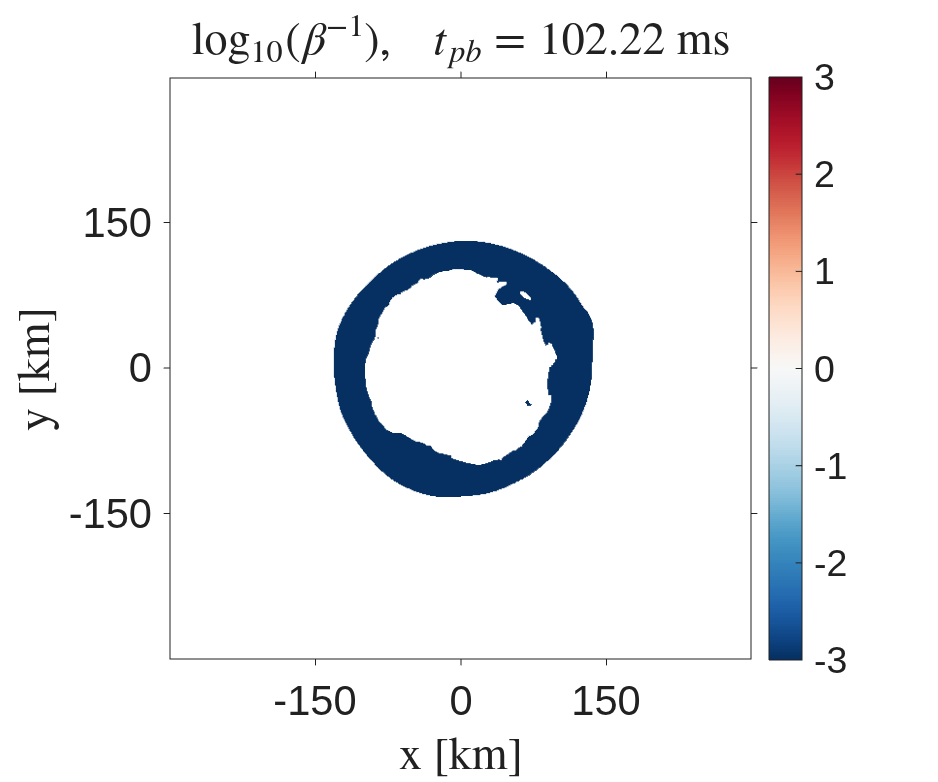}\hspace*{0.0cm} &
\includegraphics[trim=0.0cm 0.1cm 1.4cm 0.1cm, clip,width=4.3cm]{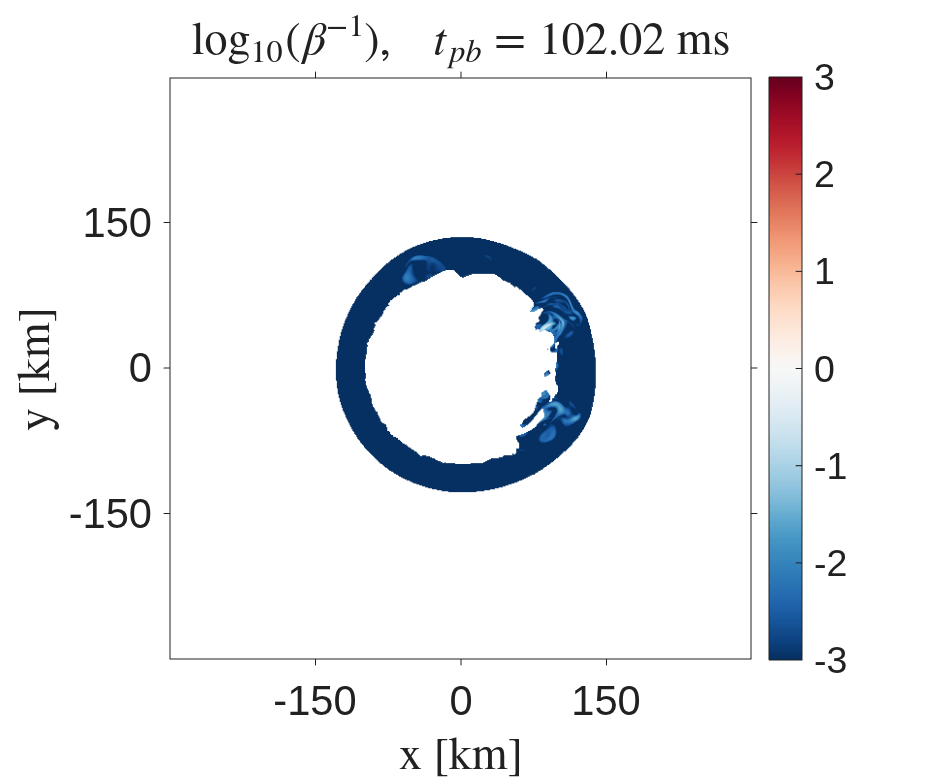}\\
\includegraphics[trim=0.0cm 0.1cm 1.4cm 0.1cm, clip,width=4.3cm]{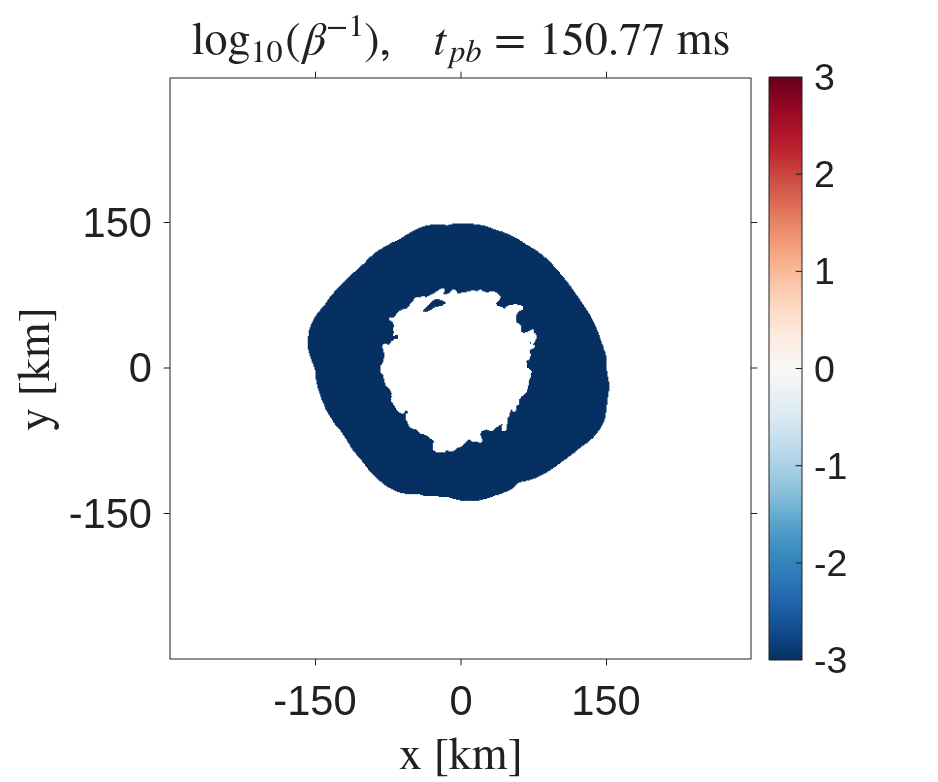}\hspace*{0.0cm}&
\includegraphics[trim=0.0cm 0.1cm 1.4cm 0.1cm, clip,width=4.3cm]{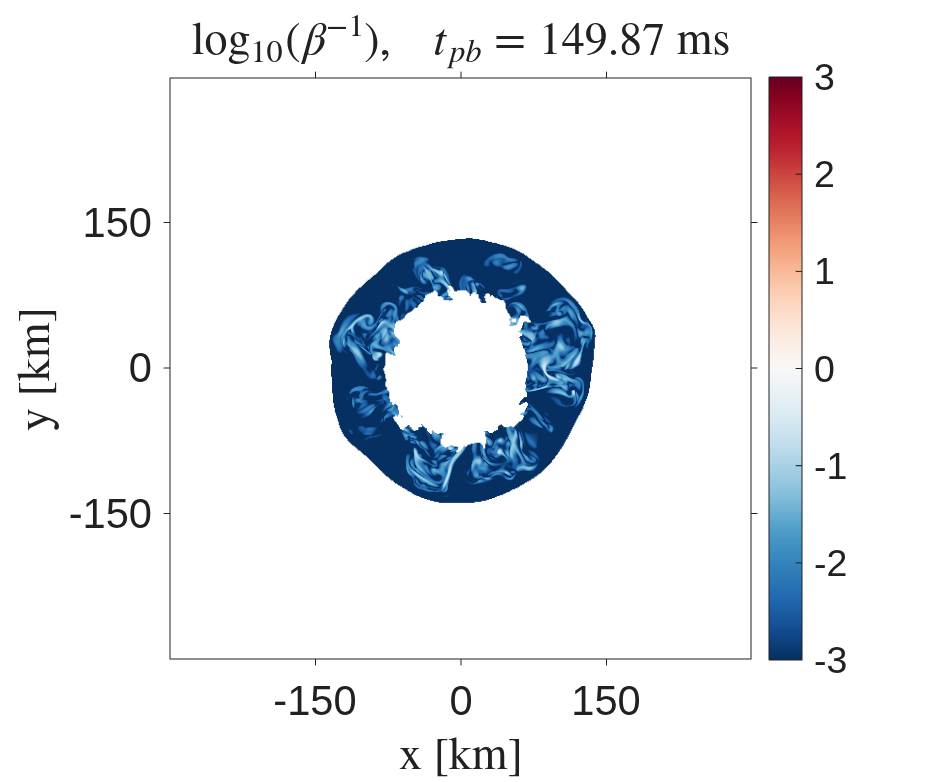}\\
\includegraphics[trim=0.0cm 0.1cm 1.4cm 0.1cm, clip,width=4.3cm]{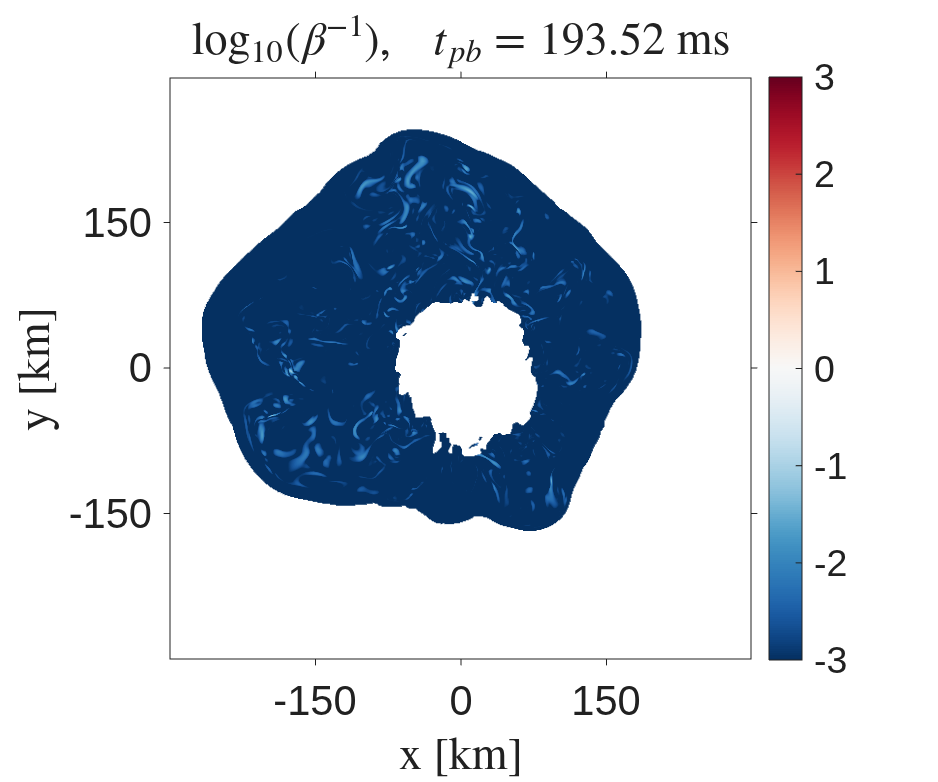}\hspace*{0.0cm}&
\includegraphics[trim=0.0cm 0.1cm 1.4cm 0.1cm, clip,width=4.3cm]{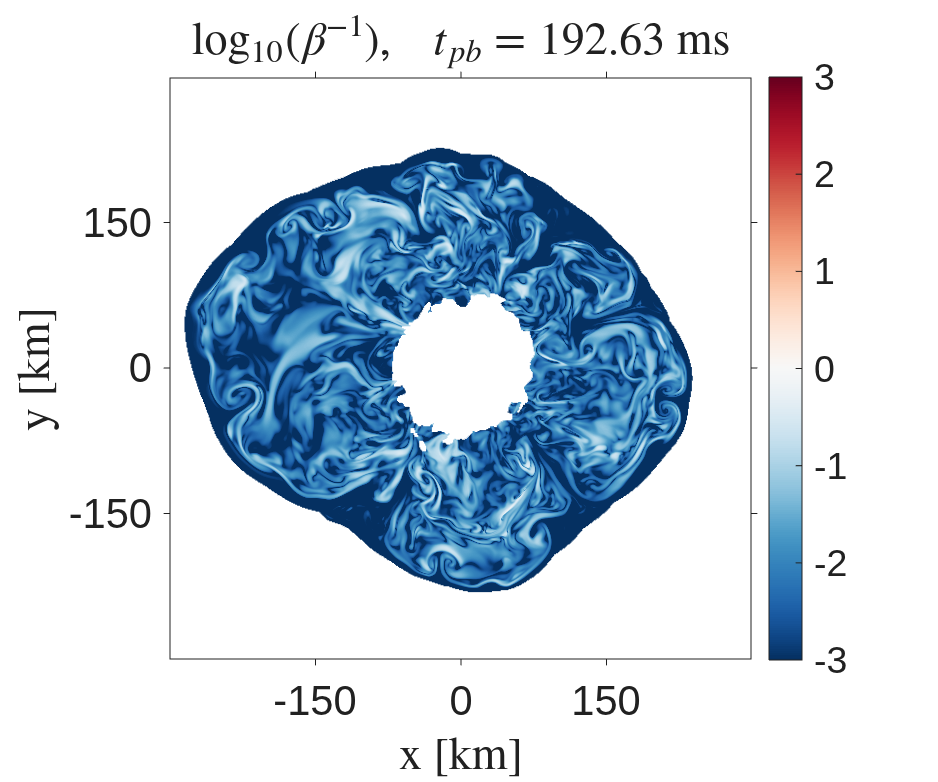}\\
\includegraphics[trim=0.0cm 0.1cm 1.4cm 0.1cm, clip,width=4.3cm]{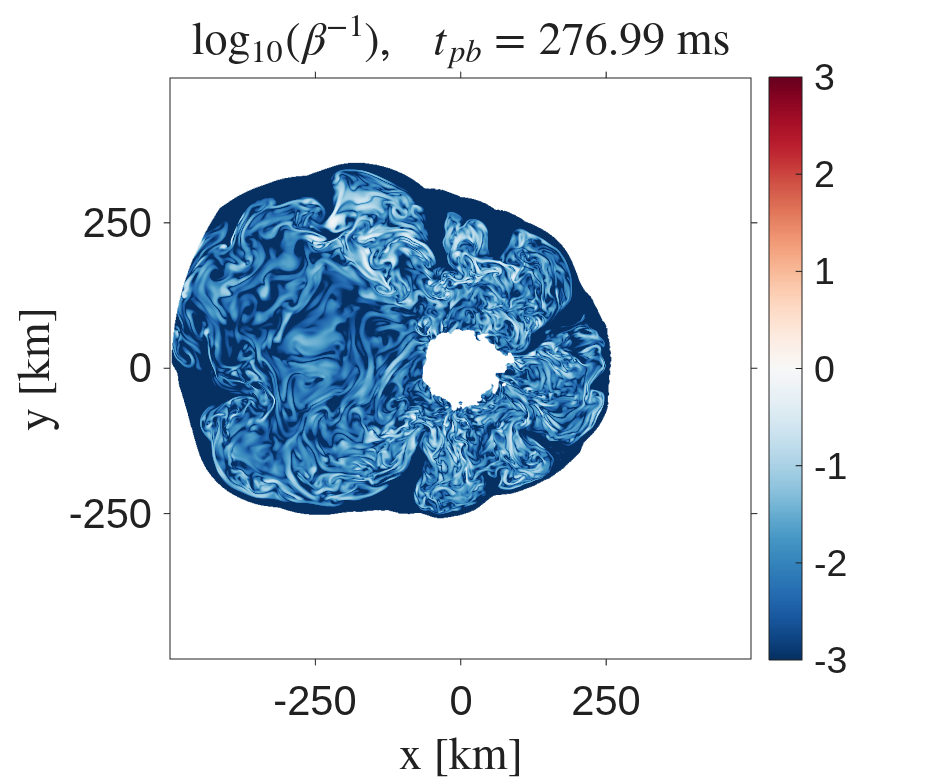}\hspace*{0.0cm}&
\includegraphics[trim=0.0cm 0.1cm 1.4cm 0.1cm, clip,width=4.3cm]{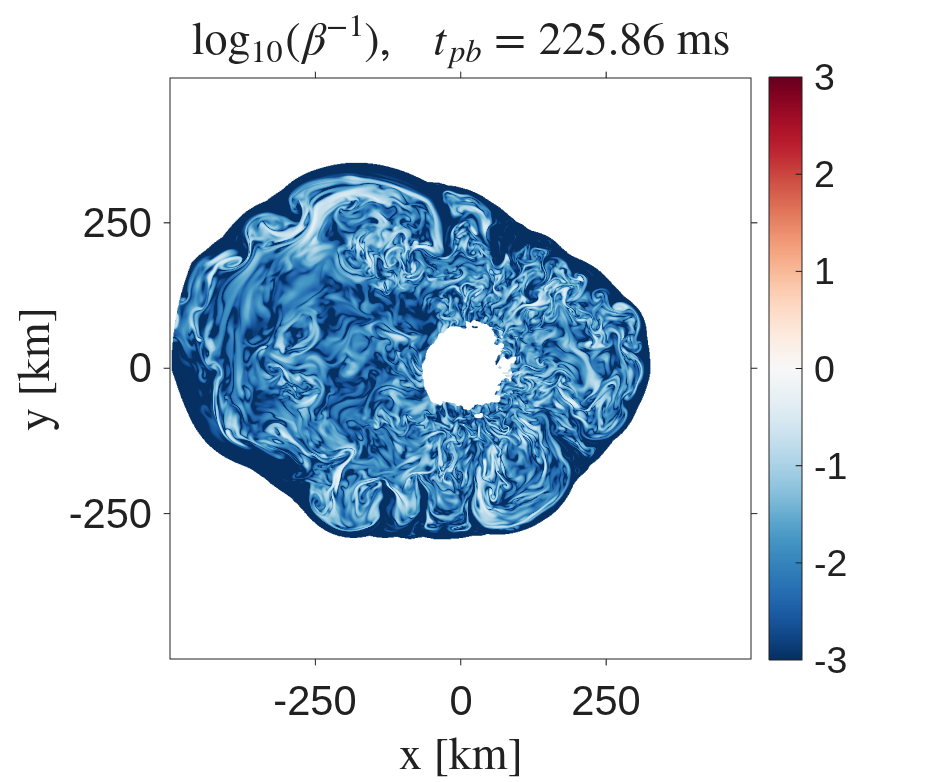}\\
\end{tabular}
\end{center}
\caption{Maps of inverse of plasma beta, scaled logarithmically, for the 
s20-T10 run (left column) and 
s20-T12 run (right column) 
at the post-bounce times (from top to bottom) of $\sim 100$~ms, $\sim 150$~ms, $\sim193$~ms, and at the final simulation time. 
Note the change of domain size in the last row as the shock expands.
\label{fig:plasma_beta_s20} 
}
\end{figure}

\subsection{The Spatial Structure of the Magnetic Field}
\label{sec:Bfield-structure}

As Figure \ref{fig:Eb} shows, globally throughout the gain region, the total magnetic field energy in the gain region is always significantly less than the total kinetic energy in the gain region. However, locally we find pockets of material where the magnetic field energy density is similar to the local kinetic energy density, or similarly, where the magnetic field pressure is similar to the thermal pressure. Such local enhancements of the field pressure were also seen in \cite{Endeve_2012}. 
We adopt the parameter `plasma beta', defined as $1/\beta=P_{\mathrm{mag}}/P_{\mathrm{th}}$, to quantify the strength of the magnetic pressure relative to the thermal pressure. 
In Figure \ref{fig:plasma_beta_s20} we show spatial maps of plasma beta for four select time slices 
of the s20-T10 and 
of the s20-T12 simulation as an example. 
The results for the two s15 simulations are similar.  
We find that at $t_{\rm pb} \approx 100\;{\rm ms}$, the s20-T10 simulation has a negligible contribution from the field pressure to the total pressure, and even by $t_{\rm pb} \approx 150\;{\rm ms}$, there are only faint hints that $1/\beta$ is not zero in small regions of the gain region.
At the time when the s20-T10 simulation was terminated, the field pressure grew to be as large (or even larger) than the thermal pressure in some places. In the s20-T12 simulation, the contribution from the magnetic pressure to the total pressure is apparent 
even at $t_{\rm pb} \approx 100\;{\rm ms}$  
and by $t_{\rm pb} \approx 150\;{\rm ms}$ one can find places where $1/\beta$ is $\mathcal{O}(1)$.

Generally, $1/\beta$ is smallest immediately below the shock and in the low-entropy accretion downflows which thread the gain region, though the correspondence is not exact. Within the high-entropy heated fluid there is a great deal of intermittency of $1/\beta$ with large variations over small distances i.e. over scales much smaller than the size of the gain region. This suggests that the field is generated at the inner boundary of the gain region and then the tangled field is transported towards the shock in the convective plumes.

\section{The Fluid Kinetics}
\label{sec:fluid_energies}


\subsection{Turbulence} 
\label{sec:turbulence}

\begin{figure}
    \includegraphics[width=0.48\textwidth]{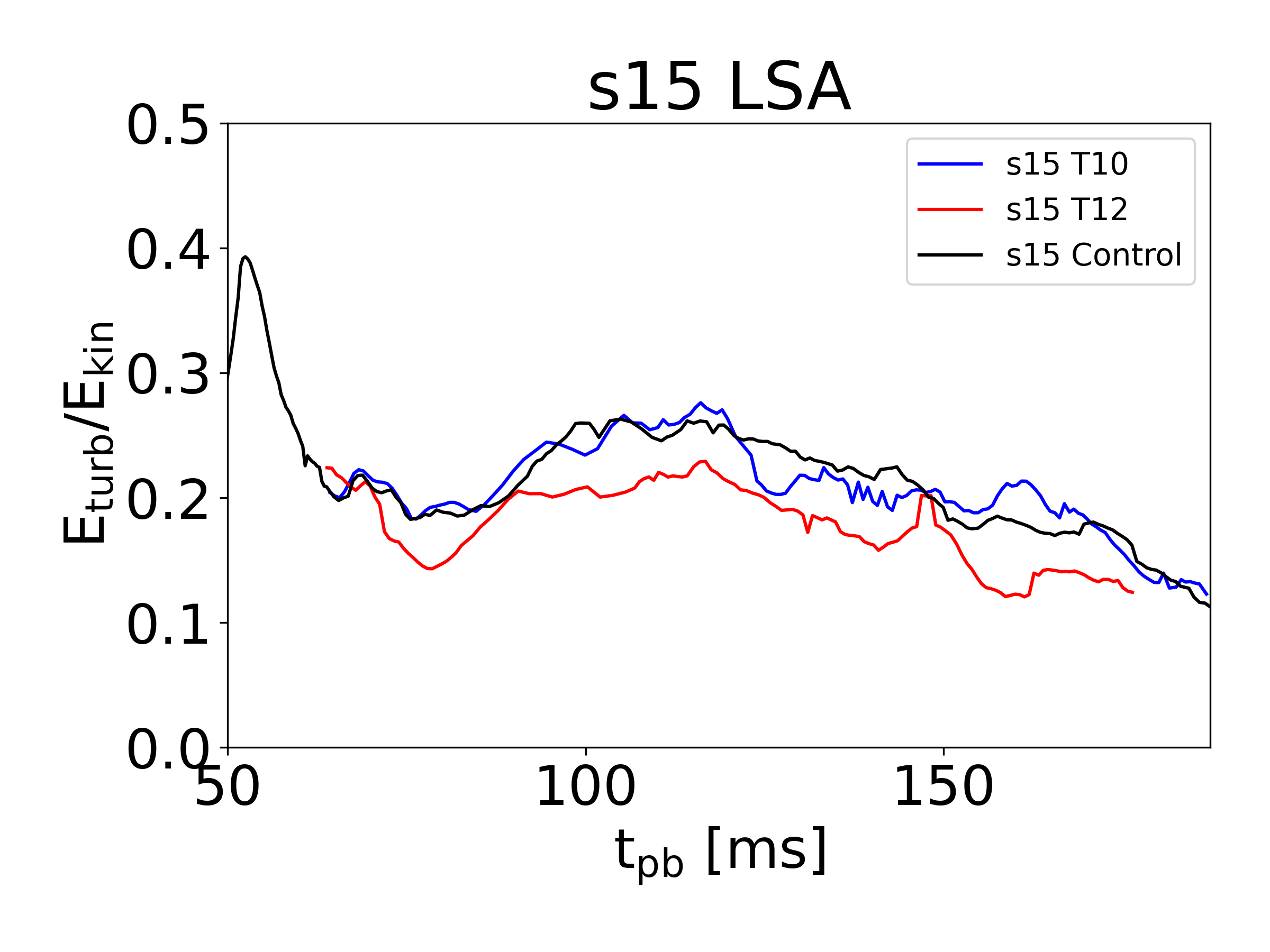}
    \includegraphics[width=0.48\textwidth]{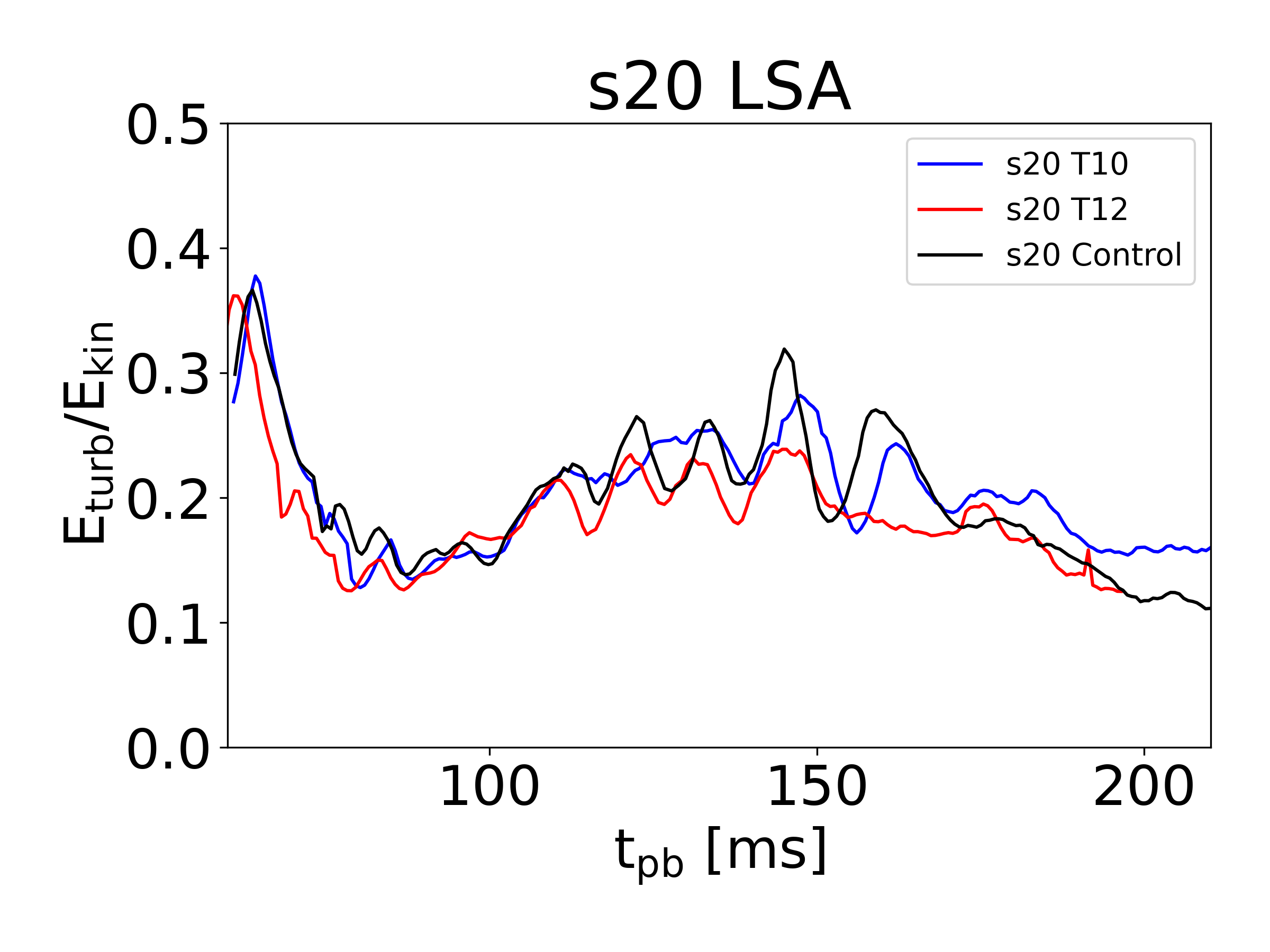}
    \caption{The ratio of turbulent kinetic energy to total kinetic energy (both taken in the gain layer only) as a function of post-bounce time. Top: s15 simulations. Bottom: s20 simulations.
    \label{fig:TKE}
    }
\end{figure}

The kinetic energy of the fluid may be decomposed into two\footnote{Some authors divide the kinetic energy into three parts: the turbulent kinetic energy, the non-turbulent kinetic energy, and the turbulent momentum flux.} parts: the turbulent kinetic energy density (TKE), $E_{\turb}$, and (for lack of a better term) the non-turbulent kinetic energy\footnote{We avoid using the term `laminar kinetic energy' since this is often used in the literature with a different meaning than we use here.} density. By non-turbulent kinetic energy we mean the kinetic energy density associated with fluid flow on the largest scales, in our case, on the scale of the gain region; by turbulent kinetic energy we mean the energy associated with the fluid motion on much smaller scales i.e. the scale of the turbulent eddies. These two quantities need to be suitably defined, a task which is not trivial in the case of core-collapse supernovae. As in \cite{2025ApJ...995..109C}, we undertake a Reynold's decomposition of the fluid velocity so $\vec{v} = \langle \vec{v}\,\rangle + {\vec{v}}\,'$ where $\langle \vec{v}\,\rangle$ is an expectation value and ${\vec{v}}\,'$ is the deviation (fluctuation) from $\langle \vec{v}\,\rangle$. 
In a fluid which has reached a statistical steady state the expectation value of the velocity is the time-average of $\vec{v}$: in situations where the fluid is not in a statistical steady state, the expectation value is often computed as a spatial (volume) average. Of the two methods of this type considered in \cite{2025ApJ...995..109C}, we shall consider here only the Local Spatial Average (LSA) method: an approach that has been seen to yield maps of TKE which are a better match to other indicators of the turbulence. 
In the LSA method we define the expectation value of the velocity, $\langle \vec{v}\,\rangle$, for a particular grid cell to be the mass-weighted volume average of the velocity in a cube  of side length $L=20\;{\rm km}$ surrounding the cell (see \citet{2025ApJ...995..109C}). 
In practice, we only include those cells in the volume which are found to be inside the gain region, hence the averaging volume is not always a full cube. 
 Once computed, the local turbulent kinetic energy density is thus $E_{\turb} = \rho\,|{\vec{v}}\,'|^2 / 2$. 

The TKE relative to the total kinetic energy in the gain region as a function of post-bounce time is shown in Figure \ref{fig:TKE} for all six simulations (top: s15; bottom: s20). 
First, we note that the ratio of kinetic energies after $t_{pb} = 100\;{\rm ms}$ is approximately $0.2$. This is to be compared with the ratio of magnetic field energy relative the total kinetic energy seen in Figure \ref{fig:Eb}. With the caveat mentioned previously about measuring the TKE, the comparison indicates that the amount of TKE is larger than the magnetic field energy by a factor of order 2-5, similar to the ratios seen by Cho \emph{et al.} \cite{Cho_2009} and Haugen \emph{et al.} \cite{2003ApJ...597L.141H}. 
Looking more closely at the individual simulations, we see that for s15, there is no apparent difference in the ratio of TKE to total kinetic energy between the s15-control run and the s15-T10. 
In contrast, the s15-T12 simulation appears to show a somewhat smaller ratio of turbulent to total kinetic energy starting at $t_{\rm pb} \approx 80\;{\rm ms}$ when convection begins. 
A similar, although less pronounced, behavior can be seen in the s20 simulations, where the s20-T12 also hints at a small reduction of this ratio relative to the s20-control and s20-T10 runs.
However, we again caution the reader that the turbulent kinetic energy is a difficult quantity to measure, and the LSA method we use in this paper is known to yield values which depend upon the size of the averaging volume; see \cite{2025ApJ...995..109C}. Therefore the $5-10\%$ reduction of the ratio of the ratio of TKE to total kinetic energy shown in Figure \ref{fig:TKE} must not be regarded as being definitive and needs to be corroborated by other evidence.


\subsection{Vorticity and Enstrophy}
\label{sec:vorticity}

The presence of vorticity is often regarded as indicating the presence of turbulence in a fluid. Like the magnetic field\footnote{The similarity between turbulent hydrodynamics and electromagnetism has been noted on several occasions \cite{1998PhFl...10.1428M,2010FlDyR..42e5502K,2012PhPl...19a0702T}.}, the vorticity, $\vec{\omega} = \nabla \times \vec{v}$, is divergence free, i.e. $\nabla \cdot \vec{\omega} = 0$. From equations (\ref{eq:mass}) and (\ref{eq:momentum}) we find the vorticity evolves according to 
\begin{equation}
\frac{\partial\,\vec{\omega}}{\partial t} = \nabla \times ( \vec{v} \times \vec{\omega}) + \vec{m},
\label{eq:domegadt}
\end{equation}
where $\vec{m}$ is the baroclinic vector. 
The quantity $\vec{\omega} \times \vec{v} \def \vec{l}$ is known as the Lamb vector or Lamb force. 
Using a vector calculus identity and that $\nabla \cdot \vec{\omega} =0$, equation (\ref{eq:domegadt}) can be written as 
\begin{equation}
\frac{\partial\,\vec{\omega}}{\partial t} + (\vec{v} \cdot \nabla)\,\vec{\omega} = (\vec{\omega} \cdot \nabla)\,\vec{v} - \vec{\omega}\, (\nabla \cdot \vec{v})  + \vec{m}.
\label{eq:domegadt2}
\end{equation}
The left-hand of this equation is seen to be the convective derivative, and the first two terms on the right-hand side describe the effect of stretching and compression, respectively, and depend only on the gradients of the fluid velocity, not the gradients of the vorticity. 
Including the magnetic field contribution, the expression for the baroclinic vector $\vec{m}$ is given  
\begin{equation}
\vec{m} =  \frac{1}{\rho^2}\,\left( \nabla \rho \times \left[\nabla 
P + \vec{\mathcal{S}}_M \right]
\right)  + \frac{1}{\rho} \,\nabla \times \left( \frac{ (\vec{b} \cdot \nabla ) \, \vec{b} }{\rho} + \vec{\mathcal{S}}_M \right).
\label{eq:baroclinic}
\end{equation}
The first term on the right hand side of equation (\ref{eq:baroclinic}) includes the contribution due to the gradient of the thermal plus magnetic pressures, while the second term on the right hand side of equation (\ref{eq:baroclinic}) includes the contribution due to the magnetic tension. 
The presence of the magnetic field in both terms of the baroclinic vector indicates that the magnetic field affects the vorticity, and we note that the baroclinic vector depends upon the \emph{derivative} of the field. 
We do not require that the magnetic field dominate the pressure i.e. that $1/\beta = P_{\mathrm{mag}}/P_{\mathrm{th}} \geq 1$. It is possible that even in places where $1/\beta$ is small, the contribution of the magnetic field to this term in the baroclinic vector could be significant if the gradients of the field are large. The dependence upon the gradient of the field tension means the more tangled the magnetic field, the larger this term contributing to the baroclinic vector. 

As with the magnetic field, one can derive from equation (\ref{eq:domegadt}) and equation (\ref{eq:mass}) that the ratio $\vec{\omega} / \rho$ evolves according to 
\begin{equation}
\frac{\partial}{\partial t} \left(\frac{\vec{\omega}}{\rho} \right) + \left( \vec{v} \cdot \nabla \right) \left(\frac{\vec{\omega}}{\rho} \right) - \left(\frac{\vec{\omega}}{\rho} \cdot \nabla \right)\;\vec{v} = \frac{\vec{m}}{\rho}.
\label{eq:omegaoverrho}
\end{equation}
Again, the left hand side of this equation defines the fluid derivative of $\vec{\omega} / \rho$, the interpretation of which is the same as that of the combination $\vec{b}/\rho$. 
In the absence of a magnetic field and neutrino scattering / absorption / emission, and when the fluid is barotropic, the baroclinic vector vanishes and $\vec{\omega}/\rho$ remains fixed between any two fluid elements connected along a vortex line. Whether the vorticity increases or decreases as the distance $\Delta\vec{x}$ between the fluid elements changes depends upon how the density behaves. But in the presence of a magnetic field, it is possible for the right-hand side of equation (\ref{eq:omegaoverrho}) to be nonzero even if the fluid is barotropic, and the relationship between $\vec{\omega}/\rho$ and the separation $\Delta\vec{x}$ is severed. 

\begin{figure*}
\begin{tabular}{ccc}
\textbf{s20-Control} & \textbf{s20-T10} & \textbf{s20-T12} \\

\includegraphics[trim=0.0cm 1.1cm 1.0cm 0.1cm, clip, width=6cm]{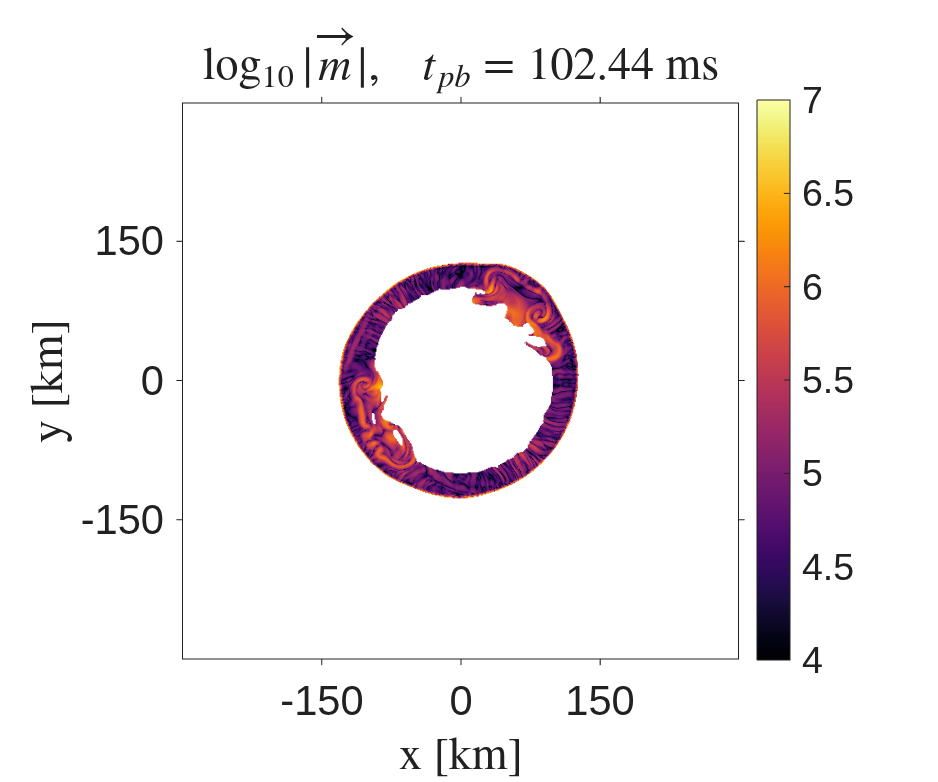}\hspace*{0.0cm}&
\includegraphics[trim=0.0cm 1.1cm 1.0cm 0.1cm, clip, width=6cm]{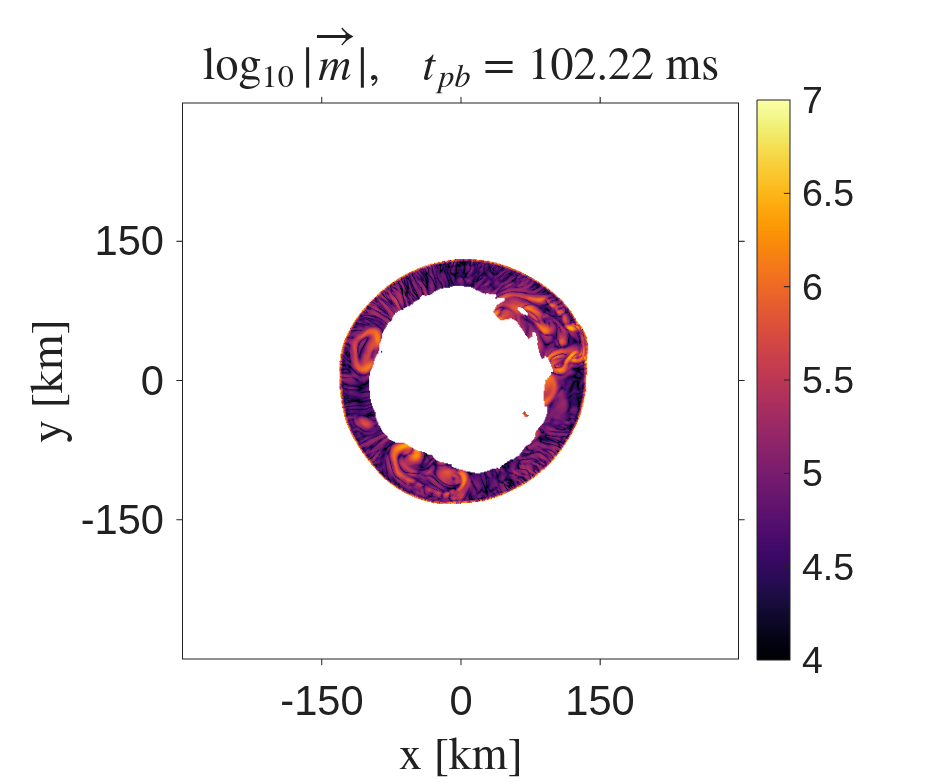}&
\includegraphics[trim=0.0cm 1.1cm 1.0cm 0.1cm, clip, width=6cm]{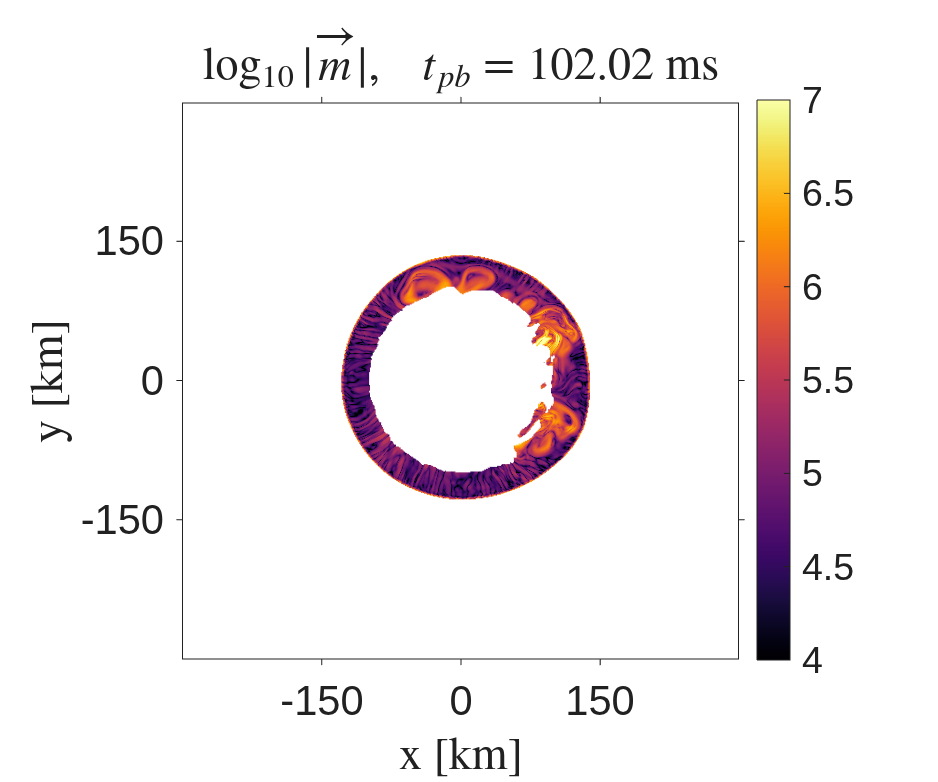}\\
\includegraphics[trim=0.0cm 1.1cm 1.0cm 0.1cm, clip, width=6cm]{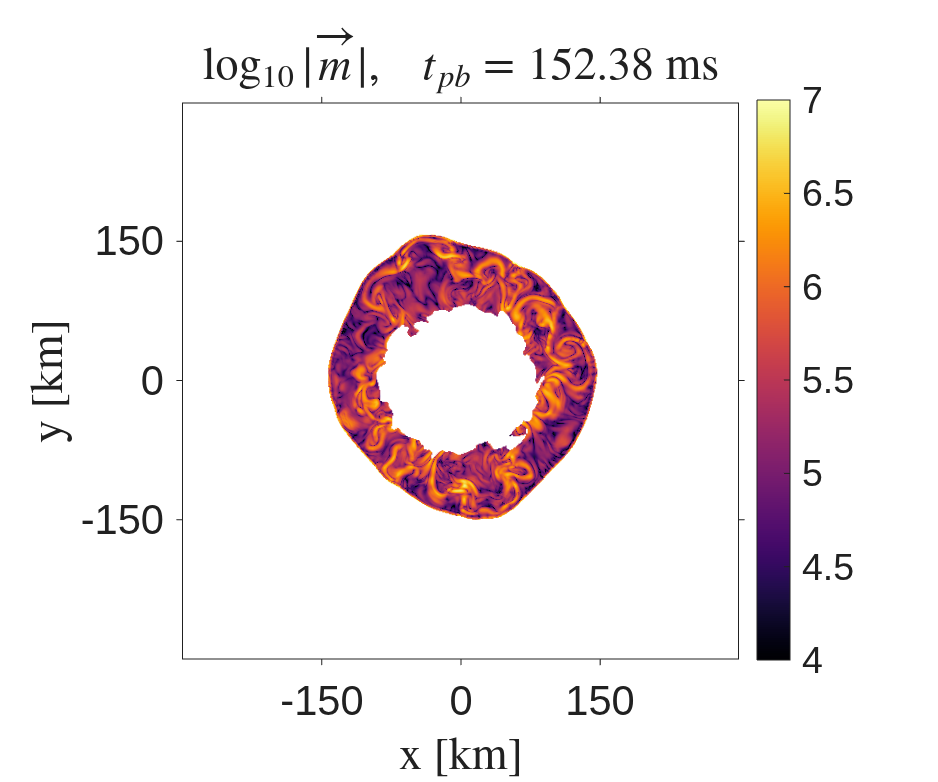}\hspace*{0.0cm}&
\includegraphics[trim=0.0cm 1.1cm 1.0cm 0.1cm, clip, width=6cm]{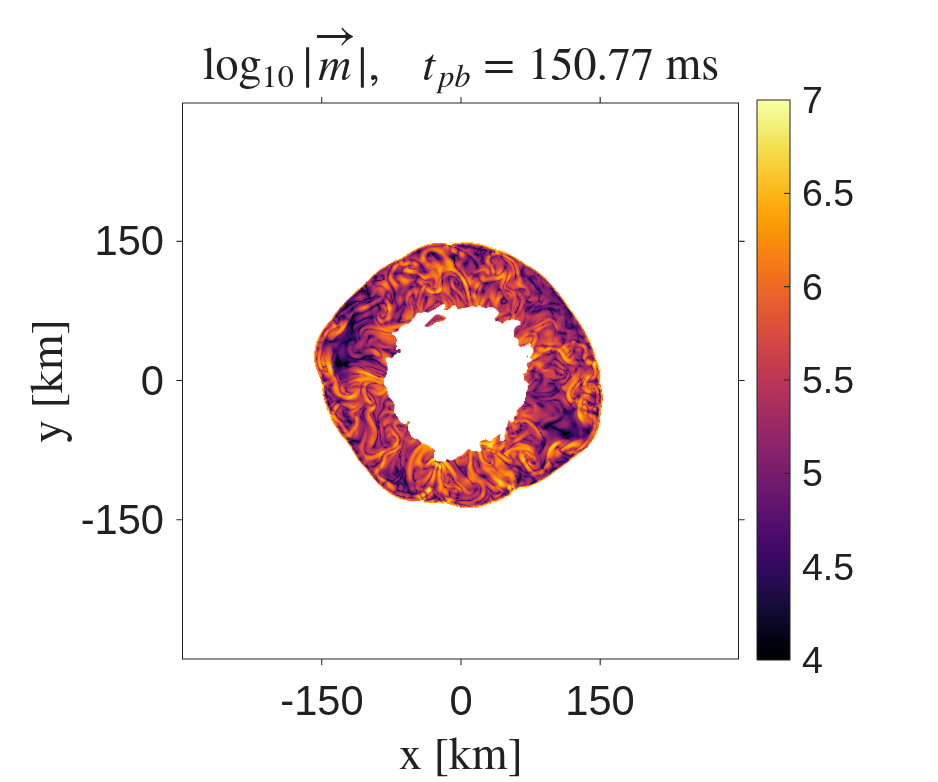}\hspace*{0.0cm}&
\includegraphics[trim=0.0cm 1.1cm 1.0cm 0.1cm, clip, width=6cm]{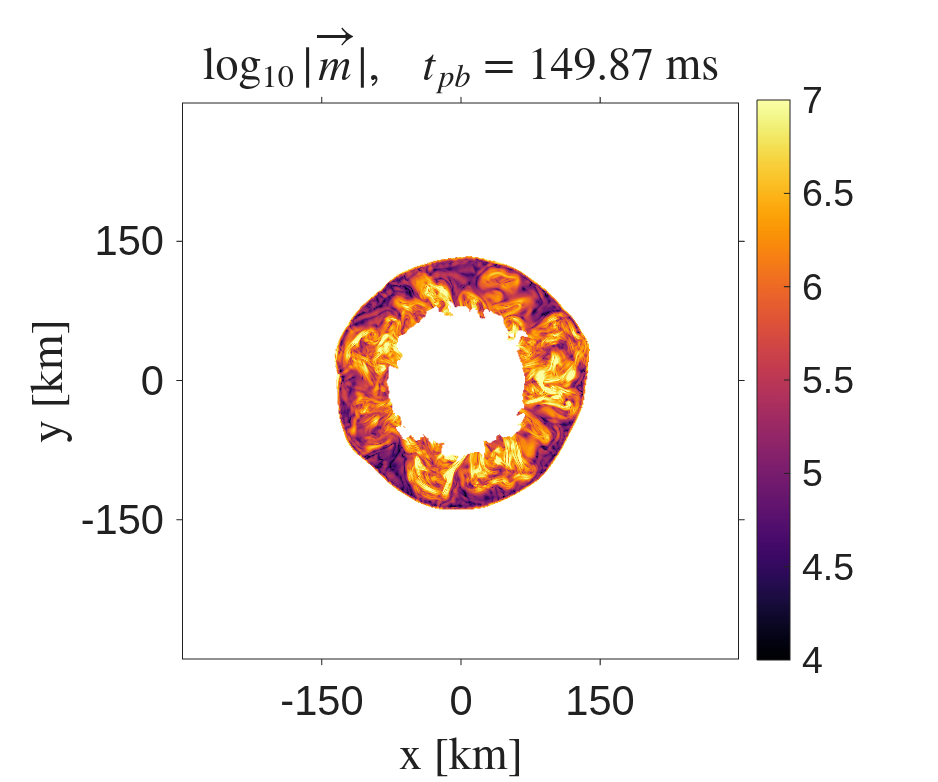}\\
\includegraphics[trim=0.0cm 1.1cm 1.0cm 0.1cm, clip, width=6cm]{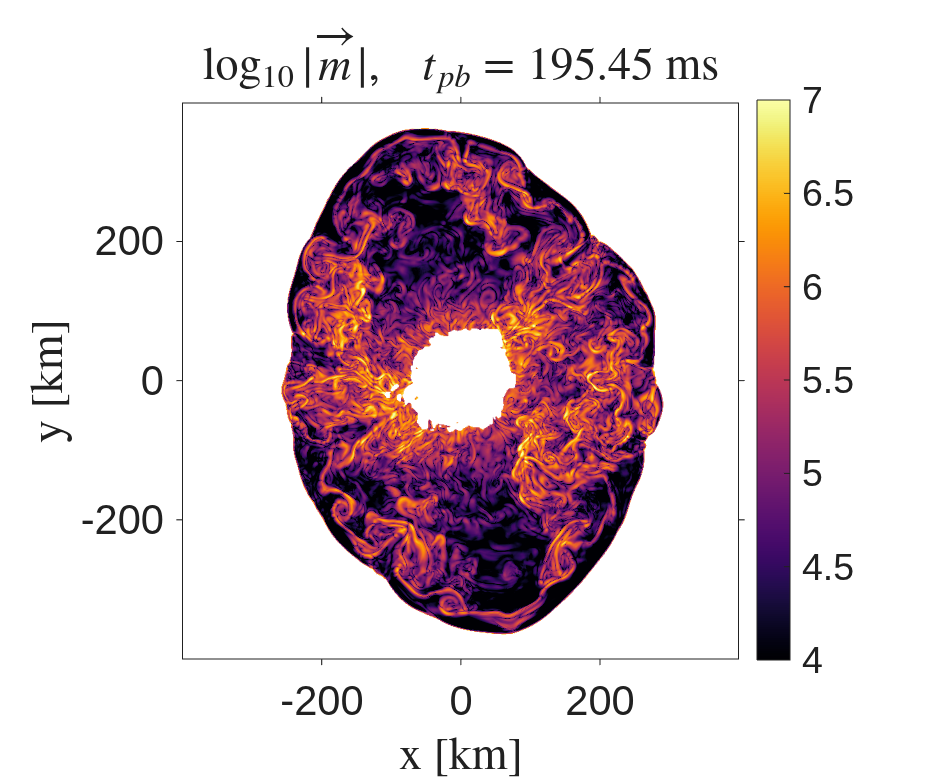}\hspace*{0.0cm}&
\includegraphics[trim=0.0cm 1.1cm 1.0cm 0.1cm, clip, width=6cm]{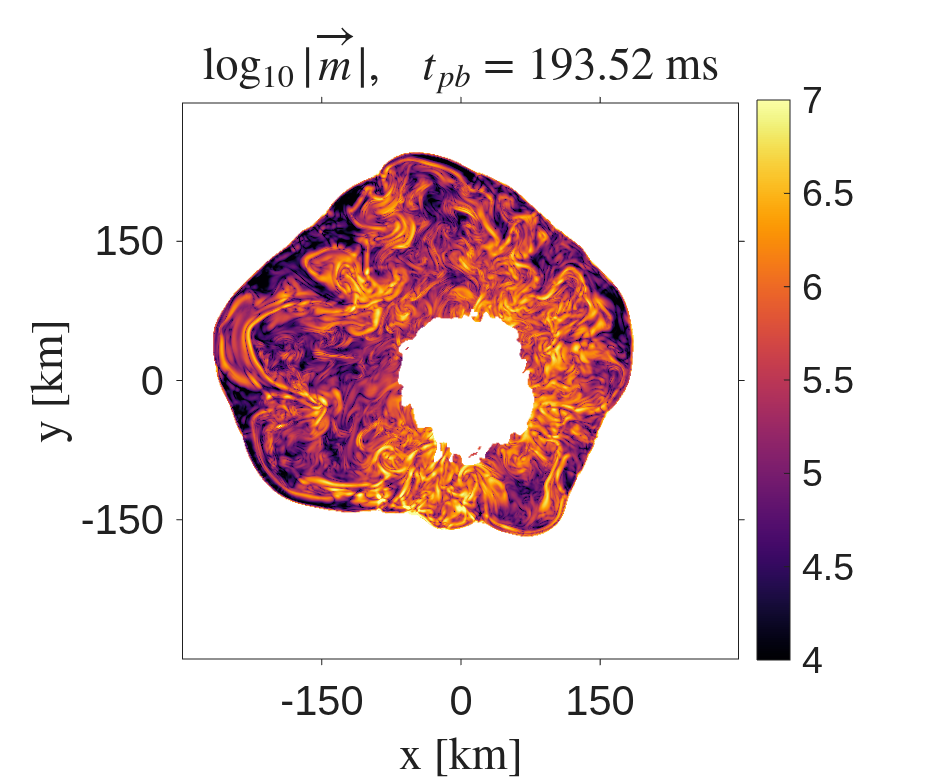}\hspace*{0.0cm}&
\includegraphics[trim=0.0cm 1.1cm 1.0cm 0.1cm, clip, width=6cm]{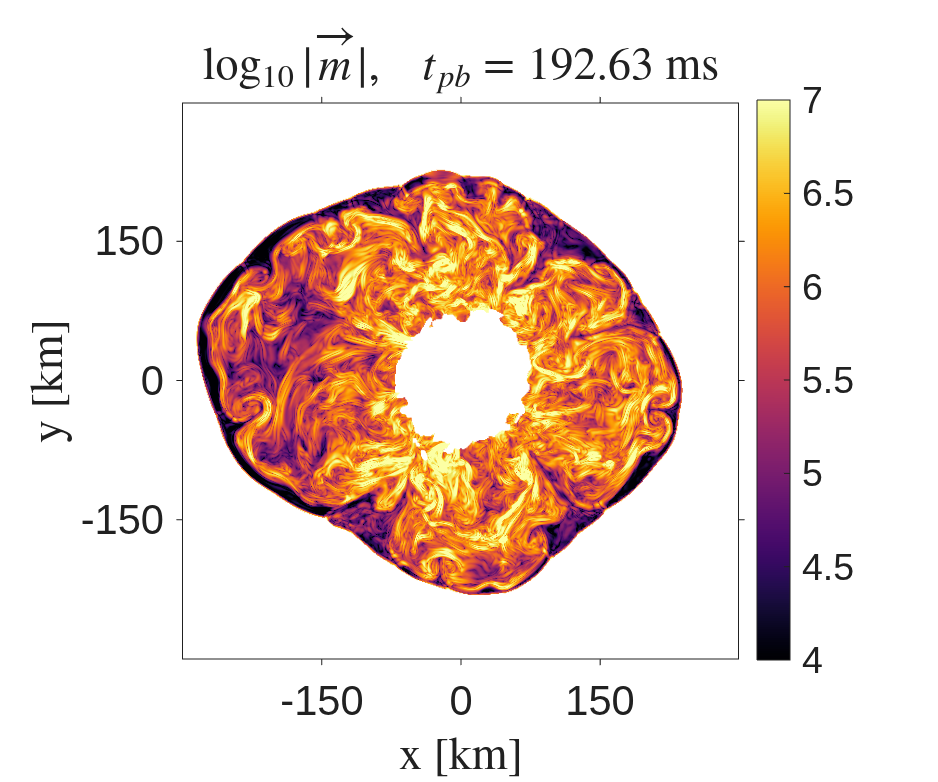}\\
\includegraphics[trim=0.0cm 0.1cm 1.0cm 0.1cm, clip, width=6cm]{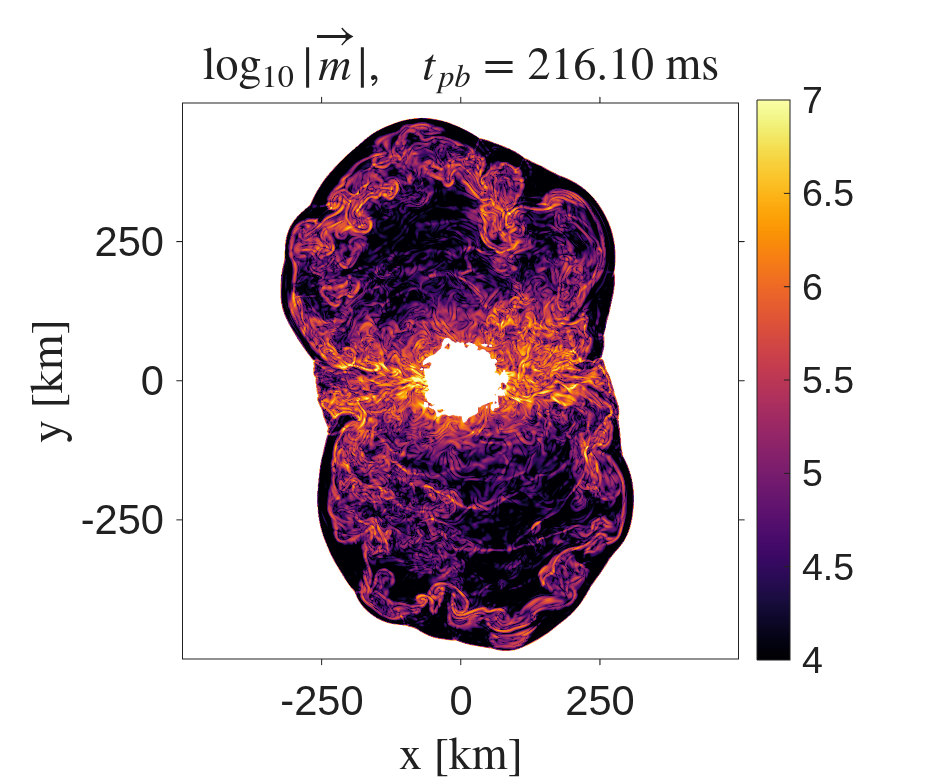}\hspace*{0.0cm}&
\includegraphics[trim=0.0cm 0.1cm 1.0cm 0.1cm, clip, width=6cm]{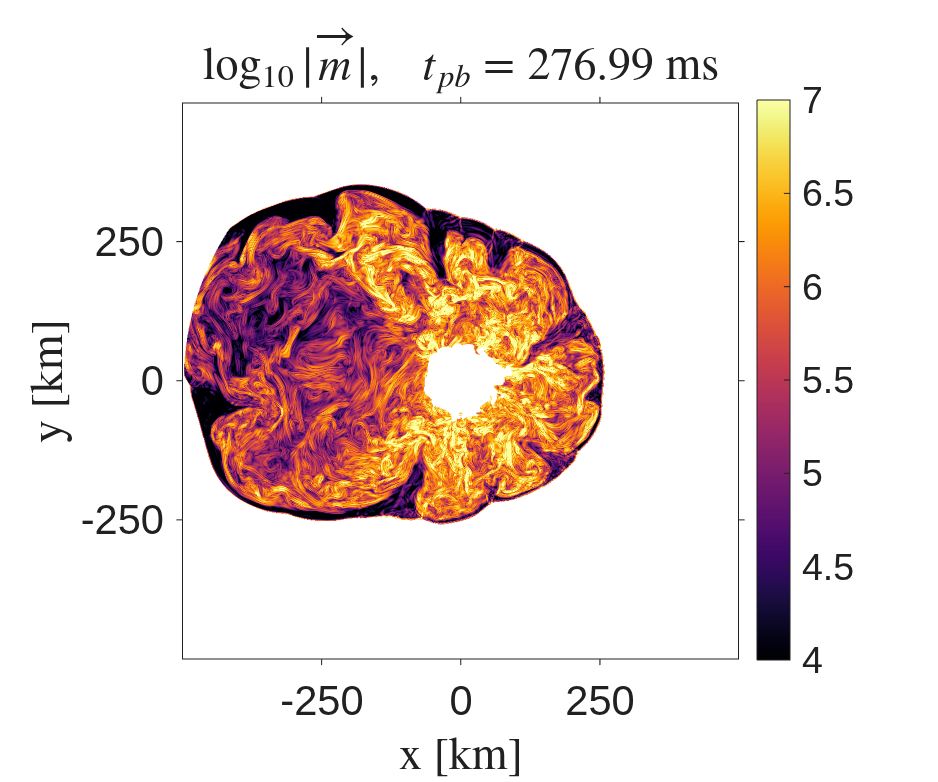}\hspace*{0.0cm}&
\includegraphics[trim=0.0cm 0.1cm 1.0cm 0.1cm, clip, width=6cm]{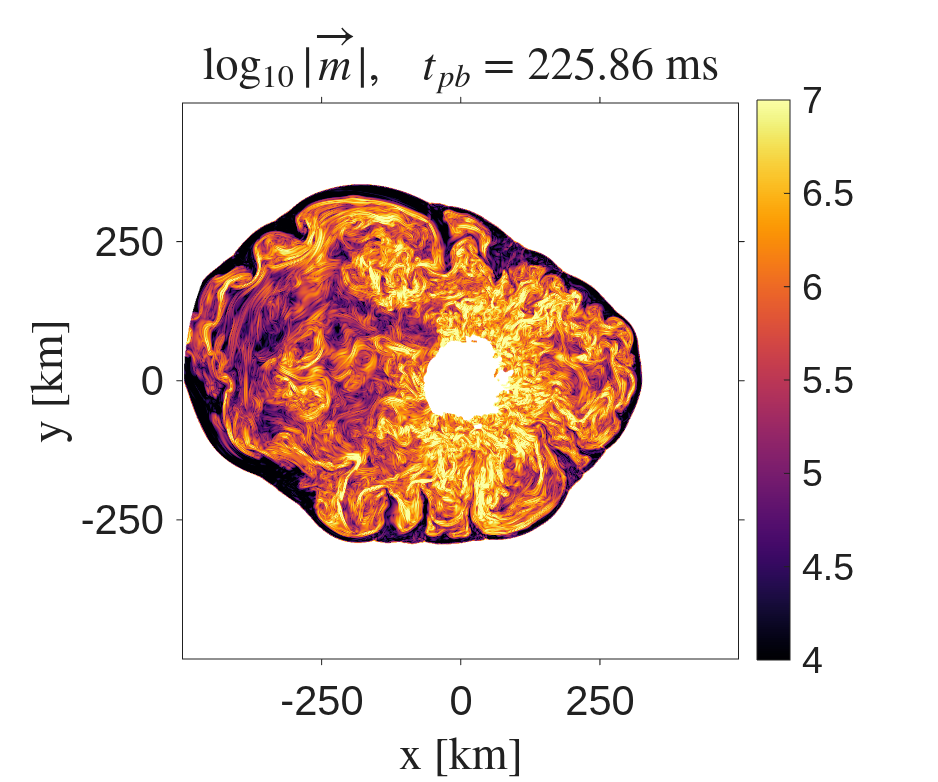}\\
\end{tabular}
\caption{Maps of the magnitude of the baroclinic vector (thermal and magnetic field contributions of equation \eqref{eq:baroclinic}) at different post-bounce times for the s20-control (left), s20-T10 (middle), and s20-T12 (right) simulations. 
\label{fig:Baro_Con_s20}
}
\end{figure*}

In Figure \ref{fig:Baro_Con_s20} we show 2D maps of the baroclinic vector (thermal and magnetic field contributions of equation \eqref{eq:baroclinic}) at various snapshots through the three s20 simulations, which the reader should compare with the maps of $1/\beta$ in Figure \ref{fig:plasma_beta_s20}, and the entropy in Figure \ref{fig:entropy_s20}. 
At $t_{\rm pb}\sim 100\;{\rm ms}$ there is little difference between the control, the T10, and the T12 simulations. By $t_{\rm pb} \gtrsim 150\;{\rm ms}$, the control and the T10 simulations still appear quite similar, while the magnitude of the baroclinic vector is almost an order of magnitude larger in the T12 simulation. 
Another $\sim 40\;{\rm ms}$ later (at $t_{\rm pb} \gtrsim 190\;{\rm ms}$), the s20-T10 is now noticeably different from the s20-control with a much larger volume of the gain region dominated by a large baroclinic vector, but not as large as in the T12. 
Finally, at the end of our simulations, the baroclinic vector has reached similar magnitude in the s20-T10 and in the s20-T12 simulations. 
In the control, the baroclinic vector is largest around downflows (compare with the same snapshot shown in Figure \ref{fig:entropy_s20}); while in the T10 and T12 simulations, it is largest at the inner surface of the gain region and in the spaces between downflows where $1/\beta$ shows the field is highly tangled (see Figure \ref{fig:plasma_beta_s20}). At the end of our simulations the baroclinic vector in the T10 and T12 simulations is smallest in the downflows and in the centers of the large convective plumes. 
These figures show that the principle effect of the magnetic field in the simulations is to change the vorticity of the fluid in the gain region. The effect of the field upon the fluid vorticity is much more significant than the changes due to the energy exchange between field and fluid.

\begin{figure}
\includegraphics[width=\linewidth]{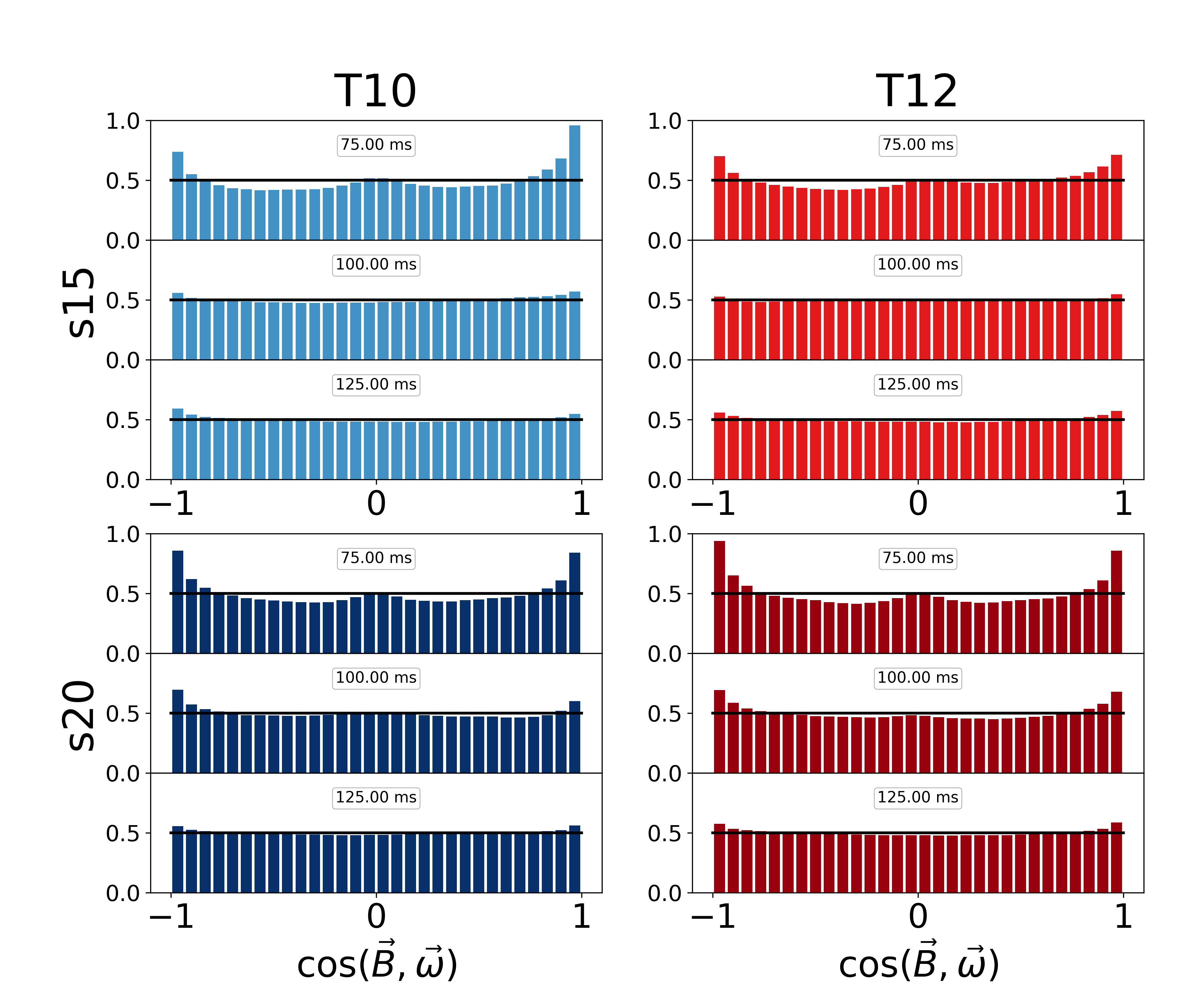}
\caption{Distribution of the angle between $\vec{B}$ and $\vec{\omega}$ for select post bounce times. The black horizontal line in each subpanel represents a uniform distribution. 
Top row: s15 simulations. Bottom row: s20 simulations. 
Left column: T10 simulations. Right column: T12 simulations.
\label{fig:vortangle}
}
\end{figure}

It has previously been suggested that the similarity of the equations (\ref{eq:boverrho}) and (\ref{eq:omegaoverrho}) for the magnetic field and vorticity might indicate they evolve similarly. In the simulations by Endeve \emph{et al.} \cite{Endeve_2012} the authors found the distribution of the angle between the vorticity and the magnetic field had two peaks at zero and $\pi$ supporting this conjecture---see also \cite{1996JFM...306..325B}. 
In Figure \ref{fig:vortangle} we show the results from our investigation of whether the magnetic field and vorticity align in the gain region as a function of post-bounce time in our simulations.
The black horizontal line in the figure is the uniform distribution that one would expect if the two vectors were random and uncorrelated. Our results indicate there is only a slight preference for alignment and anti-alignment of the vorticity and magnetic fields when considered over the entire gain region in all our simulations. 
There is no noticeable difference between the T10 simulations and T12. 
In Figure \ref{fig:Eb} we saw that some sort of equilibrium of the magnetic field with the fluid after $t_{\rm pb} \approx 125\;{\rm ms}$ (for the T12 simulations) / around $t_{\rm pb} \approx 277\;{\rm ms}$ (for the s20-T10 simulation). Thus, it is perhaps not surprising that the vorticity and magnetic field orientations are not correlated after those times. 
It is interesting to note that there is also no strong alignment before this equilibrium is reached (see e.g.\ at $t_{\rm pb} \approx 100\;{\rm ms}$ in the T12 simulations or any time before $t_{\rm pb} \approx 277\;{\rm ms}$ in the T10 simulations), supporting the notion that the vorticity and the magnetic field do not grow together.


\begin{figure*}
\includegraphics[width=0.8\textwidth]{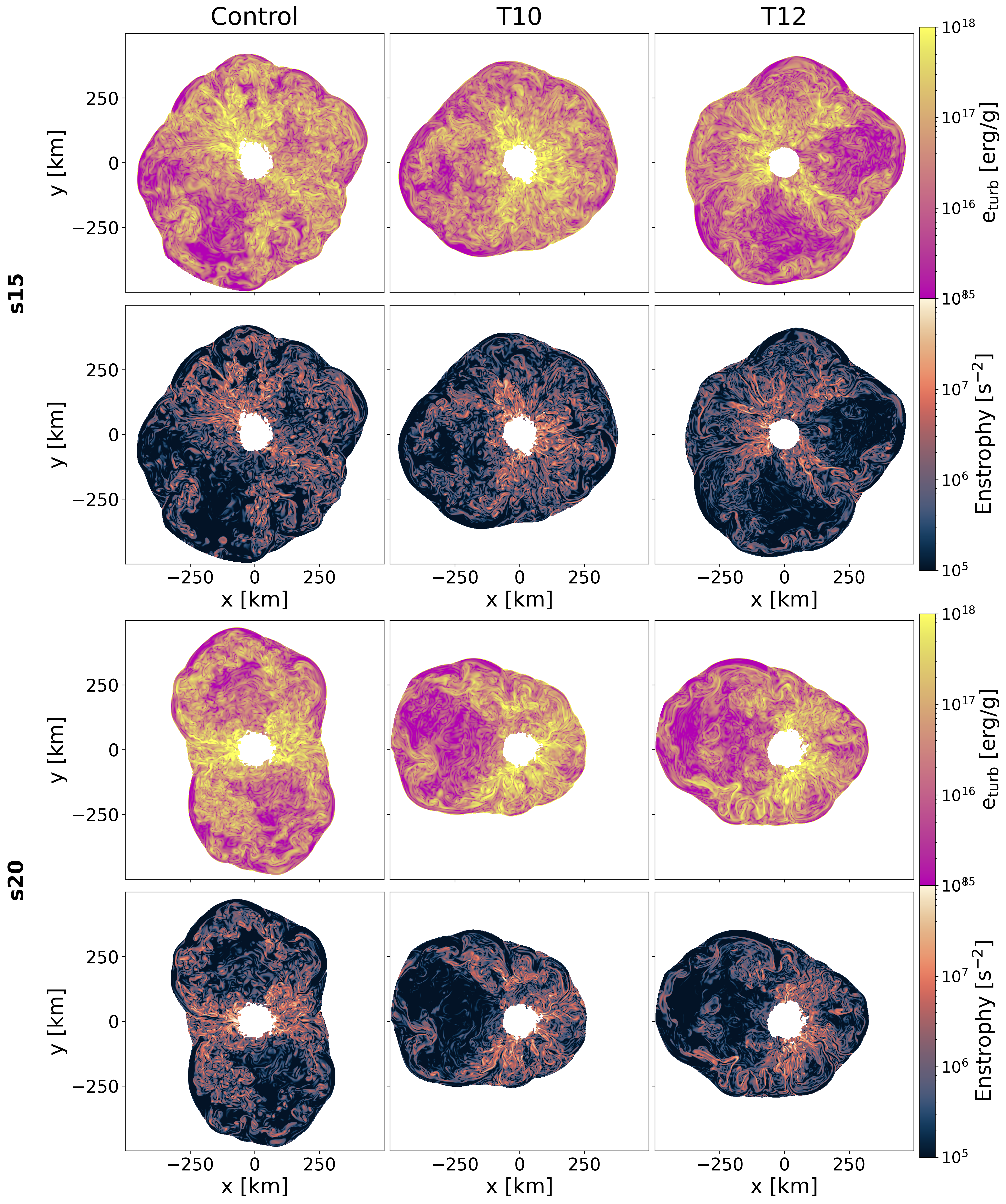}
\caption{Maps of specific turbulent kinetic energy (first and third rows) and enstrophy (second and fourth rows) at the final simulation time of each our six simulations. 
Left column: Control simulations without magnetic field and rotation. 
Middle column: T10 simulations. 
Right column: T12 simulations. 
Top two rows: s15 simulations. 
Bottom two rows: s20 simulations. 
\label{fig:TKE-enstrophy-lastslices}
}
\end{figure*}

The enstrophy of the fluid, $\epsilon_{\omega}$, is defined to be $\epsilon_{\omega} = ( \vec{\omega} \cdot \vec{\omega} )/2$. 
In Figure \ref{fig:TKE-enstrophy-lastslices} we show 2D maps of specific turbulent kinetic energy and of the enstrophy for the last time slices of all six simulations. 
The reader will observe that the maps of enstrophy closely resemble the maps of the specific TKE, with regions of little enstrophy corresponding to the upflows. Indeed, in \cite{2025ApJ...995..109C} we used the good visual agreement between the specific TKE and enstrophy to argue that the LSA method for computing the TKE was a good one.

We can go beyond a visual comparison and compute the correlation between the enstrophy and specific TKE. 
All four simulations show a similar evolution of the correlation coefficient between the specific TKE and enstrophy in the gain regions. At $\sim 100$~ms p.b. the correlation coefficient is $\sim 0.4$ and it rises to $0.6$ at the end of each simulation ($\sim 190$~ms p.b. for the s15 simulations, $\sim 275$~ms p.b. for s20-T10 simulation, and $\sim 220$~ms p.b. for the other two s20 simulations) 
This shows that enstrophy is a good, but not perfect, proxy for the turbulence.

\begin{figure}
    \includegraphics[width=0.8\linewidth]{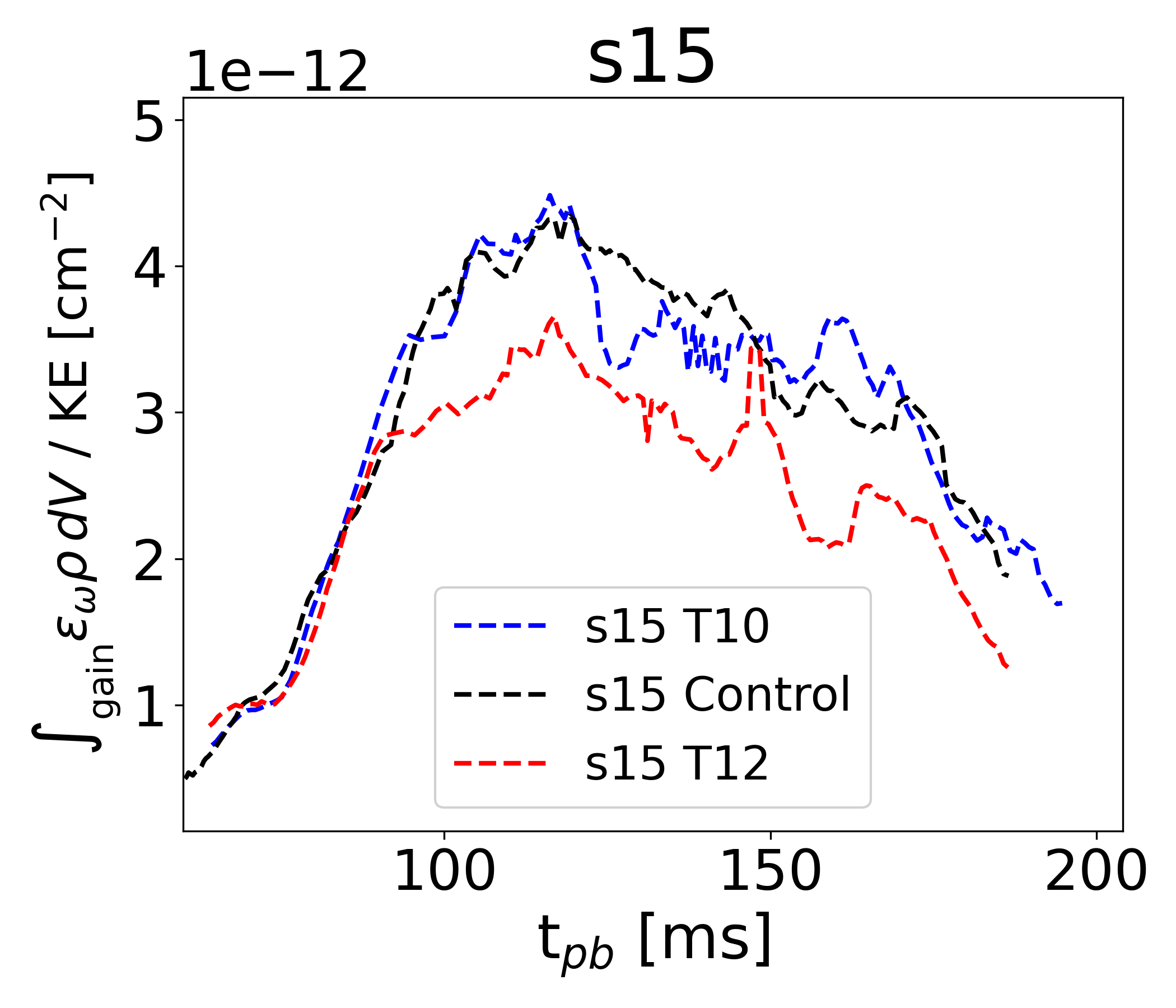}
    \includegraphics[width=0.8\linewidth]{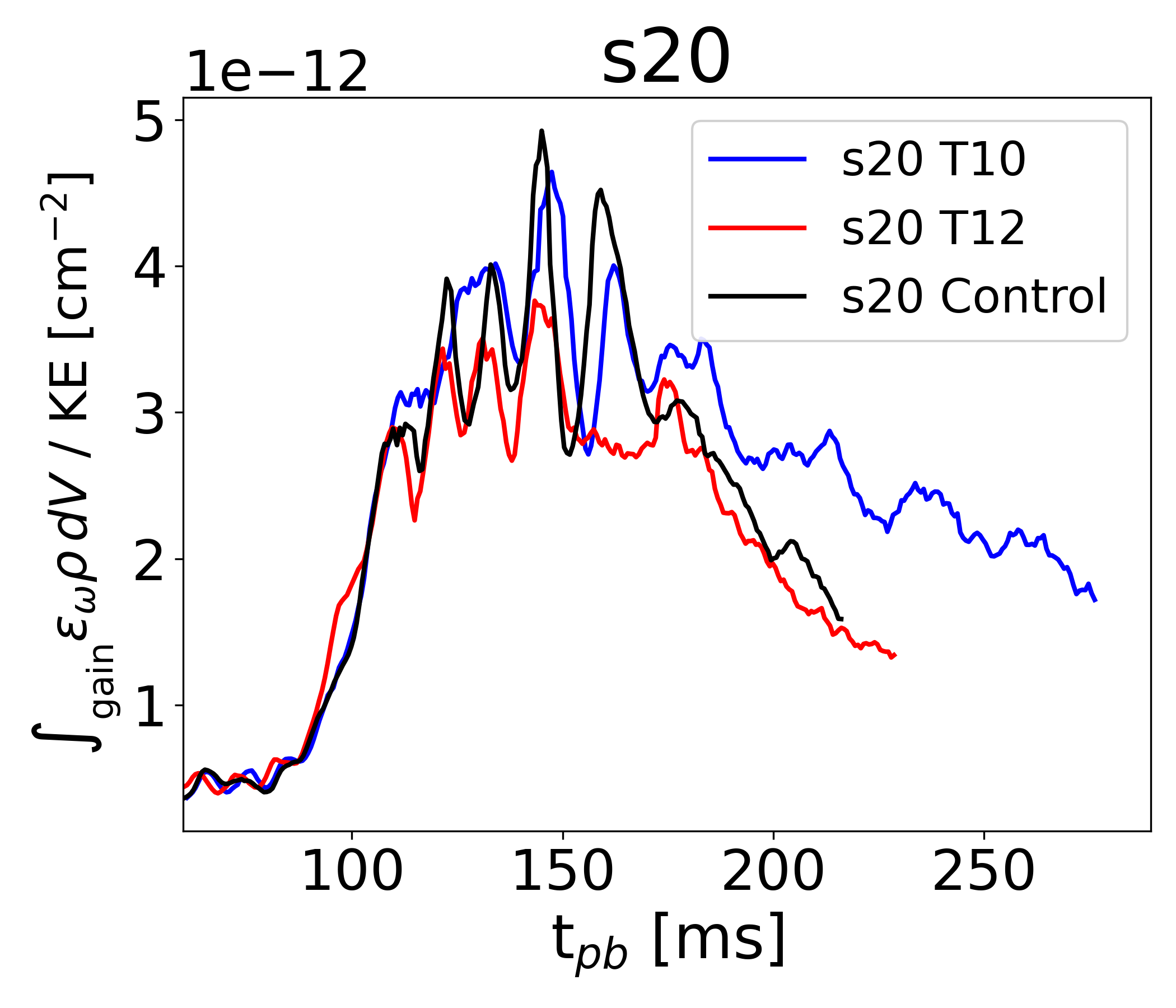}
    \caption{The ratio of enstrophy (computed as enstrophy times density, integrated over the gain region) to total kinetic energy in the gain region as a function of post-bounce time for all six simulations of this study. 
    Top: s15 simulations. 
    Bottom: s20 simulations.}
    \label{fig:enstrophy_ke}
\end{figure}

If so, then from the previously suggested reduction of the TKE in simulations with a magnetic field, we should expect that the growth of the magnetic field energy should also lead to a decrease in the integrated enstrophy over the gain region. In Figure \ref{fig:enstrophy_ke} we show the ratio of the integral of $\rho\, \epsilon_{\omega}$ over the gain region to the integrated kinetic energy of the gain region for all six simulations. We add the additional factor of the density in the enstrophy integral 
in order to compensate for the compression of the fluid towards the inner boundary of the gain region. The figure shows that for both the s15 and s20 progenitors, the control and T10 simulations have ratios which are very similar until $t_{\rm pb} \approx 200\;{\rm ms}$ when the bottom panel of Figure \ref{fig:Eb} indicates the magnetic field approaches an equilibrium with the fluid. In contrast, the T12 simulations have a ratio which is smaller by $\sim 30\%$ relative to the other two simulations with the same progenitor. Thus, when the initial magnetic field is strong, the  field appears to have the effect of reducing the net enstrophy in the gain region supporting the previous suggestion that the field reduces the amount of turbulence.

\section{Conclusions and Discussion} \label{sec:conclusions}

The magnetic field, the turbulence and the vorticity/enstrophy in a core-collapse supernova are interlinked, as demonstrated in equations (\ref{eq:dEmagdt})
and (\ref{eq:domegadt}) -- (\ref{eq:omegaoverrho}). 
To explore the connection between these quantities we have undertaken six 3D MHD simulations using two progenitors and three different initial magnetic field strengths. Our main findings are the following. 
\begin{itemize}
\item The results of our simulations indicate the magnetic field energy relative to the fluid kinetic energy in the gain region grows exponentially over time until it reaches a quasi-stationary value in the range of $5-10\%$. For the two T12 simulations this equilibrium is reached well before the shock is revived, in the two T10 simulations the equilibrium is reached well after the shock revival. The field energy does not appear to greatly affect the  dynamics of the supernova or the shock revival time. 
\item The energy of the magnetic field in the gain region grows at the expense of the fluid's kinetic plus internal energy, but the \emph{net} energy flow into the field is a small fraction of the energy turnover. Using the local spatial average (LSA) method, we find evidence for a reduction of the turbulent kinetic energy of the fluid in the gain region. 
\item The structure of the magnetic field in the gain region quickly becomes a tangled mass of flux ropes. Within this mass, we find isolated pockets of fluid where the field pressure equals, and even exceeds, the thermal pressure. 
\item The highly-tangled field structure means the field is the dominant contribution to the baroclinic vector, the source of fluid vorticity. In the simulations with a magnetic field, the baroclinic vector is large over a much greater volume of the gain region compared to the simulations without a field. 
The vorticity/enstrophy in the fluid is concentrated around the downflows that thread the gain region. We confirm that the enstrophy and turbulent kinetic energy show good, but not perfect, correlation over the gain region allowing enstrophy to be used as a proxy for the turbulence.
\item Although governed by very similar transport/evolution equations to the magnetic field, the fluid vorticity and magnetic field in the gain region become randomly aligned very soon (within $\approx 20\;{\rm ms}$) after convection begins at $80 - 100\;{\rm ms}$ post-bounce. This random alignment occurs in both the T12 simulations \emph{and} both the T10 simulations.
\item We find that the integral of the fluid density times the enstrophy over the gain region decreases relative to the total kinetic energy in the gain region in simulations with magnetic fields. This reinforces our claim that the magnetic field reduces the amount of turbulent kinetic energy in the gain region. 
\end{itemize}
Our findings should be compared to previous studies. Like us, both Sykes and M{\"u}ller \cite{2025PhRvD.111f3042S} and Varma and M{\"u}ller \cite{2026MNRAS.tmp..593V} also saw that the magnetic field energy density was only a few percent of kinetic equipartition in their simulations and did not greatly affect the shock revival time. However our results differ from those by Matsumoto, Takiwaki and Kotake \cite{2024MNRAS.528L..96M} who concluded that the magnetic field had a significant effect in their simulation and was responsible for the earlier shock revival in their rotating+magnetized simulation compared to the non-rotating/non-magnetized control. Our results can also be compared with Endeve \emph{et al.} \cite{Endeve_2012,2013PhST..155a4022E} although these simulations are not of a supernova because their simulations did not include neutrino heating and thus do not explode. 
Nevertheless these authors found the magnetic field does not noticeably change the dynamics of the simulations, and we similarly find that in the gain region the turbulent kinetic energy is only $\sim 25\%$ of the total kinetic energy (although this value is dependent upon the non-physical parameter choices made in the methods used to calculate the TKE), and also find the magnetic field is highly structured with localized pockets of high magnetic field energy. However we do not observe that the magnetic field and vorticity are preferentially aligned or anti-aligned indicating that something in our simulations --- presumably the neutrino heating / momentum exchange --- breaks the connection. 

In summary, the magnetic field, turbulence and vorticity/enstrophy exhibit a dynamic interaction in the gain region of a core-collapse supernova. Exactly how the field is amplified is unclear although ultimately the energy in the field comes at the expense of the fluid's kinetic plus internal energy. As net field energy in the gain region is growing, locally there is a constant energy exchange back-and-forth that we expect is due to a combination of Alfv\'en, magnetoacoustic, and gravity waves, which we plan to explore in future work. 

The data behind the figures and the scripts to make the figures are available on zenodo \url{10.5281/zenodo.21900438}.

\begin{acknowledgments}
The work at NC State was supported by United States Department of Energy, Office of Science, Office of Nuclear Physics (award number DE-FG02-02ER41216). The work of Michael Redle was partially funded by German Research Foundation (DFG) Research Unit FOR5409 ``Structure-Preserving Numerical Methods for Bulk- and Interface-Coupling of Heterogeneous Models (SNuBIC)" (grant \#463312734). 
This study used Bridges-2 at Pittsburgh Supercomputing Center through allocation PHY200074 from the Advanced Cyberinfrastructure Coordination Ecosystem: Services \& Support (ACCESS) program, which is supported by National Science Foundation grants \#2138259, \#2138286, \#2138307, \#2137603, and \#2138296. This study also used the Extreme Science and Engineering Discovery Environment (XSEDE), which is supported by National Science Foundation grant number ACI-1548562. Specifically, it used the Bridges-2 system, which is supported by NSF award number ACI-1928147, at the Pittsburgh Supercomputing Center (PSC).
Finally, the authors also acknowledge the Texas Advanced Computing Center (TACC) at The University of Texas at Austin for providing computational resources that have contributed to the research results reported within this paper. URL: http://www.tacc.utexas.edu
\end{acknowledgments}


\bibliography{references}


\end{document}